\documentclass[final]{elsarticle}

\usepackage{amsmath}
\usepackage{amssymb}
\usepackage{graphicx}
\usepackage{bm}

\journal{Physics Reports}

\begin{document}

\begin{frontmatter}

\title{Imbalanced Fermi Gases at Unitarity}

\author[ITP]{K.B.~Gubbels}
\ead{K.B.Gubbels@uu.nl}

\author[ITF]{H.T.C.~Stoof}
\ead{H.T.C.Stoof@uu.nl}

\address[ITP]{Institute of Theoretical Physics, University of Cologne,\\ Z\"{u}lpicher Str. 77, 50937 Cologne, Germany}

\address[ITF]{Institute for Theoretical Physics, Utrecht University,\\
Leuvenlaan 4, 3584 CE Utrecht, The Netherlands}

\begin{abstract}

We consider imbalanced Fermi gases with strong attractive interactions, for which Cooper-pair formation plays an important role. The two-component mixtures consist either of identical fermionic atoms in two different hyperfine states, or of two different atomic species both occupying only a single hyperfine state. In both cases, the number of atoms for each component is allowed to be different, which leads to a spin imbalance, or spin polarization. Two different atomic species also lead to a mass imbalance. Imbalanced Fermi gases are relevant to condensed-matter physics, nuclear physics and astroparticle physics. They have been studied intensively in recent years, following their experimental realization in ultracold atomic Fermi gases. The experimental control in such a system allows for a systematic study of the equation of state and the phase diagram as a function of temperature, spin polarization and interaction strength. In this review, we discuss the progress in understanding strongly-interacting imbalanced Fermi gases, where a main goal is to describe the results of the highly controlled experiments.  We start by discussing Feshbach resonances, after which we treat the imbalanced Fermi gas in mean-field theory to give an introduction to the relevant physics. We encounter several unusual superfluid phases, including phase-separation, gapless Sarma superfluidity, and supersolidity. To obtain a more quantitative description of the experiments, we review also more sophisticated techniques, such as diagrammatic methods and the renormalization-group theory. We end the review by discussing two theoretical approaches to treat the inhomogeneous imbalanced Fermi gas, namely the Landau-Ginzburg theory and the Bogoliubov-de Gennes approach.

\end{abstract}


\end{frontmatter}

\tableofcontents

\section{Introduction}
\label{sec:intro}

Many fundamental discoveries have been made in the history of many-body quantum physics, which include new states of matter, unexpected phase transitions, and novel macroscopic quantum effects. These discoveries have in common that the rules of quantum mechanics are no longer restricted to the subatomic world, but also determine the collective behavior of systems that are large enough to be observed with the naked eye. The prime example that illustrates such a discovery is the experimental observation of superconductivity by Kamerlingh Onnes \cite{Onnes11a}. In 1911, he found that the resistance of mercury suddenly dropped to zero at a temperature of
about 4~K. It was not immediately realized that a macroscopic
quantum effect caused superconductivity, an important reason being
that quantum mechanics was still at its infancy. Nearly half a century later
superconductivity could be finally explained from first
principles by Bardeen, Cooper and Schrieffer (BCS) \cite{Bardee57a}, who
showed that in particular the pairing of electrons plays a crucial
role.

Pairing occurs in nature when two
particles stick together due to an attractive interaction. This
attractive force may be strong, so that the pair is deeply bound
and is hard to separate. The attraction may also be weak, so
that the paired particles are much further apart, and a slight
disruption can break the pair up. The group behavior of pairs can be
quite different from their individual behavior, which is
particularly true for fermionic many-body systems. Here, pairing
gives rise to a bosonic degree of freedom, whose statistics are
freed from the Pauli principle that governs the behavior of the
fermions. A macroscopic number of pairs can therefore occupy
the same single-pair quantum state, lifting this state
from the microscopic to the macroscopic world. The macroscopically
occupied quantum state gives rise to the state of matter known as
a Bose-Einstein condensate (BEC), which was already predicted by
Einstein in 1925 to occur in a noninteracting gas of bosons
\cite{Einstein25a}. When Kapitsa, Allen and Misener discovered
flow without friction in the strongly interacting bosonic liquid
${}^4$He in 1938 \cite{Allen38a,Kapits38a}, it took only a few
months before a qualitative link between superfluidity and
Bose-Einstein condensation was first made by London \cite{London38a}.

However, in fermionic systems it took some more time before the
link between superconductivity and condensation was noticed, since it was not until 1956 that the microscopic BCS theory of superconductivity
was finally formulated. This theory showed that a weak
attraction between electrons, mediated by the subtle mechanism of
lattice vibrations in metals (phonons), caused a formation of
loosely bound Cooper pairs, which could subsequently Bose-Einstein
condense \cite{Bardee57a}. The BCS theory of pairing in Fermi mixtures
revolutionized many-body quantum physics. It turned out to
describe a very general mechanism that arises under a wide range
of circumstances. An important reason for this is that most matter
in the world around us is made out of fermions, whether they are
quarks, electrons, protons, neutrons, or more complicated composed
particles, like fermionic atoms or molecules. Moreover, these
fermions only need very little attraction to form a Cooper paired
state. Indeed, Cooper showed that an infinitesimal amount of
attraction is enough in the presence of a Fermi sea
\cite{Cooper56a}. As a result, paired condensates have been
observed or predicted to occur not only in a wide range of
condensed-matter systems, but for example also in ultracold atomic
gases \cite{Stoof96a,Regal04a,Ingusc08a} or in the cores of neutron stars
\cite{Bailin84a,Casalb04a}.

The BCS variational wavefunction itself also turned out to be much
more general than initially expected. The many-body wavefunction
was intended to describe the case of weak interactions between
fermions that leads to the formation of loosely-bound pairs, whose
size is much larger than the average interparticle distance. This
regime is also called the BCS limit. As realized early on by Leggett \cite{Legget80a}, the BCS
Ansatz also describes the case of strong attractions between the
fermions leading to the formation of tightly bound molecules,
whose size is much smaller than the interparticle distance. This
regime is also called the BEC limit. Leggett predicted
that the two extremes are continuously connected by a crossover,
which was verified in recent years in a series of ground-breaking
cold-atom experiments by the group of Jin using $^{40}$K
\cite{Regal04a} and the groups of Ketterle, Thomas, Grimm, Salomon,
and Hulet using $^{6}$Li
\cite{Zwierl04a,Kinast04a,Barten04a,Bourde04a,Partri05a}.

The crossover experiments could be performed in the field of
atomic quantum gases, because they have the unique feature that
the effective interparticle interaction strength can be varied
over an infinite range by means of a so-called Feshbach resonance
\cite{Feshba58a,Stwall76a}. In a Feshbach-resonant collision,
two atoms collide and virtually form a long-lived molecule with a
different spin configuration than the incoming atoms, after which
the molecule ultimately decays into two atoms again. The two
different spin configurations are also called the two different
channels of the resonance. The
scattering properties of the colliding atoms depend very
sensitively on the energy difference of the molecular state with
respect to the threshold of the two-atom continuum, which can be
changed with an applied magnetic field. The theoretical prediction
for the presence of tunable resonances in alkali atoms
\cite{Tiesin93a} and their actual realization in experiments
\cite{Inouye98a} have enormously enhanced the power and scope of
the experimental possibilities, especially in the case of
ultracold Fermi mixtures. To illustrate this we notice that
already shortly after Bose-Einstein condensation (BEC) was
achieved in a dilute gas of bosonic alkali atoms by the group of
Cornell and Wieman in 1995 \cite{Anders95a}, it was predicted that
the superfluid regime could also be reached in a dilute mixture of
fermionic atoms \cite{Stoof96a}. However, since the temperature
for the BCS transition is exponentially suppressed for weak
attractive interactions, the critical temperatures initially
turned out to be beyond the reach of experiments with ultracold
Fermi mixtures. By using Feshbach resonances the interaction could
be effectively enhanced which finally allowed for the observation
of fermionic superfluidity and the BEC-BCS crossover.

The crossover physics is usually studied in a two-component Fermi
mixture with an equal number of atoms in each of the two different
spin states. Then, a maximal number of
Cooper pairs is created, which we can understand in the following
way. At low temperatures two particles predominantly interact
through low-energy collisions for which the angular momentum is
zero, also called $s$-wave collisions. Since for identical
fermions the wavefunction should be anti-symmetric upon particle
exchange, and since the two-particle wavefunction for $s$-wave
scattering is symmetric in coordinate space, we conclude that
$s$-wave interactions can only take place if the wavefunction is
anti-symmetric in spin space. This is impossible to achieve if the
particles only have access to one spin state. As a result, the
fully polarized gas can typically be considered as noninteracting
at very low temperatures, so that pairs between particles of the
same spin are not formed. However, in the case of two spin states
an anti-symmetric combination of the two spin states can be made
and $s$-wave attractive interactions are possible. In the case of
an equal amount of particles in both spin states, each particle
can find another particle of a different spin to pair with, which
leads to the formation of a superfluid after condensation of the
pairs.

\begin{figure}
\begin{center}
\includegraphics[width=0.60\columnwidth]{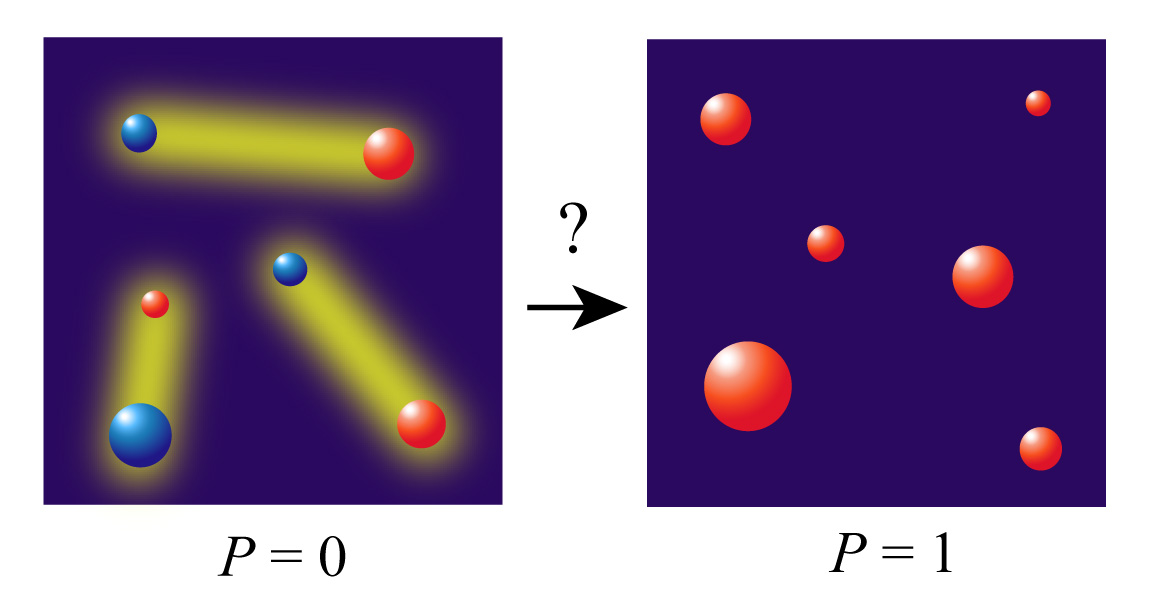}
\caption{\label{fig:sf2n} Illustration of the
superfluid-to-normal transition as a function of polarization $P$
in a two-component Fermi mixture. In the balanced case, $P=0$,
attractive interactions between particles of different spin allow
for the formation of a Cooper-pair condensate. Since the fully
polarized case, $P=1$, can be considered noninteracting, it leads
to the normal state. Obtaining a detailed understanding of the
system's evolution between these two limits is an important goal
of this review.}
\end{center}
\end{figure}

From the above discussion it becomes immediately clear that
something must happen as a function of spin imbalance
or polarization, defined as $P=(N_{+}-N_{-})/(N_{+}+N_{-})$ with
$N_{\sigma}$ the number of particles in spin state $\sigma$.
Indeed, since at ultralow temperatures the gas is in a superfluid
state for a spin-balanced mixture with attractive interactions,
while it is noninteracting and thus in the normal state for a
fully polarized mixture, it must undergo a phase transition as a
function of polarization. This transition is illustrated in Fig.\
\ref{fig:sf2n}. The ultracold atomic Fermi gas was studied experimentally as a function of spin polarization for the first
time by Zwierlein {\it et al.} \cite{Zwierl06a} and Partridge {\it et al.}  \cite{Partri06a},
after which a large amount of experimental and theoretical activity followed. A
main goal of this review is to achieve a detailed understanding of
the phase diagram for the imbalanced Fermi mixture in the strongly
interacting regime. Due to the generality of the pairing mechanism
in fermionic many-particle systems, this topic is also of direct
interest to condensed-matter, nuclear and astroparticle physics
\cite{Bedaqu03a,Casalb04a}.

Although in the past century an impressive amount of knowledge
about pairing in Fermi mixtures has been achieved, many questions
still remain. An important example is the phenomenon of
high-temperature superconductivity, which was discovered in a
certain class of ceramic materials in 1986 \cite{Bednor86a}. Here,
the microscopic nature of the pairing has turned out to be one of
nature's best kept secrets. In order to address the many remaining
fundamental open questions that involve pairing in Fermi mixtures,
it is extremely helpful to have a physical system that is
experimentally very clean, highly tunable and easy to probe. Such
a system then gives rise to a well-defined microscopic hamiltonian
where the interactions between the particles are accurately known.
It is thus an ideal starting point for calculations of
thermodynamic functions that govern the macroscopic properties of
the system. These calculations go hand in hand with detailed
experimental studies, where either experiments set benchmarks for
sophisticated theoretical studies, or where theoretical
predictions give rise to landmark experimental discoveries. In
this way, a much more detailed understanding of fermionic
superfluidity is achieved. Such a system is indeed nowadays
available, namely in the field of ultracold atomic quantum gases.
Whether the mystery of high-temperature superconductivity will
ultimately be unravelled in the world of cold atoms still remains
to be seen, but many fundamental discoveries have already followed
each other up at high speed, while many more are likely to
follow soon \cite{Ingusc08a,Bloch08a,Giorgi08a,Stoof09a}.

\subsection{Ultracold atomic quantum gases}
\label{par:uqg}

When Bose-Einstein condensation (BEC) was ultimately realized in
trapped gases of bosonic atoms \cite{Anders95a,Davis95a,Bradle95a},
a completely new category of systems for studying macroscopic
quantum effects became available. A crucial ingredient for
reaching BEC was the development of laser cooling, with
which atoms are made to absorb and emit photons such that a
momentum transfer takes place \cite{Hansch75a,Winela78a,Neuhau78a,Metcal99a}. This was used to create a friction
force for atoms, slowing them down to a near standstill and
thereby making the gas cloud enter the microKelvin regime. Equally
important was the trapping of the atoms in a magnetic trap \cite{Migdal85a}.
As a result, the gas could be held together without
making contact to material walls that cannot be cooled to such
ultralow temperatures. Moreover, the depth of the magnetic trap
was easily lowered which allows the most energetic atoms to
escape. If the remaining atoms undergo elastic collisions, then
they re-thermalize at a lower temperature, allowing for an
evaporative cooling of the trapped quantum gas \cite{Hess86a, Metcal99a}. This
mechanism is similar to the way in which a cup of coffee gets cold
during lunch time. Evaporative cooling turned out to be extremely
efficient, allowing experimentalists to reach a temperature of 170
nK at a particle density of $2.5 \times 10^{12}$ cm${}^{-3}$,
where BEC was finally observed \cite{Anders95a}. Note that this
means that an ultracold atomic quantum gas is typically about ten
million times thinner than air. Moreover, it also means that a
condensed gas of alkali atoms is strictly speaking in a metastable
state, because at these ultralow temperatures the stable state is
actually a solid. However, the solid formation takes so long
on the time scales of the experiment, that we can consider the
dilute gas to be in thermodynamic equilibrium.

Although the discovery of the Bose-Einstein condensed state of
matter was extremely exciting in itself, it turned out to be only
the beginning for the exploration of new macroscopic quantum
effects in ultracold gases. A reason for the fruitfulness of
atomic many-particle systems is their extreme cleanness, meaning
that there are essentially no impurities in the experiments. This
is in a sharp contrast to solid-state systems, where the
understanding of many-body experiments can be extremely severed
due to an uncontrolled amount of imperfections in the materials.
Disorder in atomic gases is typically negligibly small, unless it
is deliberately added
\cite{Billy08a,Roati08a,Kondov11a}, which led to a very controlled study of
Anderson localization.

This brings us automatically to a second reason for the
fruitfulness of degenerate quantum gases, namely the amazing
amount of experimental control that is currently achieved
\cite{Ingusc08a, Bloch08a}. We have already mentioned the two-channel Feshbach
resonance with which the interatomic interaction strength can be
tuned. Another important example of control is the external
trapping potential of the gas cloud, which can be precisely
tailored experimentally. A particularly convenient setup is
achieved when the atoms are optically trapped with the use of the
strong electric fields in laser beams. The laser beams can then be
made counter-propagating which leads to an intense standing wave
of light. This creates a periodic potential for the atoms due to
the Stark effect, giving rise to a so-called optical lattice
\cite{Jessen92a,Greine02a}. Optical lattices can be used to simulate ionic
lattices, which offers the opportunity to explore various aspects
of condensed-matter physics in the very clean environment of
ultracold atoms \cite{Bloch08a}. The depth of the periodic potential
is now easily tunable by varying the laser intensity, while the
period of the lattice is directly related to the frequency of the
laser light. An even more flexible possibility to create arbitrary
potential landscapes for the atoms is achieved by shining a laser
onto a holographic mask \cite{Bakr09a}.

In the so-called Hubbard model \cite{Hubbar63a}, a paradigm in
condensed-matter physics, particles are allowed to tunnel between
adjacent lattice sites and they have an on-site interaction. When
an interacting atomic gas at ultralow temperatures is loaded into
an optical lattice, then an essentially perfect manifestation of
the Hubbard model is realized. This was first realized by Jaksch
{\it et al.} \cite{Jaksch98a}, who treated repulsively interacting
bosons. Considering the case of an equal number of atoms and
lattice sites, we have that the bosons are expected to form a delocalized
superfluid state when the tunneling strength $t$ is dominant.
However, when the on-site interaction strength $U$ is dominant,
the ground state is given by the so-called Mott-insulator state,
which has precisely one localized atom at each lattice site. At
zero temperature, a transition occurs between these two phases as
a function of $t/U$, which is solely driven by quantum
fluctuations and not by thermal fluctuations \cite{Fisher89a}. This quantum phase transition
was observed in a
landmark atomic physics experiment by Greiner {\it et al.}
\cite{Greine02a}. Here, the tunneling strength was varied by
simply changing the laser intensity. Note that an analogous
control over the tunneling would be hard to achieve in a
condensed-matter system. An ultracold Fermi mixture can also be
loaded into an optical lattice, which leads to a realization of
the fermionic Hubbard model.  This has led to the observation of the Mott-insulator phase with ultracold fermions \cite{Jorden08a,Schnei08a}.  In the absence of doping the ground state of the Hubbard model is predicted to be the anti-ferromagnetic N\'{e}el state, while the doped fermionic Hubbard model has
been conjectured to encompass high-temperature $d$-wave
superconductivity \cite{Anders87a}. It would be exciting to study these phenomena in the controlled environment of ultracold atomic gases. However, to this end a further decrease in the temperature must be achieved experimentally, which is a very challenging task.

\begin{figure}
\begin{center}
\includegraphics[width=0.75\columnwidth]{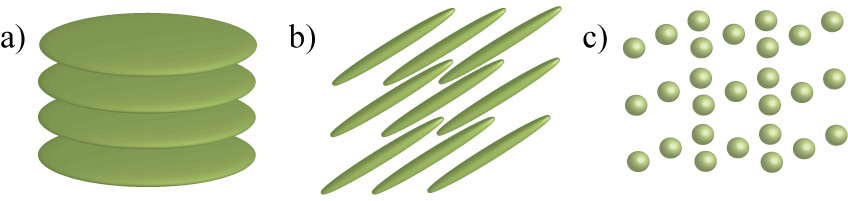}
\caption{\label{fig:optlat} Impression of several configurations that
are realizable with optical lattices. a) If an intense standing
wave of laser light is applied in one direction, then the
potential landscape for the atoms becomes a stack of
quasi-two-dimensional `pancakes'. b) If a second standing wave is
added in a perpendicular direction, then an array of
quasi-one-dimensional `cigars' arises. c) If counter-propagating
laser beams are added in the third direction, then a lattice of
point-like sites arises.}
\end{center}
\end{figure}

There are many more interesting applications of optical lattices.
For example, they can be used to create low-dimensional quantum
gases. Consider a lattice which is very steep in, let's say, the
$x$ direction, such that the ultracold particles can only occupy
the lowest-lying quantum state with an energy $\hbar\omega_x/2$ in
this direction. Then, the kinetic degrees of freedom are frozen
out in the $x$ direction, which effectively lowers the dimension
of the system. In particular, with a very steep optical lattice in
two directions it is possible to create a two-dimensional array of
effectively one-dimensional tubes, while a very steep optical
lattice in one direction leads to a one-dimensional stack of
two-dimensional pancakes, as shown in Fig.~\ref{fig:optlat}.
Low-dimensional quantum gases give rise to
rise to intriguing strongly-correlated behavior that is very
different from the three-dimensional case \cite{Giamar03a}. Experimentally observed
examples include the exotic Berezinskii-Kosterlitz-Thouless phase
transition in two dimensional gases
\cite{Hadzib06a,Schwei07a,Clade09a}, the coherence dynamics of one-dimensional Bose gases \cite{Hoffer07a}  and exotic pairing of spin-imbalanced Fermi gases in one dimension \cite{Liao10a}.

But the control in ultracold atomic gases goes further. Feshbach
resonances are for example not restricted to $s$-wave scattering,
so that also resonances of higher angular momentum have been
observed. This has led to the formation of $p$-wave Feshbach
molecules \cite{Gaeble07a, Gurari07a, Gubbel07a}, which upon condensation would give rise
to superfluidity in a nonzero angular momentum state. Analogous to
liquid ${}^3$He, this superfluid has a more complex
order-parameter structure \cite{Legget75a,Vollha90a}, which may
give rise to exotic features such as the presence of Majorana
fermions in vortex excitations, topological phase transitions and topologically protected quantum computing
\cite{Read00a,Gurari07a,Kitaev03a}. A different research direction is the
creation of condensates, where the particles interact through
long-ranged anisotropic dipole-dipole interactions. Dipolar
effects have already been observed in condensates of ${}^{52}$Cr
atoms \cite{Lahaye07a}, whose magnetic moment leads to weak dipole
interactions. The interaction is much stronger for heteronuclear
molecules that have a permanent electric dipole moment. An
ultracold gas of ground-state ${}^{40}$K${}^{87}$Rb molecules has
been achieved by associating ${}^{40}$K and ${}^{87}$Rb atoms
using a Feshbach resonance and then optically pumping the
${}^{40}$K${}^{87}$Rb dimers to their rovibrational ground state
\cite{Ni08a}. The interesting possibilities with dipolar molecules
vary all the way from the stabilization of supersolid phases
\cite{Goral02a} to the possibility of quantum computing
\cite{Demill02a}. Another promising line of research is to create effective artificial vector potentials for cold atoms using Raman transitions \cite{Lin09a}, which have among others resulted in artificial magnetic fields \cite{Lin09b} and spin-orbit coupling \cite{Lin11a}  for ultracold neutral atoms. As a result, the rich physics of gauge theories for charged particles can be studied in the versatile atomic physics environment. A final example of control that is of particular relevance to this review is the controlled preparation of an atomic gas in a selective set of internal
quantum states. In a two-component ${}^6$Li mixture it is possible
to precisely control the atom number in each of lithium's two
lowest hyperfine states by inducing nuclear spin flips
\cite{Zwierl06a, Partri06a}. As a result, the fundamental phase
diagram of the two-component Fermi mixture can be studied as a
function of temperature, polarization and interaction strength, of which the three-dimensional case is a central topic of this review.
Moreover, if two different atomic species are used, then a fourth
axis enters the phase diagram, namely that of a mass imbalance.

To make a long story short, the possibilities seem endless and the
field of ultracold atoms is able to address fundamental questions
about many-body quantum physics in great detail. Therefore, atomic
quantum gases are sometimes also referred to as ideal quantum
simulators. They allow for systematic studies of an enormous
variety of hamiltonians, ranging from weakly interacting to
strongly interacting, from one dimensional to three dimensional,
from disordered to clean, from homogeneous to periodic, where the
microscopic parameters are always precisely known and widely
tunable.

\subsection{This review}

In this review, we consider three-dimensional two-component Fermi gases with attractive interactions so that pairing is expected to occur. However, the equal-density Cooper pairing mechanism is frustrated when an increasing spin imbalance is introduced to the system. We are interested in the way the system responds to this kind of frustration, where multiple scenarios might be realized, as we discuss in the next section.  In recent years, many important experiments have been performed on the imbalanced Fermi gas, after this system was first realized experimentally in the field of ultracold quantum gases at MIT by Zwierlein {\it et al.} \cite{Zwierl06a} and at Rice University by Partridge {\it et al}. \cite{Partri06a}. The initial studies of these groups focussed on experimentally determining the phase diagram of the strongly interacting Fermi gas as a function of spin polarization and temperature \cite{Partri06b,Shin08a}. Later, the imbalanced atomic Fermi gas was also experimentally studied at ENS, where Nascimb\`{e}ne {\it et al}. accurately measured the collective modes \cite{Nascim09a} and the equation of state of the imbalanced Fermi gas \cite{Nascim10a,Nascim11a}. In most recent experimental studies, also nonequilibrium aspects of the imbalanced Fermi gas are considered, such as spin transport  \cite{Sommer11a,Sommer11b} and metastable states \cite{Liao11a}.

An early excellent review that focusses on the properties of imbalanced Fermi gases in mean-field theory at zero temperature along the BEC-BCS crossover can be found in Ref.\ \cite{Radzih07a}. Other shorter, but very insightful reviews on the imbalanced Fermi gas are Refs.\ \cite{Radzih10a, Chevy10a}. In the present long review, we mainly focus on the equilibrium properties of imbalanced Fermi gases in the unitarity limit. A central difference with Ref.\ \cite{Radzih07a} is that we spend much attention to the effects of interactions beyond mean-field theory by discussing diagrammatic methods and renormalization-group theory. Moreover, we consider the effects of the trap beyond the local-density approximation by discussing the Landau-Ginzburg approach and the Bogoliubov-de Gennes equations. Finally, we also treat nonzero temperatures and the mass-imbalanced case. The phase diagram of the imbalanced Fermi gas is a topic of fundamental interest to various areas of physics \cite{Bedaqu03a,Casalb04a},
where we can think of neutron-proton pair condensation in nuclear
physics \cite{Alm03a}, or color superconductivity of strongly
degenerate quark matter in the core of neutron stars
\cite{Bailin84a}. We start out by qualitatively discussing which kind of paired phases might occur.

\subsubsection{Unusual superfluid phases}

The `standard' way in which fermions form a superfluid is when identical particles (such as electrons) of different spin undergo $s$-wave attractive interactions and form pairs \cite{Bardee57a}. These Cooper pairs constitute a bosonic degree of freedom, and at low enough temperatures  they condense forming a superfluid.
We note here that although in three dimensions the formation of a Bose-Einstein
condensate typically goes hand in hand with the onset of
superfluidity, they are not identical phenomena. For example, the
enhancement of fluctuation effects in lower dimensions can destroy
a condensate while keeping the property of superfluidity intact.
Moreover, in three dimensions the ideal Bose gas is fully
condensed at zero temperature, but it is usually not considered to
be a superfluid \cite{Ketter08a}. The latter is because its critical
velocity, which is the highest velocity for the gas to flow
without friction, is then equal to zero. In this review we
consider the interacting three-dimensional case, so that we
use the terms Cooper-pair condensate and superfluid state
interchangeably.

When there is a population imbalance between the two
spin states of the Fermi mixture, not all particles can
find a partner to pair up with.
In this review we will study the phase
diagram of the spin-imbalanced mixture for strong attractive interactions,
both in the presence and the absence of a trapping
potential. Due to the enhanced richness of this system, we should consider  at least two
additional superfluid phases that are not present in the
spin-balanced case. First of all, we find the occurrence of a
phase-separated phase at temperatures close to zero. This phase
can arise because there is a discontinuous or first-order
transition in the system between a superfluid state that has a
small polarization and a normal state that has a high
polarization \cite{Bedaqu03a,Shin08a}. The phase-separated state then occurs when the total
polarization of the system is in between these two extremes. As a
result, the system spatially separates into a superfluid domain
and a normal domain, between which there is a first-order
interface, i.e. a domain wall. In the presence of a trap, the
superfluid domain is then found in the center of the trap, while
the normal domain surrounds it.

The second less-standard superfluid phase that we consider is the so-called Sarma phase \cite{Sarma63a,Gubbel06a}, which has a gapless excitation spectrum. As a result, this superfluid phase is polarized even at zero temperature. It is also called the interior-gap phase, or the breached-pair phase \cite{Liu03a,Forbes05a,Parish07a}. The Sarma phase is a superfluid with one or more Fermi surfaces, so that it is different from the gapped BCS phase which has no Fermi surface, and the two phases are separated by a quantum phase transition at zero temperature. For the mass-balanced case, we see later on that mean-field theory predicts
the Sarma phase to become unstable at very low
temperatures, preventing the observation of the quantum phase
transition. At nonzero temperatures, the distinction between the Sarma phase and the BCS phase is not sharp because temperature causes both phases to be polarized and to not have a sharp Fermi surface. As a result, the gapless regime and the gapped regime are then connected by a crossover \cite{Baarsm10a}. In the case of a large mass-imbalance, the Sarma phase becomes stable at zero temperature according to mean-field theory  \cite{Parish07a}. For completeness, we mention also other exotic paired phases that have been predicted to enter the phase diagram of the spin imbalanced Fermi gas, namely states with a deformed Fermi surface \cite{Sedrak07a},  $p$-wave superfluidity \cite{Bulgac06a,Patton11a}, and supersolid phases \cite{Mora05a,Bulgac08a,Radzih10a}.

Although in this review we primarily study the Fermi gas in the unitarity limit as a function of temperature and spin polarization, there are more parameters to be varied. The phase diagram can for example also be explored as a function of interaction strength, and as a function of the mass imbalance between the two different components of the mixture. The situation of a mass imbalance is currently being studied by many experimental groups, where in particular the ${}^6$Li-${}^{40}$K mixture seems promising \cite{Taglie08a,Wille08a,Voigt09a,Tiecke10a}. However, pair condensation with different fermion masses has not been observed yet. The mean-field phase diagram of the ${}^6$Li-${}^{40}$K mixture not only encompasses the rich
physics from the mass-balanced case, but it has even more structure \cite{Gubbel09a}. Namely, next to the presence of Sarma physics and phase separation, mean-field theory also predicts a Lifshitz point in the phase diagram for the unitarity limit, as we see explicitly in Section \ref{par:mi}. A Lifshitz point also enters the phase diagram in the mass-balanced case, but then at weak interaction strengths, making the corresponding Lifshitz temperature very low \cite{Sheehy06a, Son06a, Yoshid07a}. At
the Lifshitz point, the effective mass of the Cooper pairs turns negative,
which signals a transition to an inhomogeneous superfluid. This
exotic possibility was first investigated by Larkin and Ovchinnikov (LO) \cite{Larkin65a}, who considered
a superfluid with a standing-wave order parameter, namely $\Delta
({\bf R})= \Delta_{0} \cos ({\bf K}\cdot {\bf R})$. The order
parameter spontaneously breaks translational symmetry, and because
the atom densities depend on the absolute value of the superfluid
order parameter, the densities also oscillate in space. Moreover, the phase is typically energetically more favorable than a plane-wave order parameter $\Delta ({\bf R})= \Delta_{0} e^{i {\bf
K} \cdot {\bf R}}$, which was considered independently by Fulde
and Ferrell (FF) \cite{Fulde64a}. The FF phase leads to a
propagating superfluid that spontaneously beaks time-reversal
symmetry, while it does not give rise to oscillating particle
densities.

The combination of superfluidity with crystalline order in the densities makes the LO phase a particular supersolid \cite{Cheste70a,Bulgac08a}. To be more precise, the LO phase can also be called a density wave of a Cooper-pair condensate \cite{Chen04a,Agterb08a,Radzih09a,Radzih11a}, since the order parameter is a product of a pairing amplitude $\Delta_{0}$ and a standing wave $\cos ({\bf K}\cdot {\bf R})$. This is in contrast to the supersolid phases that are usually considered, for which the superfluid order and the crystalline order are completely independent of each other. A special property of the pair-density wave is that the order parameter structure not only allows for the occurrence of integer-valued vortices (as for ordinary superfluids) and the occurrence of dislocations (as for ordinary density waves), but also for composite objects that are half vortex and half dislocation \cite{Agterb08a,Radzih11a}. A careful study of the low-energy fluctuations of the LO phase, which breaks continuous rotational and translational symmetry, reveals that it is actually a phase with only algebraic long-range order, also known as a superfluid smectic liquid crystal \cite{Radzih11a}. Moreover, upon melting of the crystal other exotic phases can arise, such as fractionalized Fermi liquids and superfluids of pairs of Cooper pairs \cite{Radzih11a}. The physics of the FF and LO phases have intrigued the condensed-matter community for many decades, but it has been
experimentally very challenging to prove the existence of these phases unambiguously \cite{Casalb04a}.
The same holds for supersolidity in general, where a famous example is the heavily debated experiment
with ${}^4$He by Kim and Chan \cite{Kim04a}. A promising way to observe FFLO phases directly is to study the one-dimensional spin-imbalanced Fermi gas \cite{Liao10a}, for which the FFLO phases have been predicted to occupy large parts of the phase diagram \cite{Orso07a,Hu07b,Tezuka08a}.

\subsubsection{Outline}

Having mentioned several superfluid phases of interest that we encouter in this review, we now give a more precise outline of the topics that are to be treated. We begin in Section \ref{sec:scat} with a comprehensive discussion of two-body scattering in ultracold atomic gases. This is necessary for understanding in more detail the interaction mechanisms on the two-body level before trying to tackle the many-body properties. In particular, we focus in this section on Feshbach resonances, which are so crucial for ultracold atom experiments these days. We end the section with a qualitative discussion of an important experimental example that reveals the power of using Feshbach resonances, namely the study of the BEC-BCS crossover. Having discussed the two-body physics, we turn to the central topic of this review in Section \ref{sec:mf}, namely the imbalanced Fermi gas. We start out by giving a historical overview of the early atomic-physics experiments that have been performed with this system up to the present status of the field. This overview will give us a taste of the physical phenomena that we wish to explain.  Next, we discuss extensively the simplest theoretical approach to describe pairing in Fermi systems, namely the Bardeen-Cooper-Schrieffer (BCS) mean-field theory. We find that this simple approach can already account qualitatively for many features observed in the experiments done in the strongly-interacting regime. We calculate the mean-field phase diagram for the solely spin-imbalanced case in the unitarity limit both for the homogeneous case and for the trapped case. For the latter, the local-density approximation is applied. We end this section with the mean-field phase diagram of the mass-imbalanced ${}^6$Li-${}^{40}$K mixture.

Although mean-field theory is useful for obtaining a qualitative understanding of the relevant physics in imbalanced Fermi gases, it is not very accurate quantitatively in the unitarity regime. In Section \ref{sec:diag} we discuss diagrammatic approaches that go beyond mean-field theory and that considerably improve agreement with experiments and Monte-Carlo calculations. First of all, we treat the Gaussian fluctuations of the BCS order parameter, also known as the Nozi\`{e}res-Schmitt-Rink (NSR) approximation, which can be evaluated exactly and gives a significant improvement in the description of the BEC-BCS crossover for balanced Fermi gases compared to mean-field theory. For example, it leads to a rather accurate description of the critical temperature curve as a function of interaction strength, whereas mean-field theory fails to describe the critical temperature on the BEC side. Despite this remarkable success, the NSR method unfortunately fails upon application to the imbalanced Fermi gas. Therefore, we also discuss other techniques. First of all, we calculate the self-energy of fermions using a many-body transition-matrix approach. This self-energy can be consequently incorporated into an equation of state that is in very good agreement with experiments and Monte-Carlo simulations. Moreover, in Section \ref{sec:rg} we apply the wilsonian renormalization scheme to the imbalanced Fermi gas in two different ways. Namely, first we apply it directly to the fermionic action, in order to not only include the effect of fermionic selfenergies, but also of the screening of the interaction by particle-hole excitations. We use this method to calculate more accurately the tricritical point of the phase diagram in the unitarity limit. We also apply the wilsonian renormalization scheme to the action for the Cooper pairs, which can be exactly obtained from the fermionic action by a so-called Hubbard-Stratonovich transformation. We find that by including the interaction between the Cooper pairs, the problems encountered with the Nozi\`{e}res-Schmitt-Rink approximation are solved.

Having obtained a rather detailed understanding of the physics beyond mean-field theory for the homogeneous imbalanced Fermi gas, we turn to the inhomogeneous case. In Section \ref{sec:inh} we describe the trapped Fermi gas beyond the local-density approximation (LDA), because this approximation breaks down near the superfluid-normal interface that is observed for low-temperatures in experiments. We describe two approaches beyond LDA, namely the Landau-Ginzburg theory and the Bogoliubov-de Gennes equations. We use the Landau-Ginzburg approach to compare with experiments and to calculate the resulting surface tension. When using the Bogoliubov-de Gennes approach a proximity effect is observed, meaning that there are oscillations in the order parameter near the superfluid-normal interface, which are absent in the Landau-Ginzurg approach. Finally, we end Section \ref{sec:fin} to summarize our conclusions and to discuss some issues that are still open and that require further research.

This review is meant to be introductory and pedagogical in style, meaning that it is aimed to be comprehensible by graduate students (both theoretical and experimental) entering this particular field of research. Only Sections \ref{sec:diag} and  \ref{sec:rg}  request previous knowledge of quantum field theory in the functional formalism and are therefore mainly focussed on the theoretically interested reader. The other parts can easily be followed without reading these two sections in great detail. Whenever steps in derivations are skipped we have tried to include a reference where the corresponding steps are explicitly taken. We hope that this approach has resulted in a useful review for the reader.

\section{Scattering and Feshbach resonances}
\label{sec:scat}

We start our discussion of interacting atomic Fermi gases by describing in some detail the relevant two-body physics of these systems, which is necessary as an input for the many-body theory. To this end we introduce several useful concepts from general two-body scattering theory. The machinery to deal with scattering problems is well established and can be
found in most textbooks on quantum mechanics, see e.g. Refs.\
\cite{Landau81a,Sakura94a,Bransd00a}. Here, we only briefly
state the results that are of most relevance to us. The main goal of this section is to understand the physical properties of Feshbach resonances, which have been crucial to all experiments with imbalanced Fermi gases. More extensive reviews on Feshbach resonances can be found in Refs.
 \cite{Duine04a,Chin10a}. We end the section by briefly describing qualitatively an important application of Feshbach resonances, namely the study of the BEC-BCS crossover.

\subsection{Single-channel scattering}

Consider two particles of mass $m$ that elastically scatter from
each other under the influence of an interaction potential $V({\bf
r})$ that depends on the relative coordinate ${\bf r}$ of the
particles. If the Schr\"{o}dinger equation is separated into a
part describing the center-of-mass motion and a part describing
the relative motion of the two particles, then the center-of-mass
part behaves like a single free particle of mass $2m$, while the
relative part behaves like a single particle with reduced mass
$m/2$ moving in the potential $V({\bf r})$. Starting point
is the relative Schr\"{o}dinger equation
\begin{equation}\label{eq:scat}
\left\{ \hat{H}_{0} + \hat{V} \right\} | \psi \rangle = E | \psi
\rangle,
\end{equation}
where $\hat{H}_0 =-\hbar^{2}  \boldsymbol{\nabla}^{2}/m$ for the
relative kinetic-energy operator. We start with looking at
solutions where the particles enter the scattering region in a
plane wave state $|{\bf k}\rangle$ with energy
$E=\hbar^2k^2/m=2\epsilon_{\bf k}$, which is conserved in the
elastic collision. The state $|{\bf k}\rangle$ also solves the
Schr\"{o}dinger equation in the absence of the interaction potential. Since Eq.~(\ref{eq:scat}) is the time-independent Schr{\"o}dinger equation,
it may be interpreted as describing a steady-state solution for a
continuous stream of particles in state $|{\bf k}\rangle$
scattering from a potential. It turns out to also describe the
scattering of wavepackets, provided that the typical wavelength of
the packets is much larger than the range of the interaction
\cite{Sakura94a}. This condition is typically well satisfied for
dilute ultracold atomic gases, which are characterized by large de Broglie wavelengths and by interactions that are very short ranged compared to the average interparticle distance.

Eq.~(\ref{eq:scat}) can be formally solved by introducing
scattering states $|\psi^{(+)}_{\bf k} \rangle$ that satisfy the
following recursion relation, also known as the Lippmann-Schwinger
equation,
\begin{equation}\label{eq:lippschwing}
|\psi_{\bf k}^{(+)} \rangle = |{\bf k}\rangle+
\hat{G}_0(2\epsilon_{\bf k }) \hat{V} | \psi^{(+)}_{\bf k}
\rangle,
\end{equation}
where $\hat{G}_0(2\epsilon_{\bf k })\equiv (2\epsilon_{\bf
k}-\hat{H}_{0}+i 0)^{-1}$ with the infinitesimal positive
imaginary part being the standard way to treat the singular nature
of this operator. In a time-dependent formulation, this latter procedure
follows immediately from demanding that the particles were free in
the remote past \cite{Sakura94a}. In scattering theory a central
role is played by the two-body transition matrix, defined through
$\hat{V}|\psi^{(+)}_{\bf k}\rangle\equiv\hat{T}^{\rm 2b
}(2\epsilon_{\bf k })|{\bf k}\rangle$. As a result, Eq.\
(\ref{eq:lippschwing}) leads to
\begin{equation}
\hat{T}^{\rm 2b}(2\epsilon_{\bf k }) |{\bf k}\rangle=
\left\{\hat{V}+ \hat{V}\hat{G}_0(2\epsilon_{\bf k }) \hat{T}^{\rm
2b}(2\epsilon_{\bf k })\right\}|{\bf k}\rangle.
\end{equation}
This form of the Lippmann-Schwinger equation can also be
generalized to an operator equation, whose recursive solution
gives rise to the so-called Born series
\begin{equation}\label{eq:born}
\hat{T}^{\rm 2b}(E) = \hat{V}+  \hat{V}\hat{G}_0(E)
\hat{V}+\hat{V}\hat{G}_0(E) \hat{V}\hat{G}_0(E) \hat{V}+\ldots,
\end{equation}
for a certain energy $E$. More shortly,
\begin{equation}\label{eq:t2b}
\hat{T}^{\rm 2b}(E) = \hat{V}+  \hat{V}\hat{G}(E)\hat{V},
\end{equation}
where $\hat{G}(E)=(E-\hat{H}_{0}-\hat{V})^{-1}$. The last equation
indicates that singularities in the transition matrix are found at
the exact eigenenergies of the two-body problem.

More insight is gained by realizing that away from the scattering
region the influence of the scattering potential becomes
negligible, which leads to a noninteracting radial Schr\"{o}dinger
equation. Here, we demand that the solution describes an incoming
plane wave with wavevector ${\bf k}$ and a freely propagating
outward radial flow of probability flux. It is possible to show
that the flux-normalized solution at large ${\bf r}$ is given
by, see e.g. \cite{Sakura94a},
\begin{equation}\label{eq:scwf}
\psi^{(+)}_{\bf k} ({\bf r}) \simeq  e^{i {\bf k}\cdot {\bf r}} +
f^{\rm 2b}({\bf k}',{\bf k}) \frac{e^{i k r}}{r},
\end{equation}
where ${\bf k}'=k \hat{{\bf r}}$. The two-body scattering
amplitudes then satisfy
\begin{eqnarray}\label{eq:scamp}
f^{\rm 2b}({\bf k}',{\bf k})&=& - \frac{m}{4\pi\hbar^{2}} \langle {\bf k}'
| \hat{V} | \psi^{(+)}_{\bf k} \rangle=- \frac{m}{4\pi\hbar^{2}}
\langle {\bf k}' | \hat{T}^{\rm 2b}(2\epsilon_{\bf k}) |{\bf k}
\rangle \nonumber\\
&=&\sum_{\ell=0}^{\infty}(2
\ell+1)f^{\rm 2b}_{\ell}(k)P_{\ell}(\cos
\vartheta)\nonumber \\
&=& -\sum_{\ell=0}^{\infty}\frac{(2
\ell+1)m}{4\pi\hbar^2}T^{\rm 2b}_{\ell}(k)P_{\ell}(\cos
\vartheta).
\end{eqnarray}
In the second line the partial-wave method was employed to expand
the scattering amplitude and the transition matrix, where
$P_{\ell}(x)$ are the Legendre polynomials and $\vartheta$ is the
angle between ${\bf k}$ and ${\bf k}'$. We note that in obtaining the
expression of Eq.~(\ref{eq:scamp}) the normalization $\langle
{\bf r}|{\bf k} \rangle=e^{i {\bf k}\cdot {\bf r}}$ was used for
the plane waves. This reveals that the ket
$|{\bf k}\rangle$ itself is not dimensionless, which is a consequence
of using the continuum limit for the states in ${\bf k}$
space. Moreover, the expansion in Eq.~(\ref{eq:scamp}) is
useful for a spherically symmetric interaction, for which the
scattering problem is symmetric around the axis of incidence. From
the conservation of angular momentum and probability flux, it
can be shown that \cite{Sakura94a}
\begin{equation}\label{eq:scamp2}
f^{\rm 2b}_{\ell}(k)= \frac{1}{k \cot[ \delta_{\ell}(k)]-i k},
\end{equation}
where $\delta_{\ell}(k)$ is the phase shift of the $\ell$-th
partial wave due to the elastic scattering.

At low energies, the phase shift is governed by the expansion \cite{Landau81a}
\begin{equation}\label{eq:phaseshift}
k^{2 \ell +1} \cot[\delta_{\ell}(k)]=
-\left(\frac{1}{a_{\ell}}\right)^{2\ell+1}+\mathcal{O}(k^2),
\end{equation}
so that for small wavevectors the partial scattering amplitudes
are given by $f^{\rm 2b}_{\ell}(k)\simeq -a^{2\ell+1}_{\ell}k^{2 {\ell} }$
with $a_{\ell}$ the scattering parameter of the $\ell$-th
partial wave. For ${\ell} = 0$, $a_{0}$ is called the $s$-wave scattering length, which from now on we simply denote as $a$, since we only consider  scattering at zero angular momentum. This also avoids confusion with the same symbol often used for the Bohr radius. From the behavior of the partial scattering amplitudes, we thus see that at low momenta the $s$-wave scattering scattering with
zero angular momentum (${\ell=0}$) is dominant. The reason why collisions with higher
angular momentum are suppressed is also readily understood from a
more physical point of view. Namely, for $\ell \neq 0$ the
relative radial Schr{\"o}dinger equation contains a repulsive
centrifugal term $\hbar^2 \ell(\ell+1)/m r^2$, which acts as an
energy barrier. At low temperatures the particles do not have
enough kinetic energy to overcome this barrier and thus the
higher-order partial waves do not feel the short-ranged
interaction. These waves behave as if the interaction potential
were not there, giving rise to quickly vanishing phase shifts. For
$s$-wave scattering, we have that the scattering length is given by
\begin{equation}\label{eq:sclength}
a=-\lim_{k\rightarrow 0}\frac{\delta_0(k)}{k}.
\end{equation}
Moreover, we find from Eq.\ (\ref{eq:scamp2}) that $f_0(0)=-a$ and
from Eq.~(\ref{eq:scamp}) that $T^{\rm 2b}_0(0)=4\pi \hbar^2 a/m$. In
the specific case of a hard-core potential, the scattering problem
can easily be solved analytically, which leads to the $s$-wave
scattering length being equal to the radius of the impenetrable
core \cite{Duine04a}. Therefore, a positive $s$-wave
scattering length can be interpreted as the the effective
hard-core radius of the interaction.

\subsection{Two-channel scattering}
\label{par:fr}

We thus found that at low energies the two-body transition matrix is
dominated by the $s$-wave scattering length, which is interpretable as giving the effective interaction strength. Therefore, the scattering length is an important quantity to
know in the description of ultracold atomic gases. An accurate calculation
of the scattering length from first principles can be a challenging
task, because typically the atomic interaction
potentials are not known precisely enough. Fortunately,
the scattering length can be directly measured by experiments. This was for example done by Inouye {\it et al}. \cite{Inouye98a}, who showed experimentally that the effective interaction strength can be tuned over a large range by varying the applied magnetic field. The
mechanism that allows for this extremely useful control is
nowadays called a Feshbach resonance \cite{Feshba58a}, and recent extensive reviews about this topic can be found in Refs. \cite{Duine04a,Chin10a}. In the following discussion we mainly follow the lines of Ref.\ \cite{Duine04a}.

\begin{figure}[t]
\begin{center}
\includegraphics[width=0.5\columnwidth]{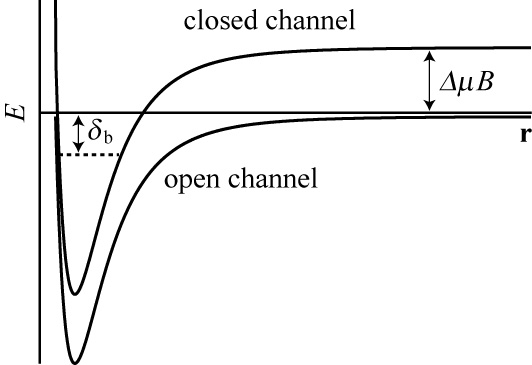}
\caption{\label{fig:fres} Illustration of the two-channel structure of
a Feshbach resonance. Shown are the atomic or open channel and the
molecular or closed channel in the absence of a coupling between
the two. The interaction potential in the open channel is called
the background interaction $V_{\rm bg} ({\bf r})$, which depends
on the interatomic separation ${\bf r}$. The relevant feature of
the closed channel is a bound state, also called the bare
molecular state, that has a small energy difference $\delta_{\rm
b}$ with the atomic continuum. Due to a different spin
configuration of the atoms in the two channels, there is a Zeeman
shift of $\varDelta \mu B$ with $\varDelta \mu$ the difference in
magnetic moments and $B$ the magnetic-field strength.}
\end{center}
\end{figure}

In an atomic Feshbach-resonant collision, the two incoming atoms virtually form a molecule with a different spin
configuration, after which this molecule
decays into two atoms again. The scattering properties
depend very sensitively on the energy
difference between the molecular state and the threshold of
the two-atom continuum. This energy difference can be controlled with an applied magnetic field, because the difference in spin states between the incoming atoms and the molecule gives rise to
a different Zeeman shift. To further study the physics, we use a two-channel model as illustrated in Fig.~\ref{fig:fres}. The atoms approach each other in the atomic channel,
while there is a bound state close to the threshold of the atomic
continuum in the molecular channel. We are interested in the
regime where the collisional energy is lower than the Zeeman
splitting. Due to energy conservation the molecular
channel is then inaccessible for the atoms at large separations from
each other after the collision. Therefore, the molecular channel
is also called the closed channel, while the atomic channel is
called the open channel. We consider only the bound state in the
molecular channel that is closest to the atomic continuum, since
this state is dominant in affecting the scattering process.

In the absence of a coupling between the channels, the molecule is
described by the wavefunction $|\psi_{\rm m}  \rangle$, which we
call the bare molecular state, and has an energy $\delta_{\rm b}$,
which we call the bare detuning. The bare molecular state and the
bare detuning are only a solution to the uncoupled Schr\"{o}dinger
equation in the closed channel. This is to be contrasted with the
so-called dressed molecular state, which is the bound-state
solution when the two channels are coupled. In the open channel
the two atoms interact through a short-ranged potential
$\hat{V}_{\rm bg}$, which depends on their relative coordinate
$\mathbf{r}$ and which is called the background interaction. The
atomic and the molecular channel are coupled by $\hat{V}_{\rm am}$
due to the possibility of electronic spin flips caused by
hyperfine interactions. To find the energy and the wavefunction of
the dressed molecule, the two-body problem is separated into a
center-of-mass part and a relative part. The center-of-mass part
gives rise to a trivially solvable free-body problem, while the
relative Schr\"{o}dinger equation becomes
\begin{equation} \label{eq:2chan}
\left[ \begin{array}{cc}
\hat{H}_0+ \hat{V}_{\rm bg} & \hat{V}_{\rm am} \\
\hat{V}_{\rm am} & \delta_{\rm b}
\end{array}
\right] \left[ \begin{array}{cc} \sqrt{1-Z_{\rm m}}|\psi_{\rm a} \rangle  \\
\sqrt{Z_{\rm m}}|\psi_{\rm m} \rangle  \end{array} \right]=
E \left[ \begin{array}{cc} \sqrt{1-Z_{\rm m}}|\psi_{\rm a} \rangle \\
\sqrt{Z_{\rm m}}|\psi_{\rm m}  \rangle \end{array} \right],
\end{equation}
where $|\psi_{\rm a} \rangle$ denotes the wavefunction in the
atomic channel, $|\psi_{\rm m} \rangle$ is the wavefunction of the
bare molecular state, $Z_{\rm m}$ is the so-called wavefunction
renormalization factor that normalizes the wavefunction of the
dressed molecule to unity, and
$\hat{H}_0=-\hbar^2\nabla^2_{\mathbf{r}}/m$ is again the relative
kinetic energy operator.

\begin{figure}[t]
\begin{center}
\includegraphics[width=1.0\columnwidth]{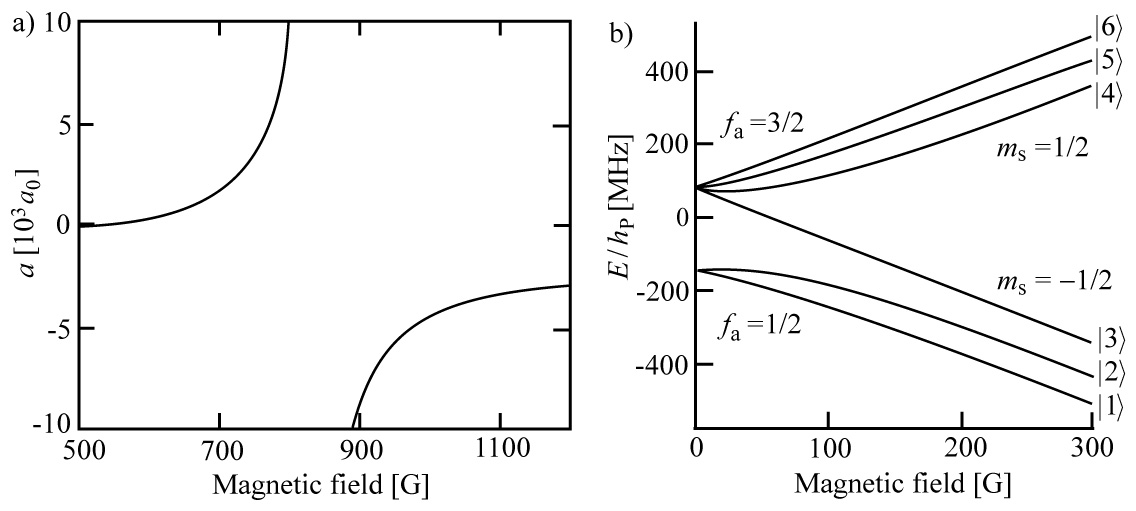}
\caption{\label{fig:lithium} a) The scattering length as a function of
magnetic-field strength near the very broad resonance of ${}^6$Li
at 834 G. It is given in terms of the Bohr radius $a_{0}$. b)
Hyperfine structure of the electronic ground state in ${}^6$Li as
a function of magnetic-field strength, where $E$ is the internal
energy and $h_{\rm P}$ is Planck's constant. The nuclear spin of ${}^6$Li
is given by $i_{\rm a}=1$, so that the spin of the outer electron gives
rise to $f_{\rm a}=3/2$ or $f_{\rm a}=1/2$ for the total angular momentum. At
large magnetic-field strength, the hyperfine energies are
predominantly determined by the projection $m_{\rm s}$ of the
electron spin. The experiments with imbalanced spin mixtures that
we consider later on in this review, are performed with states
$|1\rangle$ and $|2\rangle$.}
\end{center}
\end{figure}

\begin{figure}[t]
\begin{center}
\includegraphics[width=0.4\columnwidth]{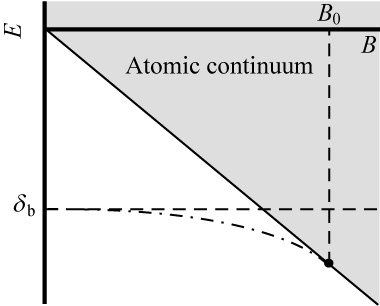}
\caption{\label{fig:continuum} Schematic representation of the physics
near the resonance at $B_0=834$ G of ${}^6$Li. The atomic
continuum is in an electronic spin triplet channel, whose energy is
lowered upon an increase of the magnetic field strength $B$. The
bare molecule is in an electronic spin singlet channel and its bare
energy is therefore not altered by the magnetic field. Due to
interactions with the atomic continuum the bare molecule becomes
dressed. When the dressed molecular energy, as sketched by the dashed-dotted line, reaches the atomic
continuum the resonance takes place. In this section, we actually
rotate the picture, namely the threshold of the atomic continuum
stays at zero energy, while $\delta_{\rm b}$ depends linearly on
$B$.}
\end{center}
\end{figure}

In this section we consider only $s$-wave scattering, which is dominant at low energies. As explained before, $s$-wave scattering of fermions only occurs
between particles in two different internal states. In the
experiments with ${}^6$Li of interest to us, a two-component
mixture is prepared in the lowest two hyperfine states of the
ground state. They are denoted by $|1\rangle$ and $|2\rangle$,
which are shown in Fig.~\ref{fig:lithium}(b). Ultracold ${}^6$Li is an
experimentalist's favorite due to the easily accessible Feshbach
resonance at a magnetic-field strength of $B_0=834$ G, which is an
extremely broad resonance. The resonance is shown in Fig.\ \ref{fig:lithium}(a).
Near 834 G, the two lowest lying hyperfine states have the single
valence electron of lithium anti-aligned with the applied magnetic
field, i.e. $m_{\rm s}=-1/2$ \cite{Ketter08a,Falco07a}. These two states then
differ in their projection of the nuclear spin. If two incoming
atoms in state $|1\rangle$ and $|2\rangle$ collide, then the
electronic spins of the atoms point in the same direction, so that
the open channel gives rise to a spin triplet potential. During
the collision an electronic spin can be flipped, which brings the
two atoms in a closed spin singlet channel. The difference in
magnetic moments between the two channels is $\varDelta\mu=2
\mu_{\rm B}$ with $\mu_{\rm B}$ the Bohr magneton. As we increase
the magnetic-field strength $B$ from the resonance strength $B_0$,
then the triplet potential is lowered in energy by an amount
$\delta=\varDelta\mu (B-B_0)$, which shifts the atomic continuum
down with respect to the bare molecular bound state as shown in
Fig.~\ref{fig:continuum}. Equivalently, we can also say that the bare
molecular bound-state energy $\delta_{\rm b}$ has been raised by
$\delta$ with respect to the atomic continuum. The difference in
Zeeman energy from the resonance position is also called the
(experimental) detuning $\delta$.  As a result, the bare detuning
$\delta_{\rm b}$ varies linearly with the experimental detuning
$\delta$, although they are not the same. As we may infer from
Fig.~\ref{fig:continuum} and as we also show next, they are separated
by a constant shift.

\subsection{Dressed molecules}
\label{par:dressmol}

We continue by looking for negative energy solutions of Eq.\
(\ref{eq:2chan}), meaning that we want to find the wavefunction and
the energy of stable dressed molecules. We start by rewriting Eq.\
(\ref{eq:2chan}) to obtain the following equation
\begin{equation} \label{eq:enmol}
\langle \psi_{\rm m}| \hat{V}_{\rm am}\hat{G}_{\rm a
}(E)\hat{V}_{\rm am}|\psi_{\rm m} \rangle = E - \delta_{\rm b},
\end{equation}
with $\hat{G}_{\rm a}^{-1}(E)=E-\hat{H}_0-\hat{V}_{\rm bg}$.
Assuming that we have been able to exactly solve the atomic part
of the scattering problem, namely $(\hat{H}_0+ \hat{V}_{\rm
bg})|\psi^{(+)}_{\bf k} \rangle=2\epsilon_{\bf k}|\psi^{(+)}_{\bf
k} \rangle$, we insert a completeness relation of the exact
scattering states in the open channel $|\psi^{(+)}_{\bf k}
\rangle$, which gives
\begin{equation} \label{eq:selfenmol}
E -\delta_{\rm b}=\int\frac{d{\bf k}}{(2\pi)^3}
\frac{|\langle\psi_{\rm m}| \hat{V}_{\rm
am}|\psi^{(+)}_{\bf k} \rangle|^2}{E-2\epsilon_{\bf k}}\equiv \hbar
\Sigma(E),
\end{equation}
where we have interpreted the expression containing the integral
in Eq.\ (\ref{eq:selfenmol}) as the molecular self-energy
$\hbar\Sigma(E)$. Note that for simplicity we have ignored the
possibility of bound states in the atomic background potential. If
they are important, then they can be easily taken into account
along the lines of Ref. \cite{Romans06a}. We thus see that the exact
equation for the molecular bound state energy has the intuitive
form $E=\delta_{\rm b} + \hbar \Sigma(E)$. As we see soon, the resonance
takes place when the energy $E$ of the dressed molecule reaches
the threshold of the atomic continuum, i.e. when $E=0$ and
$\delta_{\rm b} =- \hbar \Sigma(0)$. Therefore, we define the
(experimental) detuning as $\delta\equiv\delta_B + \hbar
\Sigma(0)$, so that we find for the energy equation $E=\delta +
\hbar \Sigma'(E)$ with $\hbar \Sigma'(E)\equiv \hbar
\Sigma(E)-\hbar\Sigma(0)$. Since the resonance now takes place for
$\delta=0$, we indeed have that $\delta=\varDelta\mu(B-B_0)$.

To make further progress with Eq.\ (\ref{eq:selfenmol}) we realize
that the matrix element behaves for low momenta as \cite{Duine04a}
\begin{eqnarray}\label{eq:amcoup}
\langle\psi_{\rm m}| \hat{V}_{\rm am}|\psi^{(+)}_{\bf k}
\rangle&=&\langle\psi_{\rm m}| \hat{V}_{\rm am}\hat{V}_{\rm bg
}^{-1}\hat{T}^{\rm 2b}_{\rm bg}|{\bf k} \rangle \nonumber \\
&=&\frac{g}{1+i k
a_{\rm bg}},
\end{eqnarray}
where $a_{\rm bg}$ is the $s$-wave scattering length of the
background interaction and $g$ is by definition the coupling
strength at zero momentum. In the limit of vanishing background
interaction, the above equation reduces to $\langle\psi_{\rm
m}|\hat{V}_{\rm am}|{\bf k} \rangle=g$, which means we have used a
momentum-independent or local atom-molecule coupling. This
approximation is appropriate, because the spatial extent of the
molecular wavefunction and the atom-molecule coupling is very
small compared to the de Broglie wavelength of the scattering
atoms. Using for the same reason also a local or
momentum-independent background interaction $\hat{V}_{\rm bg}$,
then the momentum dependence in Eq.~(\ref{eq:amcoup}) is thus caused
by the low-momentum behavior of the background transition matrix,
which according to Eqs.~(\ref{eq:scamp}), (\ref{eq:scamp2}) and
(\ref{eq:phaseshift}) is proportional to $(1+i k a_{\rm bg})^{-1}$.

Using Eqs. (\ref{eq:selfenmol}) and (\ref{eq:amcoup}), we obtain that
\begin{eqnarray} \label{eq:sqrtE}
\hbar\Sigma'(E)&\equiv&\hbar\Sigma(E)-\hbar\Sigma(0)\nonumber \\
&=&
\int\frac{d{\bf k}}{(2\pi)^3}\frac{g^2}{1+ k^2
a^2_{\rm bg}} \left(\frac{1}{E-2\epsilon_{\bf k}}
+\frac{1}{2\epsilon_{\bf
k}}\right)\nonumber\\&=&\frac{\eta\sqrt{-E}}{1-a_{\rm
bg}\sqrt{-mE}/\hbar}
\end{eqnarray}
with $\eta=g^2m^{3/2}/4\pi\hbar^3$ and where we have assumed a
negative background scattering length, which is the case
for ${}^{6}$Li. The bound-state equation is now an analytically
solvable cubic equation, although its solution is somewhat
cumbersome \cite{Falco07a}. A simpler result is obtained when we
ignore $a_{\rm bg}$, which is allowed close to resonance, where
$E$ goes to zero. Then, we find
\begin{equation}
E=\delta+\frac{\eta^2}{2}\left(\sqrt{1-\frac{4\delta}{\eta^2}}-1\right),
\end{equation}
so that for $\delta\rightarrow 0$ we have $E=-\delta^2/\eta^2$.
This implies that the bound-state energy of the dressed molecule
goes to zero quadratically with the applied magnetic-field
strength $B$, although the bare molecular energy varies linearly
with $B$. Moreover, from Eq.\ (\ref{eq:2chan}) it follows that the
dressed molecular state is given by $|\psi_{\rm
dr}\rangle=\sqrt{Z_{\rm m}}|\psi_{\rm m}\rangle +\sqrt{1-Z_{\rm m}}|\psi_{\rm
a}\rangle$, where
\begin{equation}\label{eq:Z}
\frac{1-Z_{\rm m}}{Z_{\rm m}}=\langle \psi_{\rm m} |\hat{V}_{\rm am} \hat{G}_{\rm
a }(E)^2\hat{V}_{\rm am} | \psi_{\rm m}
\rangle=-\frac{\partial\hbar\Sigma(E)}{\partial E
}.
\end{equation}
Eqs. (\ref{eq:sqrtE}) and (\ref{eq:Z}) then allow us to
calculate $Z_{\rm m}(E)$, i.e. the amplitude of the dressed molecular
state in the closed channel, where for the energy $E$ we must
substitute the solution of the bound-state equation.

\subsection{Resonant atomic interaction}

Having determined the properties of the Feshbach molecules, we now examine the
effect of the Feshbach resonance on the atomic physics in the open
channel. Solving Eq.\ (\ref{eq:2chan}) for the atomic channel, we
find that
\begin{equation}
\left\{\hat{H}_0+ \hat{V}_{\rm bg}+ \hat{V}_{\rm m}\right\}
|\psi_{\rm a} \rangle = E |\psi_{\rm a} \rangle,
\end{equation}
where $\hat{V}_{\rm m}$ is the molecule-mediated interaction given
by
\begin{equation}\label{eq:molint}
\hat{V}_{\rm m}= \frac{\hat{V}_{\rm am}|\psi_{\rm m
}\rangle\langle\psi_{\rm m }|\hat{V}_{\rm am}}{E-\delta_{\rm b}},
\end{equation}
and we also used that in our model $|\psi_{\rm m
}\rangle\langle\psi_{\rm m }|$ is the unity matrix in the closed
channel.  As a result, the total interaction in the open channel
is given by $\hat{V}=\hat{V}_{\rm bg}+\hat{V}_{\rm m}$. From Eq.\
(\ref{eq:t2b}), we have that the exact transition matrix in the open
channel is given by
\begin{eqnarray}
\hat{T}^{\rm 2b}&=&\hat{V}+\hat{V}\hat{G}\hat{V}\nonumber\\
&=&\hat{T}^{\rm 2b}_{\rm bg}+\hat{T}^{\rm 2b}_{\rm bg}\hat{V}_{\rm
bg}^{-1}\hat{V}_{\rm m}(1-\hat{G}_{\rm a}\hat{V}_{\rm m})^{-1}
\hat{V}_{\rm bg}^{-1}\hat{T}^{\rm 2b}_{\rm bg} \nonumber \\
&\equiv& \hat{T}^{\rm
2b}_{\rm bg}+\hat{T}^{\rm 2b}_{\rm res}
\end{eqnarray}
with $\hat{G}^{-1}(E)=E-\hat{H}_0-\hat{V}_{\rm bg}-\hat{V}_{\rm
m}$. The exact two-body $T$ matrix is thus seen to consist of a
purely background part $\hat{T}^{\rm 2b}_{\rm bg}$ and a resonant
part $\hat{T}^{\rm 2b}_{\rm res}$. The first is given by
$\hat{T}^{\rm 2b}_{\rm bg}=\hat{V}_{\rm bg}+\hat{V}_{\rm
bg}\hat{G}_{\rm a }\hat{V}_{\rm bg}$, which, as before, leads for
$s$-wave scattering at zero momentum to $T^{\rm 2b}_{\rm
bg}(0)=4\pi\hbar^2 a_{\rm bg}/m$.

The momentum dependence of the resonant part may be further
evaluated by
\begin{eqnarray}\label{eq:kdept2b}
 \langle{\bf k}'|\hat{T}^{\rm 2b}_{\rm res}(2\epsilon_{\bf k})|{\bf
k }\rangle= \frac{\langle {\bf k}' |\hat{T}^{\rm 2b}_{\rm
bg}\hat{V}_{\rm bg}^{-1}\hat{V}_{\rm am}|\psi_{\rm
m}\rangle\langle \psi_{\rm m} |\hat{V}_{\rm am}\hat{V}_{\rm
bg}^{-1}\hat{T}^{\rm 2b}_{\rm bg}|{\bf k}\rangle}{2\epsilon_{\bf
k}-\delta_{\rm b} -\hbar\Sigma(2\epsilon_{\bf k})},
\end{eqnarray}
where we used Eq.\ (\ref{eq:molint}). Moreover, we used that $\hbar\Sigma(E)$ is given by the left-hand side of Eq.\ (\ref{eq:enmol}), and that $|\psi_{\rm m}\rangle\langle \psi_{\rm m} |$ is the unity matrix in the molecular channel. Combining Eqs.\ (\ref{eq:amcoup}) and (\ref{eq:kdept2b}), we find that at zero momentum and energy $T^{\rm 2b}_{\rm res}(0)=-g^2/(\hbar\Sigma(0)+\delta_{\rm
B})=-g^2/\delta$. This indeed shows that for $\delta=0$ the
transition matrix in the open channel goes to infinity at zero
energy, implying a diverging scattering length. We thus have that
\begin{eqnarray}\label{eq:t2b0}
T^{\rm 2b}(0)&=&\frac{4 \pi \hbar^2 a}{m}=\frac{4 \pi \hbar^2 a_{\rm
bg }}{m}-\frac{g^2}{\delta} \nonumber\\
&=&\frac{4 \pi \hbar^2a_{\rm bg
}}{m}\left(1-\frac{\varDelta B}{B-B_0}\right),
\end{eqnarray}
where the width of the resonance $\varDelta B$, the location $B_0$
and the background scattering length are readily determined
experimentally by fitting to the data. As a result, the
atom-molecule coupling constant $g$ can be expressed in terms of
the experimentally known parameters as $g=\sqrt{4\pi \hbar^2
a_{\rm bg}\varDelta B\varDelta\mu/m}$.

\begin{figure}[t]
\begin{center}
\includegraphics[width=0.65\columnwidth]{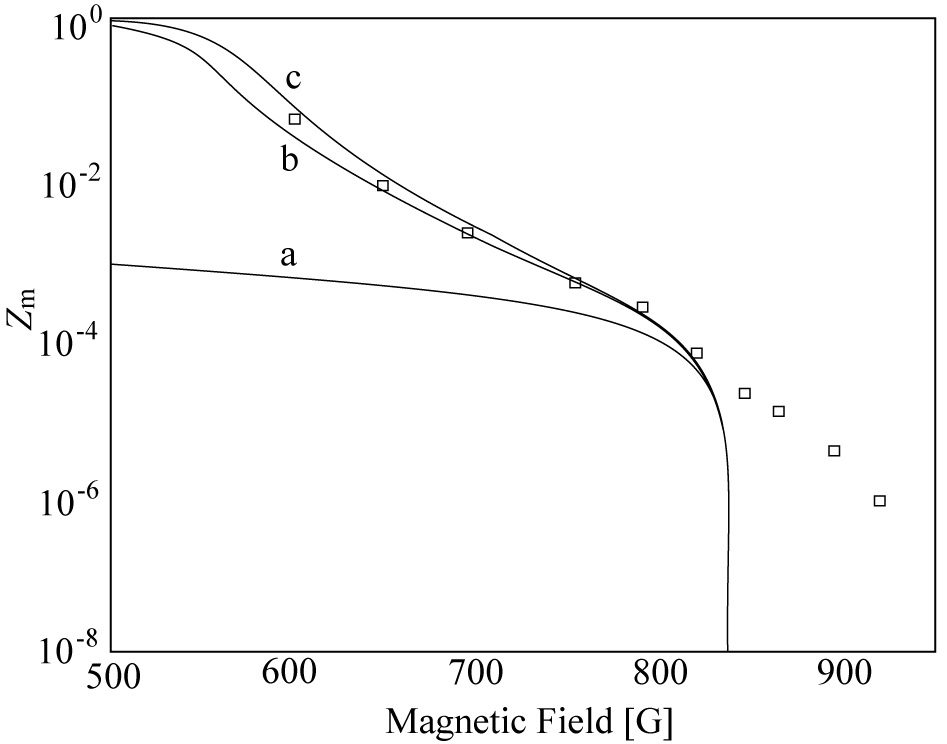}
\caption{\label{fig:Z} Wavefunction renormalization factor $Z_{\rm m}$ as a
function of magnetic field strength, as adapted from Ref.
\cite{Romans06a}. Curve (a) is calculated without background
interactions, curve (b) with a constant background scattering
length $a_{\rm bg}$ and (c) with a magnetic-field dependent
$a_{\rm bg}$ as determined in Ref. \cite{Falco05a}. The experimental
data (squares) is from Ref. \cite{Partri05a}. At resonance, $Z_{\rm m}$ goes to zero,
while for positive detuning, the molecule is not stable in the
two-body case. To get agreement with the experimental data also
for positive detuning, a many-body calculation is needed
\cite{Romans05a}. }
\end{center}
\end{figure}

To summarize, we have been able to solve for the properties of the
dressed molecular state near a Feshbach resonance. Moreover, we
have studied what the effect of this molecular state is on the
transition matrix in the open channel. In particular, we have seen
when the resonant part of the transition matrix diverges and gives
rise to an infinite scattering length. Since the location of the
resonance, its width, and the background scattering length are
easily determined by fitting to the experimental data, we have
access to all relevant parameters for a quantitative study of the
molecular self-energy from Eq.\ (\ref{eq:sqrtE}) and the
wavefunction renormalization factor $Z_{\rm m}$ from Eq.\ (\ref{eq:Z}). The
results for $Z_{\rm m}$ are shown in Fig.~\ref{fig:Z} for the broad
resonance of ${}^6$Li near $B_0=834$ G, as measured in Ref.\ \cite{Partri05a} and as determined theoretically in Ref. \cite{Romans06a}. It shows that the amplitude of the dressed molecule
in the closed channel is very small over a very wide range near
resonance. As a result, all the action happens in the open
channel, where, apart from allowing the resonance to actually take
place, the bare molecular state hardly plays any role. This
observation then leads to the so-called single-channel model of a
Feshbach resonance, where only the atomic channel is taken into
account with the transition matrix given by Eq.\ (\ref{eq:t2b0}). The
single-channel model is valid for wide resonances, where $Z_{\rm m}$ is
small. It is thus particularly valid for the extremely broad
resonance of ${}^6$Li near $B_0=834$ G, which is used in the experiments on the imbalanced Fermi gas that are discussed later on. Therefore, we use the single-channel model in several theoretical many-body calculations throughout this review. Next, we use the single-channel model to qualitatively discuss an important application of Feshbach resonances, namely the study of the BEC-BCS crossover.

\subsection{Pairing in the BEC-BCS crossover}\label{par:becbcs}

If fermions of different spin attract
each other, they have the tendency to form pairs, which may result
in a paired condensate at ultralow temperatures
\cite{Bardee57a,Stoof96a}. However, if these fermions repel each other,
then they have the tendency to align in spin space, favoring
(itinerant) ferromagnetism \cite{Duine05a, Jo09a, Condui09a}. Although close to resonance these two instabilities compete \cite{Pekker11a}, we consider in this review solely pair condensation. Since in ultracold atomic
gases the range of the interaction is typically very small
compared to the interparticle distance, it is convenient to use a
contact potential of strength $V_0$ as interaction potential $V({\bf r})$, i.e., $V({\bf r})=V_0 \delta({\bf r})$, to incorporate
interaction effects. The Fourier transform of the potential is
then a constant, which leads to a particularly simple form of the
Lippmann-Schwinger equation in momentum space. For
$s$-wave collisions, the transition matrix at zero momentum and energy becomes \cite{Stoof09a}
\begin{equation}\label{eq:t2b-v0}
\frac{1}{T^{\rm 2b}(0)}=
\frac{1}{V_0}+\frac{1}{\mathcal{V}}\sum_{\bf
k}\frac{1}{2\epsilon_{\bf k}},
\end{equation}
where the kinetic energy of a particle with mass $m$ and
wavevector ${\bf k}$ equals $\epsilon_{\bf k}=\hbar^2k^2/2m$ and
$\mathcal{V}$ is the volume of the system.\footnote{In actual
analytical or numerical evaluations of the sum over states in
${\bf k}$-space, we make use of the continuum limit $\sum_{\bf
k}1/\mathcal{V}\rightarrow\int d{\bf k}/(2\pi)^3$.} Note that the
sum on the right-hand side of Eq.~(\ref{eq:t2b-v0}) is actually not
convergent, which comes from the anomalous behavior of the contact
potential at high momenta. This is usually not a problem, since we
are only interested in the low-energy properties of the
interaction. As we see explicitly in Section \ref{par:nsr}, we can then use Eq.\ (\ref{eq:t2b-v0}) to
eliminate the contact potential from our theory in favor of the two-body
$T$ matrix, which at zero energy is related to the $s$-wave
scattering length $a$ through $T^{\rm 2b}(0)=4 \pi \hbar^2 a/m$.
This scattering length can be directly extracted from experiment.

\begin{figure}[t]
\begin{center}
\includegraphics[width=0.7\columnwidth]{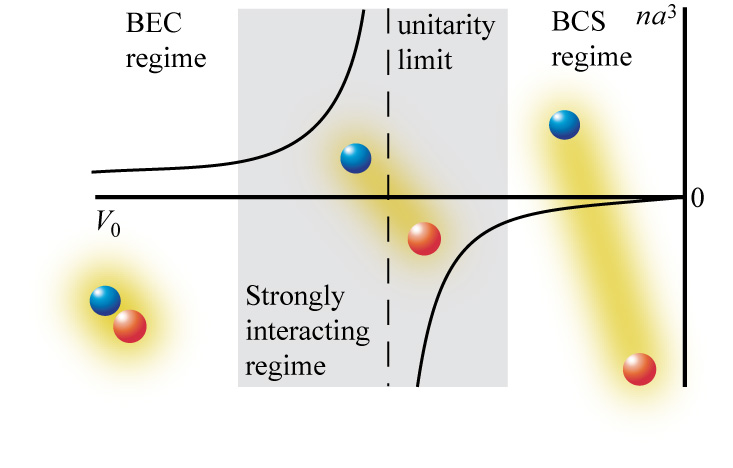}
\caption{\label{fig:va} Schematic representation of the BEC-BCS
crossover. In the BCS regime, the microscopic interaction
strength, given by $V_0$, and the effective interacting strength,
given by the scattering length $a$, are both negative and small,
namely $|a|< n^{-1/3}$ with $n$ the total particle density.
Pairing is then a many-body effect and the size of the Cooper pairs is
much larger than the average interparticle distance. When $|a|> n^{-1/3}$,
we enter the strongly interacting regime. At
$V_0=-2\pi^2\hbar^2/m\Lambda_0$ (see text), the scattering length diverges,
which is also called the unitarity limit. Here, the pair size is
comparable to the average interparticle distance. For stronger microscopic
attractions, a two-body molecular bound state enters the
interaction potential. When $|a|< n^{-1/3}$, we enter the BEC
regime. The size of the molecules is here much smaller than the average
interparticle distance and the ground state of the system is a
weakly interacting molecular Bose gas.}
\end{center}
\end{figure}

Another approach to regularize the high-momentum behavior of the contact potential is to
consider only momenta up to a certain cut-off $\Lambda_0$ for the
sum in Eq.~(\ref{eq:t2b-v0}), after which we find \cite{Bijlsm96a}
\begin{equation}\label{eq:intvscut}
V_0=\frac{4\pi\hbar^2 a}{m}\frac{\pi}{\pi-2 a \Lambda_0},
\end{equation}
whose result is plotted in Fig.\ \ref{fig:va}. The weakly
interacting regime, where both $a$ and $V_0$ are small and
negative, is also called the BCS regime, since it is analogous to
the weakly attractive case known from ordinary superconductors. In
the BCS regime, the two-body potential is too weak to support a
bound state, so that Cooper pairing is truly a many-body effect
that occurs only in the presence of a Fermi sea \cite{Cooper56a}.
The size of these Cooper pairs then turns out to be much larger
than the average interparticle distance. When
$V_0=-2\pi^2\hbar^2/m\Lambda_0$, we see that the scattering length
diverges. This resonance in the cross section is physically caused by
a bound state that enters the attractive potential
\cite{Busch98a}. To see the latter more clearly, we can consider a
slightly more realistic interaction potential, namely a square
well. The exact solution to the resulting problem shows that, upon
increase of the well depth, the scattering length diverges each
time a new bound state enters the square-well potential
\cite{Duine04a}. Such resonances are called single-channel shape
resonances. Note that it is not so clear how to tune a shape
resonance experimentally, because it is usually not possible to
precisely control the interatomic interaction potential. However,
in the previous paragraphs, we discussed the more flexible mechanism of
a two-channel Feshbach resonance, which is nowadays routinely used
by experimentalists to vary the scattering length at will.

The region where the scattering length becomes infinite is
commonly referred to as the unitarity limit. Here, the size of the
Cooper pairs turns out to be comparable to the average
interparticle distance. The unitary regime is also called the
strongly interacting limit. This as opposed to the regime where
the microscopic attraction has become so strong that there is a
deep bound state in the interaction potential. As a result,
bosonic molecules are formed, whose size is much smaller than the
average interparticle distance. This regime is then called the BEC
regime. It is unique that the complete evolution from the BEC
regime to the BCS regime can be explored in ultracold quantum
gases \cite{Ingusc08a}. From
Fig.\ \ref{fig:va} we might be surprised that the evolution merely
leads to a crossover. Namely, the system is seen to evolve through
a true resonance in the effective interaction strength, which we
might have expected to profoundly influence the thermodynamics of
the mixture. Moreover, theoretical predictions based on
perturbative approaches are expected to fail near resonance. This
particularly holds for the mean-field BCS theory, which was used
to predict the crossover \cite{Legget80a}. As a result, there has been much
theoretical and experimental interest in the BEC-BCS crossover,
and in particular on the behavior of the Fermi mixture in the
unitary regime \cite{Ingusc08a,Giorgi08a}.

A second reason for the appeal of the crossover is the
unification of BEC-like superfluidity and BCS-like superfluidity
as two-sides of the same coin. Both can be viewed as a coherent
state (or Bose-Einstein condensate) of fermionic pairs, which is
signalled by a nonzero expectation value of the pair annihilation
operator, i.e. $\langle\hat{\psi}_{-}({\bf R}+{\bf r}/2)
\hat{\psi}_{+}({\bf R}-{\bf r}/2)\rangle$. Here, we introduced the
annihilation operator $\hat{\psi}_{\sigma}({\bf x})$ for a single
atom with spin $\sigma$ at position ${\bf x}$, while ${\bf R}$ is
the center-of-mass coordinate and ${\bf r}$ the relative
coordinate of the pair. This puts us in the position to define the equilibrium
BCS order parameter $\langle\Delta\rangle$, given by
\begin{eqnarray}
\langle\Delta\rangle ({\bf R})&=&\int d{\bf r}~ V({\bf r}) \left\langle
\hat{\psi}_{-}\left({\bf R} + \frac{{\bf r}}{2}\right)
\hat{\psi}_{+} \left({\bf R}-\frac{{\bf r}}{2}\right)
\right\rangle \nonumber\\
&=& V_0\langle \hat{\psi}_{-}({\bf
R}) \hat{\psi}_{+} ({\bf R}) \rangle \nonumber\\
&=& \frac{V_0}{\mathcal{V}} \sum_{\bf q, k}\langle
\hat{\psi}_{-,\frac{{\bf q}}{2} + {\bf k}} \hat{\psi}_{+,\frac{{\bf
q}}{2}-{\bf k}} \rangle e^{i {\bf q}\cdot {\bf R}},
\end{eqnarray}
where $\hbar {\bf q}$ is the center-of-mass momentum and $\hbar
{\bf k}$ is the relative momentum. Because the equation above is
seen to involve an integral over the attractive interaction, the
order parameter also describes the energy cost to break up a
Cooper pair. For this reason, $\langle\Delta\rangle({\bf R})$ is also referred
to as the (local) pairing gap. Usually, the paired state only occurs at at zero momentum, $\hbar {\bf q}=0$, which upon pair
condensation gives rise to an order parameter that is spatially
independent, namely
\begin{eqnarray}
\langle\Delta\rangle &=& \frac{V_0}{\mathcal{V}} \sum_{\bf k}\langle
\hat{\psi}_{ -,{\bf k}} \hat{\psi}_{+,-{\bf k}}\rangle.
\end{eqnarray}
The above equation also reveals that in this case the Cooper pairing
occurs in momentum space between particles of opposite spin and
momentum.

Using BCS mean-field theory it is only possible to study the crossover qualitatively at zero temperature, where it describes the evolution from a condensate of loosely-bound Cooper pairs to a condensate of tightly bound molecules \cite{Legget80a,Giorgi08a}. For the critical temperature curve, BCS mean-field theory fails even on a qualitative level. This is because the BCS critical temperature describes physically the effect of pair-breaking, which is not the correct mechanism for the phase transition in the BEC regime, where superfluidity is lost to the thermal occupation of nonzero momentum states by tightly-bound pairs. The simplest theory that also describes these pairs with nonzero momentum is called the Nozi\`{e}res-Schmitt-Rink approximation \cite{Nozier85a}, which we discuss in more detail in Section \ref{par:nsr}. It turns out to give a rather accurate description of the critical temperature throughout the crossover \cite{Sademe93a}. Although the BEC-BCS crossover was predicted already a long time ago, it was hard to experimentally verify the smooth evolution from BCS to BEC superfluidity using condensed-matter systems, since the attractive phonon-mediated electron-electron interaction is not so easily tunable. Therefore, as mentioned in the Introduction, the BEC-BCS crossover has only been observed experimentally in recent years \cite{Zwierl04a,Kinast04a,Barten04a,Bourde04a,Partri05a}, profiting from the extreme tunability of ultracold quantum gases.  The experiments on imbalanced Fermi gases that we discuss more extensively in this review were performed mainly about halfway the crossover, where the scattering length diverges. This is also called the unitarity regime, where the Fermi gas has an exceptionally large critical temperature that is about one-tenth of the Fermi energy \cite{Ingusc08a}, and where the Fermi gas also shows remarkable universal properties \cite{Ho04a,Thomas05a,Tan08a,Tan08b}.

\section{Mean-field theory}
\label{sec:mf}

In this section, we use BCS mean-field theory to study the strongly interacting two-component Fermi mixture with a population imbalance. We start out by giving a comprehensive historical overview of the experiments that have been performed since Zwierlein {\it et al}.\ \cite{Zwierl06a} and Partridge {\it et al}.\ \cite{Partri06a} achieved control over the spin polarization of the Fermi gas. After this experimental overview, we calculate the mean-field phase diagram for the homogeneous case. Although perturbative approaches are not expected to be quantitatively
correct for strong interactions, mean-field theory turns out to be useful for
explaining the relevant physics in the system. In order to also
arrive at the phase diagram for the trapped case, we use the
local-density approximation, which assumes that the gas behaves
locally in the trap as if it were homogeneous. From the mean-field
phase diagram, we can understand the qualitative aspects of the
early experiments exploring the imbalanced Fermi mixture. We end this section on mean-field theory with a calculation of the phase diagram of the ${}^6$Li-${}^{40}$K mixture.

\subsection{Experimental overview}\label{par:exp}

In the beginning of 2006, two experimental groups obtained a full control over the spin imbalance or polarization $P$ in an ultracold atomic Fermi mixture. This
polarization is defined by $P=(N_{+}-N_{-})/(N_{+}+N_{-})$ with
$N_{\sigma}$ the total number of particles in the hyperfine state
$\sigma$. These first experiments at MIT by Zwierlein {\it et
al}.\ \cite{Zwierl06a} and at Rice University by Partridge {\it
et al}.\ \cite{Partri06a} realized a strongly interacting mixture
of ${}^6$Li atoms in the two lowest hyperfine states of the ground
state (see Fig.\ \ref{fig:lithium}), where the atom number in each of
the two states could be precisely tuned. Within a few months time these experiments were followed by a large amount of theoretical studies,
which studied the imbalanced Fermi gas using mean-field approaches
\cite{Sheehy06a,Duan06a,Haque06a,Chien06a,Martik06a,Gubbel06a},
using thermodynamic approaches \cite{Chevy06a, Bulgac07a}, using diagrammatic approaches \cite{Pieri06a}, using the Bogoliubov-de Gennes approach \cite{Kinnun06a,Machid06a}, including the effects of surface tension \cite{Silva06a}, including the gradient energy of the density profiles  \cite{Imambe06a}, and including Gaussian order parameter fluctuations \cite{Parish07b}.
The interesting aspect of a polarization in the system is that not all
atoms can find a partner to pair up with, because the fermions
only have attractive $s$-wave interactions when they are in
different hyperfine states. Since pairing is the mechanism behind
fermionic superfluidity, the following fundamental question
arises: what happens to a superfluid paired Fermi mixture upon
increasing the polarization?

The early $^6$Li experiments of Zwierlein {\it et al}.\
\cite{Zwierl06a,Zwierl06b} and Partridge {\it et al}.\
\cite{Partri06a} were performed in a trap to confine the atomic
clouds in space. Both experiments revealed that superfluidity in
an ultracold Fermi mixture with attractive interactions is in
first instance maintained upon going to an imbalance in spin
populations. However, on a more detailed level
contradictory results were obtained for the behavior of
the trapped mixture as a function of polarization. Namely,
Zwierlein {\it et al}.\ observed a rather smooth phase transition
between a phase with a superfluid core and a phase that was fully
normal at a high critical polarization of about 0.7. However,
Partridge {\it et al}.\ seemed to observe a transition between two
different trapped superfluid phases at a low critical polarization
of about 0.1. Moreover, Partridge {\it et al}.\ did not observe a vanishing
of the superfluid core in their experiments, where even at their
highest imbalances ($P>0.90$) the core seemed fully paired.

\begin{figure}[t]
\begin{center}
\includegraphics[width=0.8\columnwidth]{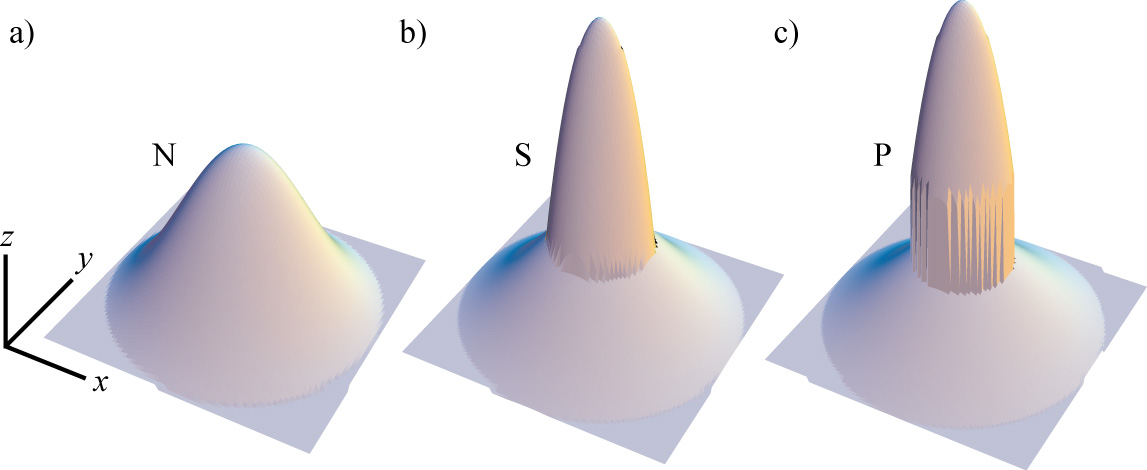}
\caption{\label{fig:3d} Sketchy impression of three different phases that
can be present for a strongly interacting Fermi mixture in the
trap. The $x$ and $y$ coordinates correspond to the axial and the radial direction in the trap,
while the $z$ coordinate represents the total density of
atoms. Although the axial direction is elongated in the actual experiments, we have drawn here a more spherically symmetric situation.
a) At high temperatures, the gas is in
the normal phase (N), leading to a thermal distribution of atoms
in the trap. b)-c) At low temperatures, a condensate of Cooper
pairs appears in the center of the trap, where the particle
densities are highest. The condensate leads to a pronounced
enhancement of the central densities. The superfluid core and the
normal outer region in the trap can be separated by b) a
continuous second-order transition (S) or c) a discontinuous
first-order transition (P). Note that we have exaggerated the `bump' in the profile due to the Cooper-pair condensate
and the jump in the density due to the first-order phase transition for illustrative reasons. }
\end{center}
\end{figure}

More experimental and theoretical studies followed in order to try
to understand these differences. In a later study at MIT, Shin
{\it et al}. showed that their trapped Fermi mixture was described
by a shell structure consisting of an equal-density superfluid
core, surrounded by a partially polarized normal shell and a fully
polarized outer region \cite{Shin06a}. As a function of
polarization, the MIT data showed a smooth onset of the superfluid
condensate in the center of the trap, which agreed with a
continuous, or second-order, phase transition. Also at Rice
University a further study was performed, where Partridge {\it et
al}.\ observed a deformation of the superfluid core at their
lowest temperatures \cite{Partri06b}. This deformation was explained
in terms of a first-order interface between the superfluid core
and the normal outer region in the trap \cite{Silva06a,Haque07a}, see
also Fig.\ \ref{fig:3d}(c). Namely, a first-order interface in
general gives rise to a surface tension, which energetically favors
a minimal area of the interface, i.e. a spherical shape. However,
the trap that was used in the experiments at Rice university was
actually highly elongated, so that the superfluid core was
consequently deformed from the trap shape by the surface tension.
At higher temperatures the deformation disappeared, although the
core seemed to remain paired. The natural explanation of this
behavior is that at higher temperatures the superfluid-normal
transition in the trap is of second order, as in Fig.\
\ref{fig:3d}(b), so that there is no surface tension and
deformation.

Since the higher-temperature results at Rice University resembled
qualitatively the behavior that was seen in the MIT experiments,
it seemed like a difference in temperature would be the most
natural way to explain both experiments in a single theoretical
picture \cite{Gubbel06a}. Moreover, in another experiment at
MIT, Shin {\it et al}.\ performed precise measurements as a
function of position in the trap and temperature \cite{Shin08a}.
Because the local-density approximation applies for the MIT
experiments, this actually means that Shin {\it et al}.\ were able
to experimentally map out the homogeneous phase diagram of the
spin-imbalanced Fermi mixture in the unitarity limit. They now
observed both a second-order phase transition in the trap at
higher temperatures as well as a first-order transition at their
lowest temperatures. On a qualitative level, all experimental
results then seem to fit in a phase diagram that is governed by a
tricritical point \cite{Gubbel06a,Parish07a}.

However, on the quantitative level there remained striking
differences between the two experimental groups.
First of all, there was the difference in the critical imbalance
$P_{\rm c}$, at which the trapped gas becomes completely normal.
Namely, Shin {\it et al}.\ found at their lowest temperatures
$P_{\rm c}< 0.8$ \cite{Zwierl06b,Shin08a}, while Partridge {\it
et al}.\ found $P_{\rm c}> 0.9$ \cite{Partri06b}. Although the first
result agrees with Monte-Carlo calculations in the local-density
approximation (LDA) \cite{Lobo06a}, the latter result was not understood.
The second issue involves the normal region that
surrounds the superfluid core. While the MIT experiments found that
the normal state is partially polarized close to the superfluid
interface, the Rice experiments only observed a fully polarized
normal state. The absence of the partially polarized normal region in the
Rice experiment was also not understood. The last difference is the observation of deformation by the Rice experiment, whereas in the MIT experiment such
deformations were not seen. The high value
of the surface tension that is needed to explain the deformation
observed at Rice University could also not be
microscopically accounted for.

More recently, the imbalanced Fermi gas was also experimentally realized at ENS by Nascimb\`{e}ne {\it et al.}, who found very similar results to the MIT experiments \cite{Nascim09a,Nascim10a,Nascim11a}. The latest experiments by the group at Rice indicate that their deviating experimental results are caused by spin currents in the trap that create a long lived metastable state with a deformed superfluid core \cite{Liao11a}. These experiments confirm the theoretical proposal of Parish and Huse \cite{Parish09a}, who showed that this metastable state could arise due to the elongated trap geometry, in which evaporation occurs predominantly from the trap center. Once phase separation occurs in this geometry, particle and heat transport are suppressed at the narrow interface between the gapped superfluid core and the normal outer region \cite{Schaey07a, Schaey09a, Parish09a}. It turns out that the resulting evaporative depolarization of the central core stabilizes superfluidity, and thus favors the superfluid phase over the normal phase, even when the overall polarization of the gas is above the expected critical imbalance. The suppression of transport across the interface allows the non-equilibrium density distribution to remain metastable with observed lifetimes of multiple seconds \cite{Liao11a}. Although spin transport is an interesting topic that is currently actively being studied experimentally for imbalanced Fermi gases \cite{Sommer11a,Sommer11b}, we treat in this review only the equilibrium case. This also means that in the following we mainly compare the theoretical results with the experiments performed initially at MIT \cite{Zwierl06a,Zwierl06b,Shin06a,Shin08a}.

\subsection{Mean-field grand potential}\label{par:mftp}

We start our theoretical discussion of the strongly interacting
Fermi mixture at the mean-field level, which is useful for
qualitatively explaining the relevant physics. The derivation of
the grand-canonical thermodynamic potential density that follows from mean-field
BCS theory can be found in several textbooks, see for example Ref. \cite{Fetter71a} using operator methods or Ref.\
\cite{Stoof09a} using functional methods. In Ref.\ \cite{Baarsm10a}, the case for both a spin- and mass-imbalance is explicitly derived.  It yields for the solely population-imbalanced case at
unitarity that
\begin{eqnarray} \label{eq:tpdBCS}
\omega_{\rm BCS}[\Delta;T,\mu_{\sigma}]&=& \frac{1}{\mathcal{V}}
\sum_{\mathbf{k}}\left\{ \epsilon_{\mathbf{k}} -\mu - \hbar
\omega_{\mathbf{k}}+\frac{|\Delta|^2}{2
\epsilon_{\mathbf{k}}}\right\}
\nonumber\\
&&-\frac{1}{\beta \mathcal{V}}\sum_{\sigma,{\bf k}}\ln(1+e^{-\beta
\hbar\omega_{\sigma,\mathbf{k}}}),
\end{eqnarray}
where the kinetic energy of the atoms is given by
$\epsilon_{\mathbf{k}}=\hbar^2 \mathbf{k}^2/2m$, $m$ is the mass
of the fermions, $\mathcal{V}$ is the volume, $T$
is the temperature and $\beta=1/k_{\rm B} T$. The physical meaning
of the BCS order parameter $\Delta$ was explained in section
\ref{par:becbcs}. The index $\sigma=\pm$ specifies the hyperfine
state and is also called the (pseudo)spin of the fermions. There
are two slight differences between Eq.\ (\ref{eq:tpdBCS}) and the
standard BCS expression for the grand-canonical thermodynamic potential. First of all, there is a term $-|\Delta|^2/T^{\rm 2b}$
missing, where the two-body transition matrix is given by $T^{\rm
2b}=4\pi\hbar^2 a/m$ with $a$ the $s$-wave scattering length. This
is because the scattering length diverges in the unitarity limit,
which is the regime where the experiments operate. Second, we
allow the chemical potentials for the two hyperfine states
$\mu_{\sigma}$ to be unequal, since this takes into account the
population imbalance in the two spin species that is realized by
the experiments. The average chemical potential $\mu$ is given by
$\mu=(\mu_{+}+\mu_{-})/2$, while half the difference is denoted as
$h=(\mu_{+}-\mu_{-})/2$. In the case of nonzero $h$, the
dispersions of the Bogoliubov quasiparticles
$\hbar\omega_{\sigma,\mathbf{k}}$ are spin-dependent. Namely, we
have that $\hbar\omega_{\sigma,\mathbf{k}}= -\sigma h +
\hbar\omega_{\mathbf{k}}$ with $\hbar\omega_{\mathbf{k}}=
\sqrt{(\epsilon_{\mathbf{k}}-\mu)^2+|\Delta|^2}$ \cite{Houbie97a}.
This follows from the usual Bogoliubov diagonalization of the
mean-field hamiltonian \cite{DeGenn89a}. The logarithms in Eq.\ (\ref{eq:tpdBCS})
describe an ideal gas of fermionic quasiparticles with dispersion
$\hbar\omega_{\sigma,\mathbf{k}}$, while the other terms describe
the equal-density Cooper pair condensate.

To determine the atomic density $n_{\sigma}$ in spin state
$\sigma$, we use the relation $n_{\sigma}=-\partial \omega_{\rm
BCS }[\Delta;T,\mu_{\sigma}]/\partial \mu_{\sigma}$,
which leads to
\begin{eqnarray}\label{eq:nsf}
&& n_{\sigma}[\Delta;T,\mu_{\sigma}] = \frac{1}{\mathcal{V}}\sum_{\mathbf{k}} \left\{ |u_{\mathbf{k}}|^2
f(\hbar\omega_{\sigma,\mathbf{k}}) +|v_{\mathbf{k}}|^2 [1-f
(\hbar\omega_{-\sigma,\mathbf{k}})] \right\},\qquad
\end{eqnarray}
where  $f(x)=1/(e^{\beta x}+1)$ are the Fermi
distributions. The BCS coherence factors $|u_{\mathbf{k}}|^2$ and
$|v_{\mathbf{k}}|^2$ are determined by the relations
$|u_{\mathbf{k}}|^2 = [1+ (\epsilon_{\mathbf{k}}-\mu)/\hbar
\omega_{\mathbf{k}}]/2$ and
$|u_{\mathbf{k}}|^2+|v_{\mathbf{k}}|^2=1$ \cite{Bardee57a,DeGenn89a}.

\begin{figure}[t!]
\begin{center}
\includegraphics[width=0.8\columnwidth]{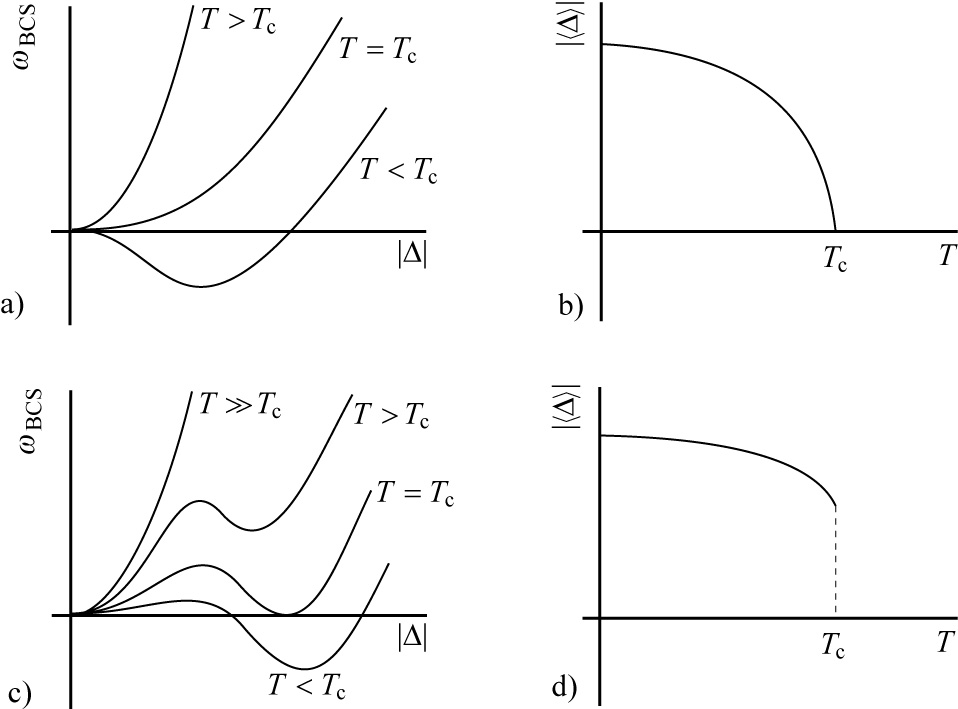}
\caption{\label{fig:landau} a) The behavior of the grand-canonical thermodynamic
potential density $\omega_{\rm BCS}$ as a function of the BCS
order parameter $|\Delta|$ for a second-order phase transition. For
high temperatures $T$, the global minimum is at
$|\langle\Delta\rangle|=0$, i.e. the normal state. At the critical
temperature $T_{\rm c}$ the minimum becomes a maximum. For
$T<T_{\rm c}$ the system is in the superfluid state. b)
Corresponding behavior of the expectation value of the order
parameter $|\langle\Delta\rangle|$. At $T_{\rm c}$,
$|\langle\Delta\rangle|$ goes continuously to zero. c) Same as a),
but now for a first-order transition. At the critical temperature
$T_{\rm c}$, the separated minima of the normal and the superfluid
state have the same pressure, given by $p_{\rm g}=-\omega_{\rm BCS}$, and
$|\langle\Delta\rangle|$ makes a jump. This discontinuity is shown
in panel d).}
\end{center}
\end{figure}

Depending on the temperature $T$ and the chemical potentials
$\mu_{\sigma}$, the thermodynamic potential density $\omega_{\rm
BCS}$ can give rise to either one, two or three extremal points.
At high temperatures there is only one extremum, namely a global
minimum at $\Delta=0$, so that the system is in the normal phase.
For the well-studied balanced case, $h=0$, BCS theory predicts
that at a certain critical temperature $T_{\rm c}$ the extremum at
$\Delta=0$ becomes a local maximum. The global minimum then
continuously shifts away to a nonzero value of $\Delta$, and the
system enters the superfluid phase. Since the transition evolves
continuously as a function of temperature, it is called a
continuous, or second-order, phase transition and it is shown in
Figs.~\ref{fig:landau}(a) and (b). For $h\neq 0$, the minimum at $\Delta
= 0$ can also be a local minimum, so that there is both a local
maximum and a global minimum at values of $\Delta$ unequal to
zero. As is seen in Figs.~\ref{fig:landau}(c) and (d), this can cause a
discontinuous, or first-order, phase transition. The extrema of
the thermodynamic potential density can be found by
differentiating with respect to $\Delta^*$ and equating the result
to zero. As the above discussion implies, there is always one
solution given by $\Delta =0$. The other solutions are found from
the so-called BCS gap equation
\begin{equation} \label{eq:gap}
\frac{1}{\mathcal{V}}\sum_{\mathbf{k}}\left[
\frac{1-f(\hbar\omega_{+,\mathbf{k}})-f(\hbar\omega_{-,\mathbf{k}})}{
2\hbar\omega_{\mathbf{k}}} -\frac{1}{2\epsilon_{\mathbf{k}}}
\right] = 0~,
\end{equation}
which thus has either one or two solutions. The study of the
extrema of the thermodynamic potential allows for a determination
of the phase diagram as a function of the chemical potentials and
the temperature, which we perform in Section \ref{par:hompd}.

\subsection{Sarma phase}

\begin{figure}[t]
\begin{center}
\includegraphics[width=1.0\columnwidth]{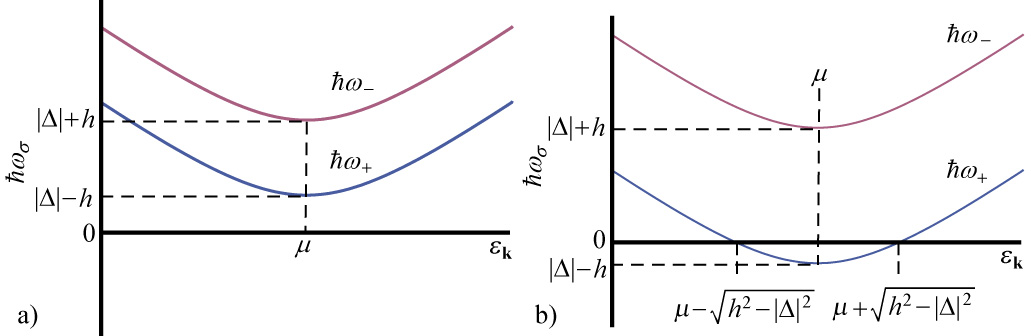}
\caption{\label{fig:disp} Quasiparticle dispersions for a) the BCS
superfluid phase and b) the Sarma superfluid phase. The upper
branch gives the dispersion for the spin-down quasiparticles, that
consist of spin-down particles and spin-up holes. The lower branch
gives the dispersion for the spin-up quasiparticles, that consist
of spin-up particles and spin-down holes. These dispersions have
their minima given by $|\Delta|\pm h$ at wavevectors for which
$\epsilon_{\bf k}=\mu$. In the BCS case the quasiparticle spectra
are gapped. In the Sarma case, a part of the spin-up quasiparticle
branch is below zero, such that its filling lowers the
ground-state energy. As a result, additional spin-down holes and
spin-up particles enter the ground state leading to a polarized
superfluid.}
\end{center}
\end{figure}

But first, let us briefly discuss in more detail the homogeneous
superfluid phases that we encounter in the spin-imbalanced case. Below the critical temperature $T_{\rm c}$, we have that
$|\Delta|\neq 0$, in which case we distinguish between two
possibilities. Namely, we have either that $h<|\Delta|$, or that
$h>|\Delta|$. The first case we call a BCS superfluid, because, as
we see next, it corresponds to the fully-gapped situation known
from ordinary BCS superconductivity in metals \cite{Bardee57a}. The
second case leads to a so-called Sarma superfluid, which gives
rise to a gapless quasiparticle dispersion for the majority spin
species $\hbar\omega_{{\bf k},+}$, as was first discussed by Sarma
\cite{Sarma63a}. The Sarma phase is sometimes also referred to as the
interior-gap phase or the breached-pair phase
\cite{Liu03a,Parish07a}. To see the difference between the two
phases more clearly, we discuss the notion of a Bogoliubov quasiparticle and
the meaning of its dispersion in more detail. A spin-up (down)
quasi-particle is a linear combination of a spin-up (down)
particle and a spin-down (up) hole, see e.g. \cite{Bruus04a}, namely
\begin{eqnarray}
\hat{b}^{\dagger}_{+,{\bf k}} &=& u_{\bf k}
\hat{a}^{\dagger}_{+,{\bf k}} -v_{\bf k}\hat{a}_{-,-{\bf k}},\\
\hat{b}^{\dagger}_{-{\bf k},-} &=& u_{\bf k}
\hat{a}_{+,{\bf k}} +v_{\bf k} \hat{a}^{\dagger}_{-,-{\bf k}},
\end{eqnarray}
where $\hat{a}^{\dagger}_{\sigma,{\bf k}}$ creates a particle with
wavevector ${\bf k}$ and spin $\sigma$, while
$\hat{b}^{\dagger}_{\sigma,{\bf k}}$ creates a quasiparticle. Note
that we have chosen real coherence factors $u_{\bf k}$ and $v_{\bf
k}$. Quasiparticles behave like ordinary particles in the sense
that they can be assigned an energy, wavevector and spin.

Physically, a quasiparticle excitation describes a single-particle
excitation on top of the Cooper pair condensate. To make such an
excitation, first a Cooper pair has to be separated into two
uncorrelated atoms, after which one of the particles is taken out
of the system, while the other remains. The remaining particle
then determines the spin of the quasiparticle excitation. The
energy difference between the excitations of different spin is
thus $2 h$, namely the difference in the chemical potentials
between the two spin states. The quasiparticle dispersions are
shown in Fig.~\ref{fig:disp}, where panel (a) corresponds to the
BCS case and panel (b) to the Sarma case. The upper curve in
Fig.~\ref{fig:disp}(a) shows the spin-down quasiparticle spectrum
$\hbar\omega_{-,{\bf k}}$ and the lower curve shows the spin-up
quasiparticle spectrum $\hbar\omega_{+,{\bf k}}$. It is seen that
it cost a certain nonzero amount of energy to make an excitation
in one of the two branches, which are therefore said to be fully
gapped. At zero temperature, both branches are completely empty,
so that only the equal-density BCS ground state, describing the
condensate of Cooper pairs, remains. In Fig.~\ref{fig:disp}(b), the
same quasiparticle dispersions are drawn for the gapless Sarma
phase, which arises when $h>|\Delta|$. The curves have the same
interpretation as for the gapped BCS case, however, now the branch
for the majority spin-up quasiparticles $\hbar\omega_{+,{\bf k}}$
goes through zero at the wavevectors ${\bf k}$ satisfying $\epsilon_{\bf
k}=\mu\pm \sqrt{h^2-|\Delta|^2}$. As a result, this additional
part of the quasiparticle branch below zero is filled at zero
temperature, because in this way the ground-state energy is
lowered. The spin-up quasiparticles make the Cooper pairs vanish
in the corresponding momentum range, so that here only spin-up
particles remain and the superfluid becomes polarized.

\begin{figure}[t]
\begin{center}
\includegraphics[width=0.8\columnwidth]{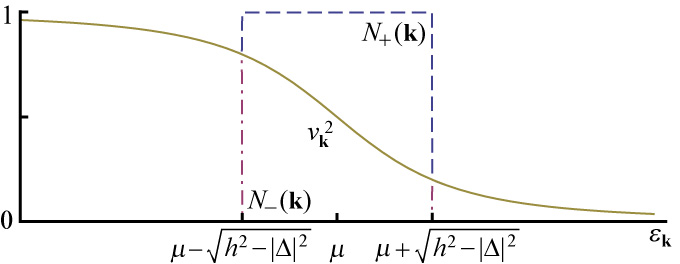}
\caption{\label{fig:occnr} Average occupation numbers $N_{\sigma}({\bf k})$ of single-particle quantum states with momentum
${\bf k}$ and spin $\sigma$ for the Sarma phase. At wavevectors
for which $\epsilon_{\bf k}<\mu - \sqrt{h^2-|\Delta|^2}$ or
$\epsilon_{\bf k}>\mu + \sqrt{h^2-|\Delta|^2}$, $N_{\sigma}({\bf k})$ equals $v^2_{\bf k}$ corresponding to the occupation numbers of an equal-density BCS superfluid. For wavevectors in
between, $N_{+}({\bf k})=1$ (dashed line) and $N_{-}({\bf k})=0$ (dash-dotted line), so that the
superfluid becomes polarized.}
\end{center}
\end{figure}

We see this more clearly if we explicitly calculate the average
occupation number for a single-particle quantum state that is
specified by the wavevector ${\bf k}$ and spin $\sigma$. We
designate this occupation number with $N_{\sigma}({\bf k})$ and it
is given by the expression that is summed over in Eq.\
(\ref{eq:nsf}), namely
\begin{equation}\label{eq:occnr}
N_{\sigma}({\bf k})= |u_{\mathbf{k}}|^2
f(\hbar\omega_{\sigma,\mathbf{k}}) +|v_{\mathbf{k}}|^2 [1-f
(\hbar\omega_{-\sigma,\mathbf{k}})].
\end{equation}
Indeed, summation over all occupation numbers $N_{\sigma}({\bf k})$
yields the total number of particles in state $\sigma$. Due to the
Pauli principle, $N_{\sigma}({\bf k})$ can be maximally unity. In
Fig.\ \ref{fig:occnr}, we show the result for $N_{\sigma}({\bf k})$
both in the case of a BCS superfluid and a Sarma superfluid at
zero temperature. In the BCS case of Fig.\ \ref{fig:disp}(a), we
obtain from Eq.\ (\ref{eq:nsf}) that the occupation for both spin
states is given by $N_{\pm}({\bf k})=v_{\bf k}^2$ at zero
temperature, so that the particle numbers are the same and the
superfluid is fully balanced. However, in the Sarma case of Fig.\
\ref{fig:disp}(b), the additional spin-up quasiparticles, which
create extra spin-down holes and spin-up particles, are seen to
cause a full polarization near the average chemical potential
$\mu$. For the wavevectors leading to a positive quasiparticle
dispersion $\hbar\omega_{+,{\bf k}}$, the BCS behavior is
recovered. The resulting configuration for the Sarma phase is
sometimes also referred to as `phase separation' in momentum
space, because for small $|{\bf k}|$ and for large $|{\bf k}|$ the system has the superfluid-state occupation numbers from BCS theory, while for $|{\bf k}|$ around $(2 m \mu)^{1/2}/\hbar$ the states are occupied as a fully polarized normal state. Fig.\ \ref{fig:occnr} shows that the Sarma phase can also be defined at zero temperature as a superfluid with a Fermi surface, or with multiple Fermi surfaces. At nonzero temperatures, the sudden rising and lowering of
the occupation numbers in Fig.~\ref{fig:occnr} becomes smoother \cite{Yi06a},
until at high temperatures this nonmonotonic behavior completely disappears.

If the Sarma phase were stable at zero temperature, it would be
separated from the BCS phase by a true quantum phase transition,
meaning that non-analytic behavior in thermodynamic quantities
would be observed as the majority quasiparticle spectrum became
gapless. However, as we find in the next section, this
gapped-gapless transition is preempted by a first-order phase
transition to the normal state, so that the situation
$|\Delta|>h$ is found to be never stable at unitarity. More
precisely, if we study  at zero tempertaure the behavior of the equilibrium order parameter $\langle\Delta\rangle$, corresponding to the global minimum of the
thermodynamic potential, then we find that when $|\langle\Delta
\rangle| = 0.70 h$, a first-order transition to the normal state takes
place, where the latter remains the global minimum for larger $h$.
However, at nonzero temperatures, the situation is very different. Namely, close to a second-order phase transition $\langle \Delta \rangle$ becomes arbitrarily small, and the
condition for stable gapless superfluidity $|\langle \Delta
\rangle| < h$ is readily satisfied. Moreover, in the case of a
large mass imbalance between the fermionic species, the Breached-Pair 1 phase (Sarma phase with a single Fermi surface) is expected to be stable all the way down to zero temperature \cite{Parish07a}. The same holds for the BEC side of the resonance where the Sarma or Breached-Pair 1 phase is stable at zero temperature according to mean-field theory \cite{Sheehy06a}. The sharp Fermi surfaces in the momentum distribution of Fig. \ref{fig:occnr}  can be used to detect the Sarma phase experimentally and distinguish it from the BCS phase. A precise scheme to measure such momentum distributions through time-of flight Raman imaging was proposed by Yi and Duan \cite{Yi06a}. At nonzero temperatures, the Fermi surfaces of the Sarma phase are smoothened, so that the evolution from the gapped to the gapless regime is no longer a phase
transition, but rather turns into a smooth crossover. As a result, there is
no non-analytic thermodynamic behavior during this evolution anymore. The sharp Fermi surfaces have now become continuous, but still there are characteristic bumps and dips in the momentum distributions. These non-monotonic distributions could be a clear experimental sign of being in the Sarma regime \cite{Yi06a}. Whether or not this non-monotonic behavior can be observed experimentally at unitarity for the mass-balanced case is an open question, both theoretically and experimentally. For the mass-imbalanced case and on the BEC side of the resonance, the Sarma phase is stable on the mean-field level down to very low, even zero, temperature, resulting in observable Fermi surfaces.

\subsection{Homogeneous phase diagram}\label{par:hompd}

Having looked in more detail at two homogeneous superfluid phases that can occur in the imbalanced system, we are now in the position to determine the
phase diagram for the two-component Fermi mixture in the unitarity
limit. In first instance, we calculate this diagram as a function
of the temperature and the chemical potentials. This means that for any
given combination of $T$ and $\mu_{\sigma}$, we specify whether
the global minimum of $\omega_{\rm BCS}$ is realized at a zero or
a nonzero order parameter $\langle\Delta\rangle$. This determines
whether we are in a superfluid or in a normal phase. For the
superfluid phase we also specify whether $h$ is larger than
$|\langle\Delta\rangle|$ or not. The first case then leads to the
gapless Sarma regime, while the second leads to the gapped BCS
regime. To calculate the phase
diagram we thus need conditions that specify the phase boundary between
the superfluid and normal phases, for which there are two
possibilities. Namely the transition can be either continuous or
discontinuous. This was illustrated in Fig.\ \ref{fig:landau}.

\begin{figure}
\begin{center}
\includegraphics[width=0.6\columnwidth]{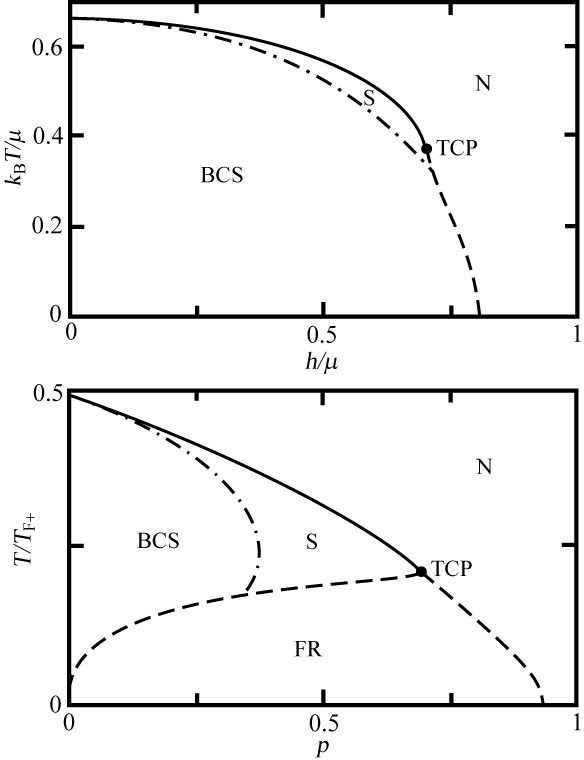}
\caption{\label{fig:hpdmf} a) The phase diagram of the homogeneous
two-component Fermi mixture in the unitarity limit, consisting of
the gapless superfluid Sarma phase (S), the gapped superfluid BCS
phase and the normal phase (N). The transition from superfluid to
normal can be either continuous (full line) or discontinuous
(dashed line), and the two possibilities meet at the tricritical
point (TCP). Between the BCS regime and the Sarma regime of
superfluidity there is a crossover (dash-dotted line). Both the
temperature $T$ and half the chemical potential difference $h$ are
scaled with the average chemical potential $\mu$. b) The same
diagram but now as a function of the polarization
$p=(n_{+}-n_{-})/(n_{+}+n_{-})$ and with the temperature scaled by
the Fermi temperature of the majority species $T_{\rm
F+}=\epsilon_{{\rm F}+}/k_{\rm B}$. Due to the discontinuous
nature of the transition below the tricritical point there is a
jump in the polarization, causing a forbidden region (FR) in the
phase diagram where the gas is unstable against phase separation.}
\end{center}
\end{figure}

For the continuous phase transition, the condition is given by the
minimum at $\Delta=0$ becoming a maximum, or equivalently, the
second derivative of $\omega_{\rm BCS}$ changing sign from
positive to negative. The criterion thus becomes
\begin{eqnarray}\label{eq:alphaL}
\alpha_{\rm L}(T_{\rm c})&=&\left.\frac{\partial \omega_{\rm
BCS}[\Delta;T_{\rm c},\mu_{\sigma}]}{\partial
|\Delta|^2}\right|_{|\Delta|=0}\\
&=& \frac{1}{\mathcal{V}}\sum_{\mathbf{k}}\left[
\frac{1-f_+({\mathbf{k}})-f_-({\mathbf{k}})}{
2(\epsilon_{\mathbf{k}}-\mu)} -\frac{1}{2\epsilon_{\mathbf{k}}}
\right] = 0~,\label{lanalp}\nonumber
\end{eqnarray}
where we introduced the shorthand notation $f_{\sigma}({\bf
k})=1/[e^{\beta(\epsilon_{\mathbf{k}}-\mu_{\sigma})}+1]$. Note
that this condition is only valid when the minimum of $\omega_{\rm
BCS}$ at $|\Delta|=0$ is a global minimum before turning into a
maximum, as in Fig.\ \ref{fig:landau}(a). This is something that we have
to check. It is certainly the case if, at the critical temperature
$T_{\rm c}$, $\omega_{\rm BCS}$ only has nonnegative and positive
coefficients for its expansion in $|\Delta|^2$ , which is true for
the balanced case, $h=0$. Following the line of second-order phase
transitions for increasing $h/\mu$, see the solid line in
Fig.~\ref{fig:hpdmf}(a), then the first higher-order coefficient of
$\omega_{\rm BCS}$ that turns negative is the fourth-order one,
which is proportional to
\begin{eqnarray}\label{eq:betaL}
\beta_{\rm L}&=&\left.\frac{\partial^2 \omega_{\rm
BCS}[\Delta;T,\mu_{\sigma}]}{(\partial
|\Delta|^2)^2}\right|_{|\Delta|=0}\nonumber\\
&=& \frac{1}{\mathcal{V}}\sum_{\mathbf{k}}\frac{\beta}
{4(\epsilon_{\bf k}-\mu)^2}\left[
\frac{1-f_+({\mathbf{k}})-
f_-({\mathbf{k}})}{\beta(\epsilon_{\mathbf{k}}-\mu)}\right.\\
&+&\left. \sum_{\sigma}
f_{\sigma}({\mathbf{k}}) (f_{\sigma}({\mathbf{k}})-1)\right]. \nonumber
\end{eqnarray}
When both $\alpha_{\rm L}=\beta_{\rm L}=0$, then we are at the
so-called tricritical point (TCP). Here, the continuous phase
transition turns into a discontinuous one. Below the tricritical
point, the condition for the first-order transition line is given
by the disconnected normal and superfluid minima having an equal
grand-potential density, or $\omega_{\rm BCS}[0;T_{\rm c
},\mu_{\sigma}]=\omega_{\rm BCS}[\langle\Delta\rangle;T_{\rm c
},\mu_{\sigma}]$. This is illustrated in
Fig.~\ref{fig:landau}(c), where the corresponding behavior of the order
parameter as a function of temperature is sketched in panel (d).

The conditions for the phase boundaries give rise to the phase diagram
of Fig.~\ref{fig:hpdmf}(a). Also drawn is the crossover between
gapped BCS and gapless Sarma superfluidity, given by the condition
$|\langle\Delta\rangle| = h$. A special feature of the phase
diagram is that it has a certain kind of universality, due to the
divergence of the scattering length \cite{Ho04a}. Namely, because
now the interaction strength does not provide a physical energy
scale, the only energy scales left are associated with the
temperature and the particle densities in the system. The latter
scales can be characterized by the chemical potentials
$\mu_{\sigma}$ or the Fermi energies $\epsilon_{{\rm
F}\sigma}=\hbar^2(6\pi^2n_{\sigma})^{2/3}/2m$. By scaling all
energies in Fig.~\ref{fig:hpdmf}(a) with the average chemical
potential $\mu$, the phase diagram becomes only dependent on
$T/\mu$ and $h/\mu$ and not on the specific value of $\mu$
anymore. At zero temperature, the first-order superfluid-normal
transition occurs when $h_{\rm c}= 0.81 \mu$, at which the gap
jumps from $|\langle \Delta \rangle|=1.16 \mu$ to zero. This critical value for $h$ is sometimes also called the Chandrasekhar-Clogston limit \cite{Chandr62a, Clogst62a}. We thus
have that $|\langle \Delta \rangle| > h_{\rm c}$, so that the
transition is between an equal density superfluid and a polarized
normal state. This means that according to mean-field theory, the
Sarma phase does not occur at zero temperature.

In the normal region of the phase diagram, we could in principle
make a further distinction between two different normal phases,
namely a partially polarized normal phase, where the minority
species is still present, and a fully polarized normal state,
where only the majority species remains. At zero temperature, the
transition between this partially polarized phase and the fully
polarized phase happens when $h=\mu$ in mean-field theory. In the
phase diagrams that follow we typically do not explicitly make the
distinction between the two cases, so that we refer to both
phases together simply as the normal phase.

The first-order transition line in Fig.\ \ref{fig:hpdmf}(a) is
characterized by a jump in the order parameter $\langle \Delta
\rangle$. From Eq.\ (\ref{eq:nsf}), we find that this causes
also a jump in the particle densities and in the polarization
$p=(n_{+}-n_{-})/(n_{+}+n_{-})$. At zero temperature the
discontinuity in polarization is largest and mean-field theory
predicts a jump from an unpolarized superfluid, $p=0$, to a normal
state with a critical polarization of $p_{\rm c}=0.93$. When we
calculate for each point in the phase diagram of Fig.\
\ref{fig:hpdmf}(a) the corresponding particle densities, we find the
diagram of Fig.\ \ref{fig:hpdmf}(b). The phase boundaries are given
as a function of $p$ and $T/T_{\rm F +}$, where the Fermi
temperature is defined by $k_{\rm B}T_{\rm F\sigma}=\epsilon_{{\rm
F}\sigma}$. The phase diagram is again universal, so that the
phases are uniquely determined by the polarization $p$ and the
ratio $T/T_{{\rm F}\sigma}$. The diagram in Fig.\ \ref{fig:hpdmf}(b)
is seen to have a forbidden region (FR), which means that the
combinations of temperature and polarization inside that region
are not stable. At zero temperature, we see for example that the
whole region between $p=0$ and $p=0.93$ is forbidden. However, as
mentioned in the introduction to this chapter, experimentally the
polarization can be fixed at any value. If the system is forced to
be in the forbidden region at a certain polarization $p_{\rm f}$,
then phase separation occurs \cite{Bedaqu03a, Cohen05a}. It means that the system forms both
superfluid domains with low polarization and normal domains with
high polarization, so that in total $p=p_{\rm f}$ is satisfied. We end this discussion by noting that the present mean-field study so far only considers homogeneous $s$-wave superfluid phases. There have been theoretical proposals claiming that small parts of the phase diagram at unitarity are occupied by more exotic superfluid phases, such as FF or LO-like superfluids \cite{Bulgac08a}, or induced $p$-wave superfluids \cite{Patton11a}. These more exotic possibilities are not included by the present mean-field study. So far, they have also not yet been seen in experiments.

\subsection{Local-density approximation}\label{par:lda}

All experimental set-ups for ultracold atomic quantum gases invoke
an external trapping potential, which is needed to keep the atom
cloud together for the duration of the experiment. The trapping
potential is created by an external magnetic field, that acts on
the magnetic dipole moment of the atoms, or by the strong electric
field in a laser beam, which induces an electric dipole moment in
the atoms. As a result, inhomogeneous magnetic or electric fields
become potential energy landscapes for the atoms, with which the
particles can be confined in space without using material walls.
Since the cold atoms accumulate around the minimum of the
potential energy, the trap can typically be well approximated by a
harmonic potential. Thus, if we want to describe an actual quantum
gas experiment, we inevitably must study the effect of the
external potential. This can be somewhat inconvenient, because
inhomogeneous systems are typically more cumbersome to deal with theoretically.
Fortunately, if the trapping potential is locally flat enough,
then we may consider the gas to be homogeneous at that point in
the trap. This is the physical essence of the local-density
approximation (LDA). The flatness condition implies that the
trapping potential has to vary slowly compared to the typical
local de Broglie wavelength of the particles. As a result, LDA
can be seen as a WKB or a semiclassical approximation.

\begin{figure}[t]
\begin{center}
\includegraphics[width=1.0\columnwidth]{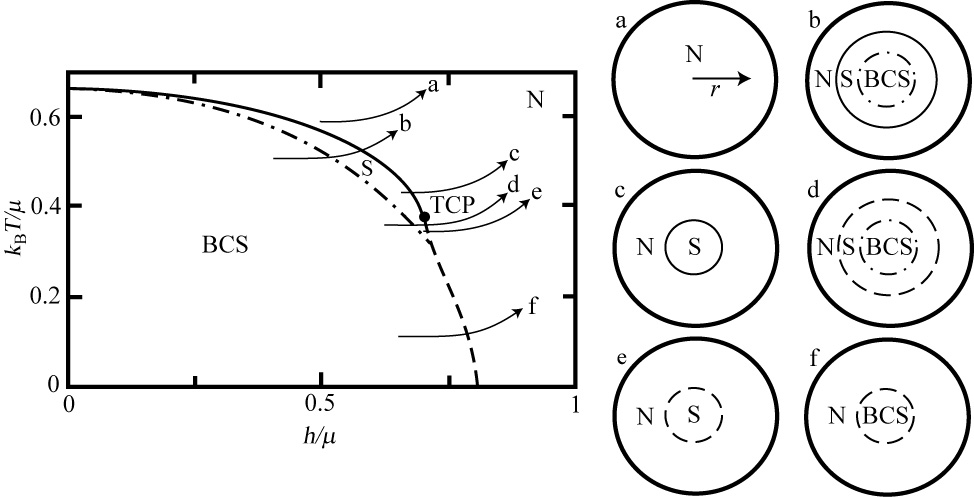}
\caption{\label{fig:lda} The homogeneous phase diagram of
Fig.~\ref{fig:hpdmf}, but now with 6 arrows (a)-(f) that represent 6
different polarized phases in a trap. The tail of the arrow
represents the center of the trap, while the decrease of $\mu({\bf
r})$ for increasing radius $r$ causes the arrow to move up and to
the right, leading the arrow to the normal phase. For each arrow,
we also draw a schematic representation of the corresponding
trapped phase. The center of the spheres correspond to the centers
of the trap, while the thick full line denotes the edge of the
cloud. The core of the cloud can be either in the normal phase
(N), the gapped superfluid phase (BCS) or the gapless superfluid
phase (S). These locally homogeneous phases are separated in the
trap by either a second-order transition (thin full line), a
first-order transition (dashed line) or a crossover (dash-dotted
line).}
\end{center}
\end{figure}

If we apply the condition for the validity of the WKB
approximation \cite{Bransd00a} along a certain direction $r_i
\in \{x,y,x\}$ in the trap, we obtain
\begin{equation}\label{eq:lda}
\frac{\hbar m \partial V^{\rm ex}({\bf r})/\partial r_i}{[2m
\epsilon_{\rm F\sigma}({\bf r})]^{3/2}}=\frac{r_i/l_{i}}{6\pi^2
n_{\sigma}({\bf r})l_{i}^3}\ll 1,
\end{equation}
where the trapping potential is given by $V^{\rm ex}(x,y,z)=(m
\omega_x^2 x^2+m \omega_y^2 y^2+m \omega_z^2 z^2)/2$, the harmonic
oscillator length in the direction of $r_i$ yields
$l_{i}=(\hbar/m\omega_{i})^{1/2}$, and where the Fermi energy
$\epsilon_{\rm F\sigma}({\bf r})=\hbar^2(6 \pi^2 n_{\sigma}({\bf
r}))^{2/3}/2m$ sets locally the typical kinetic energy scale for
the atoms. For the early MIT experiments, we have
$\omega_x=\omega_y=2\pi \times 115$ s${}^{-1}$ in the radial
direction, and $\omega_z=2\pi \times 22.8$ s${}^{-1}$ in the axial
direction. Moreover, the total number of trapped atoms is about
$10^7$, leading to central densities on the order of $10^{18}$
m${}^{-3}$ \cite{Zwierl06b}. Plugging in the numbers,
we see that the MIT experiments indeed easily satisfy the
condition for the local-density approximation in both the radial
and the axial direction. Only near the edge of the cloud, where
the atomic densities become very small, the approximation breaks
down. However, for the early Rice experiments, we have that
$\omega_x=\omega_y=2\pi \times 350$ s${}^{-1}$ and $\omega_z=2\pi
\times 7.2$ s${}^{-1}$, while their total number of particles is
about a $10^5$ \cite{Partri06a}. As a result, a question mark
arises over the validity of the local-density approximation, when
they study very large imbalances. Namely, then the left-hand side
of Eq.\ (\ref{eq:lda}) is not much smaller than unity anymore for
the minority density in the steep radial direction.

However, in the rest of this section we focus more on the MIT
experiments, namely we assume that it is valid to use the LDA. The
approximation then conveniently allows us to absorb the trapping
potential in the chemical potentials and use locally the
homogeneous theory that was developed in the previous section. Moreover, we assume that we have conveniently rescaled the coordinate $r_i$ by a factor $\bar{\omega}/\omega_i$ with $\bar{\omega}=(\omega_x\omega_y\omega_z)^{1/3}$, so that the external potential becomes spherically symmetric. As
a result, we have for the spherically symmetric trap in rescaled
units that $\mu_{\sigma}({\bf r })=\mu_{\sigma}-V^{\rm ex}({\bf
r})=\mu_{\sigma}-m\bar{\omega}^2r^2/2$, which leads for the local average
chemical potential to $\mu({\bf r })=\mu-V^{\rm ex}({\bf
r})=\mu-m\bar{\omega}^2r^2/2$, while (half) the difference remains
constant, i.e. $h({\bf r })=h$. Moreover, we can calculate the total
particle number $N_{\sigma}$ with spin $\sigma$ in the trap by
\begin{equation}\label{eq:ntot}
N_{\sigma}=\int d{\bf r}~ n_{\sigma}[\langle\Delta\rangle({\bf
r});T,\mu_{\sigma} ({\bf
r})],
\end{equation}
where the local particle densities
$n_{\sigma}[\langle\Delta\rangle({\bf
r});T,\mu_{\sigma}({\bf r})]$ are
given by Eq.\ (\ref{eq:nsf}). Since at position ${\bf r}$ in the
trap the homogeneous phase is realized that corresponds to
$T/\mu({\bf r})$ and $h/\mu({\bf r})$, we can already predict what
the trapped configurations look like by considering only the
homogeneous phase diagram of Fig.\ \ref{fig:hpdmf}(a). Namely, we
may draw an arrow in the homogeneous diagram that precisely
follows those ratios $T/{\mu ({\bf r})}$ and $h/{\mu ({\bf r})}$
that are encountered in the trap \cite{Parish07b}. We put the tail of
the arrow at $T/{\mu ({\bf 0})}$ and $h/{\mu ({\bf 0})}$
corresponding to the center of the trap. For increasing $r$ or
decreasing $\mu ({\bf r})$, both ratios increase and the arrow
consequently moves to the right and to the above in the diagram,
so that we end up in the normal phase. Only when both $h=T=0$, we
stay in the origin of the diagram and superfluidity is encountered
throughout the trap. For $h=0$ and $T\neq 0$, the arrow moves
upward along the vertical axis and we encounter a second-order
superfluid to normal transition in the trap.

However, we are interested in the polarized case, so that we
consider $h>0$. Then, the arrows can follow six different types of
paths, corresponding to six trapped configurations, as shown in Fig.\
\ref{fig:lda}. The first case is when the whole trap is in the
normal phase as follows from arrow (a). The corresponding schematic
representation of the trapped  configuration is depicted on the right in
Fig.~\ref{fig:lda}. In this representation, we schematically
show which phases are encountered as a function of the radius $r$
in the trap. The thick circle then represents the edge of the
cloud. The second case is indicated by arrow (b), where the cloud
has a BCS superfluid core. For increasing $r$, we now first
encounter a crossover to the gapless Sarma regime, after which
there is a second-order phase transition to the normal phase. The
third case, arrow (c), leads to a gapless superfluid core, which is
separated from the normal outer region by a second-order
transition. The fourth case, arrow (d), has a fully gapped
superfluid core, surrounded by a Sarma shell, which is separated
from the normal edge by a first-order transition. The fifth case,
arrow (e), has a gapless superfluid core, and the transition to the
normal state in the trap is of first order. The sixth case, arrow
(f), is similar to the fifth case, only the core is now fully
gapped.

In the next section, we present the universal phase diagram for
the strongly interacting Fermi mixture in a trap. Although we have
specified 6 trapped `phases', some of them only differed by the
presence of a crossover. Since a crossover connects different
physical regimes in a smooth manner, these regimes are strictly
speaking not different thermodynamic phases due to the absence of
a true phase transition. In that sense the Sarma phase and the BCS
phase are two different limits of the same polarized superfluid
phase at nonzero temperatures. In the BCS limit, the polarization
is caused by thermal excitation of the gapped quasiparticle
branches, whereas in the Sarma limit the gapless branch in
principle does not need temperature to be filled. If we choose to
classify the trapped phases solely in terms of the true phase
transition that occurs in the trap, only three different trapped
phases remain. First of all there is the normal phase, given by
arrow (a), for which there is no transition in the trap. The second
phase has a second-order transition in the trap, such that it
encompasses both arrows (b) and (c). Note that the phases (b) and (c)
have in common that there is always a region of Sarma
superfluidity present. Therefore, we call the phase with a
second-order transition in the trap also the Sarma phase. The
third trapped phase has a first-order transition in the trap and
thus encompasses arrows (d), (e) and (f). Due to the presence of a
first-order interface, we also call it the phase-separated phase.
The three different trapped phases are also illustrated in
Fig.~\ref{fig:3d}.

\subsection{Phase diagram in a trap}\label{par:pdtrap}

\begin{figure}[t]
\begin{center}
\includegraphics[width=0.8\columnwidth]{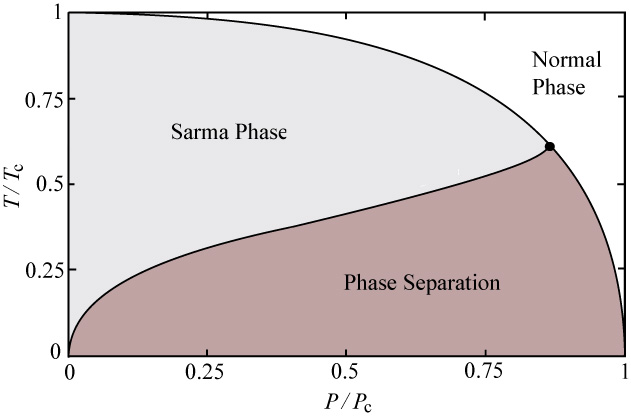}
\caption{\label{fig:pdtrap} Universal phase diagram of a trapped
imbalanced Fermi gas in the unitarity limit. The polarization $P$
is given by $(N_{+}-N_{-})/(N_{+}+N_{-})$, where $N_{\pm}$
designates the number of fermions in each hyperfine state of the
Fermi gas. The temperature $T$ is scaled with the critical
temperature $T_{\rm c}$ of the balanced trapped Fermi gas and the polarization is scaled with the critical polarization at zero temperature $P_{\rm c}$. Using mean-field theory, we have that $k_{\rm B} T_{\rm c} = 0.66 \mu$ with $\mu$ the average chemical potential at the center of the trap, while $P_c = 0.998$. }
\end{center}
\end{figure}

The main results of the mean-field calculations in the trap are
presented in Fig.~\ref{fig:pdtrap}. Here, we show the universal phase
diagram of a trapped Fermi gas in the unitarity limit as a
function of temperature and polarization \cite{Gubbel06a}.
This phase diagram is universal in the sense that it does not
depend on the total number of fermions or the trap geometry.
Fig.~\ref{fig:pdtrap} reveals that there is a tricritical point for the
trapped Fermi mixture, which was also the case for the
homogeneous gas \cite{Sarma63a,Parish07b,Combes04a}. As explained in the previous
section, the normal phase means that the gas is in its normal
state throughout the trap. In the Sarma phase the Fermi gas has a
shell structure, in which the core of the trapped gas is
superfluid, whereas the outer region is normal. Furthermore, the
normal-to-superfluid transition as a function of the position in
the trap is of second order. Since the superfluid order parameter
$\langle\Delta\rangle$ vanishes continuously at the transition, we
have for nonzero polarizations always a region in the trap where
$|\langle\Delta\rangle|({\bf r })$ is so small that it results in a gapless
superfluid with negative quasiparticle excitation energies, as
first studied by Sarma \cite{Sarma63a}. Since
$|\langle\Delta\rangle|({\bf r })$ increases towards the center of the trap,
it is also possible that the superfluid becomes gapped in the
center of the trap. This leads to a gapped BCS superfluid core
with a gapless Sarma superfluid and normal shell surrounding it
\cite{Shin06a}. Finally, in the phase-separated region of the
phase diagram, the superfluid core and the normal shell of the gas
are separated by a first-order transition as a function of
position, which implies that $\langle\Delta\rangle({\bf r })$ goes
discontinuously to zero at a certain equipotential surface in the
trap.

Fig.~\ref{fig:pdtrap} allows for a natural explanation of the
qualitative differences in the early observations by the
experimental groups at MIT \cite{Zwierl06b} and Rice university
\cite{Partri06a}. More precisely, we could argue that the initial
experiments of Zwierlein {\it et al}. have observed the smooth transition
from the normal phase to the Sarma phase at a large polarization
\cite{Zwierl06b}, implying that these early experiments have been
performed above the temperature of the tricritical point.
Moreover, we then suggest that the experiments of Partridge {\it et
al}.\ have been performed in the temperature regime below the
tricritical point, since these experiments appear to see the
transition between a non-phase-separated and a phase-separated
superfluid phase at small polarization \cite{Partri06a}. Note that this explanation based on mean-field theory can only qualitatively account for the major initial discrepancies between the Rice and the MIT experiments, but fails on a quantitative level to describe the strongly interacting experiments accurately. Morefore, as mentioned in Section \ref{par:exp}, to fully understand the early results by the Rice experiment non-equilibrium effects must be taken into account \cite{Liao11a,Parish09a}.

Now, we turn to the question how we actually obtained the phase
diagram of Fig.~\ref{fig:pdtrap}. We first determined the line between
the normal and the two superfluid trapped phases. This is achieved by chosing two central chemical potentials $\mu_{\sigma}({\bf 0})$ and
determining in the center of the trap
the temperature at which the BCS order parameter vanishes. To this end we use the homogeneous theory of Section \ref{par:hompd}, where inspection of the BCS thermodynamic potential revealed that the
vanishing of the order parameter can occur continuously or
discontinuously, i.e., by a second-order or a first-order phase
transition. If the transition is of second order in the center of the trap, we go from
the trapped normal phase to the trapped Sarma phase. If the transition is of first order in the center of the trap, we go from
the trapped normal phase to the trapped phase-separated phase. At the tricritical
point in the center of the trap these two different kinds of transitions merge. This procedure thus gives us two central chemical potentials and a critical temperature of the gas, either above or below the tricritical point. We now still have to calculate the corresponding total atom number and polarization with the use of Eq. (\ref{eq:ntot}) in order to find a point in the phase diagram of Fig.\ \ref{fig:pdtrap}, that lies on the line between the normal phase and the two superfluid trapped phases. Note that it is convenient that the phase diagram is universal, so that the diagram is independent of the total atom number. This feature saves us an additional iteration procedure compared to the case when it is necessary to calculate the phase diagram for a certain specific total atom number.

So far, we looked primarily at the center of the trap, but the tricritical
condition can also be satisfied at a certain equipotential surface
outside the center of the trap. This gives us a point on the
Sarma-to-phase-separation line \cite{Gubbel06a}. To see this, consider a point on
this line and raise the temperature slightly. This changes the
tricritical transition outside the center of the trap into a
second-order transition slightly closer to the center of the trap,
which means that the gas is in the Sarma phase. In a similar way,
a slightly lower temperature leads to a first-order transition as
a function of position in the trap, i.e. to the phase-separated
phase. So the procedure now gives us by construction a tricritical temperature and two chemical potentials at a certain radius $r_{\rm c3}$ outside of the trap center. To calculate the corresponding total atom number and polarization we use again Eq. (\ref{eq:ntot}), where we note that for $ r< r_{\rm c3}$ the gas is superfluid, while for $r>r_{\rm c3}$ the densities are normal.

\subsection{Mass-imbalanced case}\label{par:mi}

So far, we have explored the mean-field phase diagram of the strongly-interacting Fermi gas as a function of temperature and spin polarziation. Most recently, experiments have indicated that the physical
consequences of yet another parameter can be explored, namely that
of a mass imbalance between the two fermionic components. A very
promising mixture in this respect consists of ${}^6$Li and
${}^{40}$K, which has a mass ratio of 6.7. Several accessible
Feshbach resonances are identified in the mixture \cite{Wille08a,Tiecke10a},
while both species have also been simultaneously cooled into the
degenerate regime \cite{Taglie08a,Voigt09a}. In the unitarity limit, the size of the
Cooper pairs is comparable to the average interparticle distance
and the pairing is a many-body effect. The mass imbalance has a
profound effect on the pairing, because it affects the way in which the two
Fermi spheres overlap. Theoretical studies at zero temperature using mean-field theory \cite{Iskin06a} and Monte-Carlo simulations \cite{Stecher07a} considered the BEC-BCS crossover in this system. We show next that  for the large mass ratio of the ${}^6$Li-${}^{40}$K mixture, the phase diagram in the unitarity limit is even richer at nonzero temperatures than for the mass-balanced case. Similar to the solely
spin-imbalanced case is the presence of phase separation
\cite{Iskin06a,Wu06a,Parish07a}, which can occur due to the mismatch of the Fermi
surfaces. Also similar is that gapless Sarma superfluidity is
unstable at zero temperature \cite{Parish07a}, while there is a
predicted crossover to the Sarma phase at nonzero temperatures
\cite{Baarsm10a}. However, the most striking difference
is the presence of a Lifshitz point in the phase diagram \cite{Gubbel09a}.

\begin{figure}[t]
\begin{center}
\includegraphics[width=0.6\columnwidth]{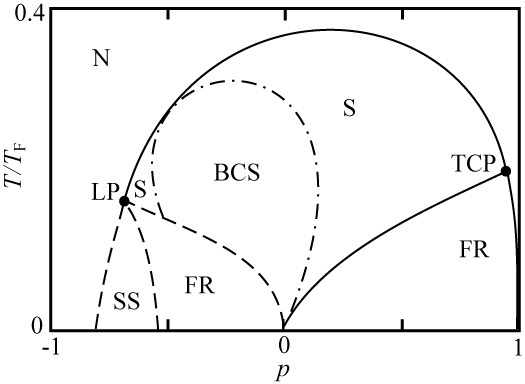}
\caption{ \label{fig:lik} Universal phase diagram for the
homogeneous ${}^6$Li-${}^{40}$K mixture in the unitarity limit as
a function of temperature $T$ and polarization $p$. The
temperature is scaled with the reduced Fermi temperature $k_{\rm
B} T_{\rm F}=\epsilon_{\rm F}=\hbar^2 (3 \pi^2 n)^{2/3}/2m$, where
$m$ is twice the reduced mass and $n$ is the total particle
density. The phase diagram is calculated with mean-field theory. For
a majority of light ${}^6$Li atoms there is a tricritical point
(TCP), at which the normal state (N), the gapless superfluid Sarma
phase (S), and the forbidden region (FR) meet each other. For a
majority of heavy ${}^{40}$K atoms there is a Lifshitz point (LP),
where there is an instability towards supersolidity (SS). The size
of the supersolid stability region is not calculated within our
theory and the dashed lines are therefore only guides to the eye. The dash-dotted line indicates the crossover from gapless Sarma to gapped BCS superfluidity.}
\end{center}
\end{figure}

At a Lifshitz point the transition to the superfluid phase
undergoes a change of character. Rather than preferring a
homogeneous order parameter, the system now becomes an
inhomogeneous superfluid. This exotic possibility was early
investigated for the weakly interacting mass-balanced case by
Larkin and Ovchinnikov (LO), who considered a superfluid with a
single standing-wave order parameter \cite{Larkin65a}. This is
energetically more favorable than the plane-wave case studied by
Fulde and Ferrell (FF) \cite{Fulde64a}. Since the LO phase results in
periodic modulations of the particle densities, it is a special kind of supersolid.
\cite{Bulgac08a,Radzih09a}. The FF and LO phases have intrigued the physics
community for many decades, but so far remained elusive in
experiments with atomic Fermi mixtures. Typically, Lifshitz points
are predicted at weak interactions where the critical temperatures
are very low \cite{Sheehy06a, Son06a, Yoshid07a,Radzih07a}. However, the phase diagram of the ${}^6$Li-${}^{40}$K mixture contains both a Lifshitz and a tricritical point in the unitarity limit, as
shown in Fig.~\ref{fig:lik}. This is in sharp contrast to the mass-balanced
case, where mean-field theory only leads to a
tricritical point at unitarity, which is in agreement with experimental observations \cite{Shin08a}.

As we discussed in Section \ref{par:hompd}, the critical properties of the
superfluid transition in a fermionic mixture are determined by the
effective Landau theory for the superfluid order parameter $\Delta
({\bf r})$. Although we consider no external potential, the order
parameter may still vary in space due to a spontaneous breaking of
translational symmetry. Close to the continuous superfluid
transition, we expand the effective grand-canonical thermodynamic potential
as
\begin{eqnarray}\label{eq:omlan}
\Omega[\Delta] &=&\int d {\bf r}\left\{\gamma_{\rm L}|\nabla \Delta|^2
+\alpha_{\rm L}| \Delta|^2+\frac{\beta_{\rm L}}{2}|\Delta |^4\right\},\quad
\end{eqnarray}
where the challenge is to express the expansion parameters in
terms of the temperature $T$ and the atomic chemical potentials
$\mu_{\pm}$ with the upper (lower) sign referring to the light ${}^6$Li
(heavy ${}^{40}$K) atoms. In this section, we use again mean-field theory to achieve this goal. A phase transition has occurred when the
global minimum of $\Omega$ is located at a nonzero order parameter
$\langle \Delta({\bf r}) \rangle$, which describes a condensate of
pairs. When $\gamma_{\rm L}$ is positive, the pairs have a positive
effective mass and their center-of-mass state of lowest energy is
at zero momentum. Then, we can consider a homogeneous pairing
field $\Delta$, for which a second-order transition occurs at a
critical temperature $T_c$ determined by $\alpha_{\rm L} (T_{\rm c })=0$, for which the expression was given in the mass-balanced case by Eq.~(\ref{eq:alphaL}). For the mass-imbalanced case, the same expression holds, but now the (average) kinetic energy is given by $\epsilon_{\mathbf{k}}=\hbar^2
\mathbf{k}^2/2m$ with $m=2 m_+ m_-/(m_++m_-)$ and the Fermi distributions become $f_{\sigma}({\bf
k})=1/[e^{\beta(\epsilon_{\sigma,\mathbf{k}}-\mu_{\sigma})}+1]$ with
$\epsilon_{\sigma,{\mathbf{k}}}=\hbar^2
\mathbf{k}^2/2m_{\sigma}$.

A continuous transition only occurs when the vanishing minimum at $\Delta = 0$ is the global minimum, which is not necessarily the case. As discussed in Section \ref{par:mftp}, the expansion of $\Omega$ may contain higher
powers of $|\Delta|^2$ that have negative coefficients, leading to
a first-order transition with a jump in the order parameter when
$\Omega[0]=\Omega[\langle \Delta \rangle]$. Second-order behavior
turns into first-order behavior when $\beta_{\rm L}$ becomes negative,
so that the temperature $T_{\rm c3}$ at the tricritical point
(TCP) is determined by $\alpha_{\rm L}(T_{\rm c3})=0$ and $\beta_{\rm L}(T_{\rm
c3})=0$. For the mass-imbalanced case, the expression for $\beta_{\rm L}$ is given by Eq.~(\ref{eq:betaL}) with the same expressions for $m$ and $f_{\sigma}({\bf
k})$ as were just introduced for $\alpha_{\rm L}$. Another possibility is that not $\beta_{\rm L}$,
but rather $\gamma_{\rm L}$ goes to zero. This leads to a Lifshitz
point (LP), which is thus determined by $\alpha_{\rm L}(T_{\rm L})=0$ and
$\gamma_{\rm L}(T_{\rm L})=0$. Since the effective mass of the Cooper
pairs becomes negative below the Lifshitz point, it is
energetically favorable for them to have kinetic energy and form a
superfluid at nonzero momentum. This can be established in many
ways, namely through a standing wave \cite{Larkin65a} or a more
complicated superposition of plane waves \cite{Casalb04a,Mora05a,Yoshid07a, Radzih10a}. Due to
the variety of possibilities it is hard to predict which lattice
structure is most favorable. However, the fact that they all
emerge from the Lifhitz point facilitates the experimental search
for supersolidity in the ${}^6$Li-${}^{40}$K mixture. The supersolid phase could be observed using Bragg spectroscopy.

The mean-field BCS thermodynamic potential density for the
${}^6$Li-${}^{40}$K mixture  is given by Eq.~(\ref{eq:tpdBCS}) with $\epsilon_{\mathbf{k}}$ the average kinetic energy, $m$ twice the reduced mass, and the Bogoliubov
quasiparticle dispersions given by
$\hbar\omega_{\sigma,\mathbf{k}}= \hbar\omega_{\mathbf{k}} -
\sigma[2h - \epsilon_{+,\mathbf{k}}+\epsilon_{-,\mathbf{k}}]/2$ with
$\hbar\omega_{\mathbf{k}}=
\sqrt{(\epsilon_{\mathbf{k}}-\mu)^2+|\Delta|^2}$. We can apply the mentioned critical
conditions to $\omega_{\rm BCS}$ and obtain the mean-field phase
diagram. Although the BCS potential neglects fluctuations in the
order parameter, it is expected that these fluctuation effects only
result in quantitative corrections. We have seen this already for
the strongly interacting experiments in the mass-balanced case,
where the mean-field diagram seems to be qualitatively correct
\cite{Shin08a}. The coefficients determining the second-order phase
transition and the tricritical point are readily calculated from the thermodynamic potential as
$\alpha_{\rm L}=\partial \omega_{\rm BCS}[0;\mu_\sigma]/\partial
|\Delta|^2$\ and $\beta_{\rm L}=\partial^2 \omega_{\rm
BCS}[0;\mu_\sigma]/\partial^2 |\Delta|^2$, resulting in the mentioned expressions. The obtained phase diagram \cite{Gubbel09a} is shown
in Fig.~\ref{fig:lik}, where the polarization is defined as
$p=(n_+-n_-)/(n_++n_-)$ with $n_+$ ($n_-$) corresponding to the Li (K) density. These particle densities are
determined by $n_{\sigma}= -\partial\omega_{\rm BCS}[\langle
\Delta \rangle;\mu_\sigma]/\partial \mu_{\sigma}$. Therefore, the
polarization is discontinuous simultaneously with the order
parameter, which gives rise to a forbidden region (FR) below the
tricritical point.

From Fig.~\ref{fig:lik}, we see that the mean-field phase diagram also
contains a Lifshitz point. It can be calculated from the
noninteracting Green's function (i.e. the propagator) for the Cooper pairs
$G_{\Delta}(i \omega_n,{\bf k})$, which is discussed in more detail in Section \ref{par:nsr}. From Eq.\ (\ref{eq:omlan}), we see that $\gamma_{\rm L}$ can be interpreted as an inverse effective mass of the Cooper pairs, while $-\alpha_{\rm L}$ corresponds to a Cooper-pair chemical potential, and $\beta_{\rm L}$ can be seen as a Cooper-pair interaction. In the normal state, the Cooper-pair
propagator is given by $\hbar G^{-1}_{\Delta}(i
\omega_n,{\bf k})=1/T^{2b}-\hbar\Xi(i \omega_n,{\bf k})$, which is derived in Section \ref{par:nsr}, leading to Eq.\ (\ref{eq:cpgf}). Here, $T^{\rm
2b}=4\pi a\hbar^2/m$ with $a$ the diverging scattering length,
while $\Xi$ is the expression for the so-called ladder diagram, shown in Fig.~\ref{fig:fdcp}. It is given by
\begin{eqnarray}\label{eq:laddermi}
\hbar \Xi(i \omega_n,{\bf k})=\int \frac{d{\bf k' }}
{(2\pi)^3}\left\{\frac{1}{2\epsilon({\bf k'})} + \frac{1-f_+({\bf k'})-f_-({\bf
k-k'})}{i\hbar\omega_n-\epsilon_{+,{\bf k'}}-\epsilon_{-,{\bf
k-k'}}+2\mu} \right\}.
\end{eqnarray}
We have that the mean-field expression for $\alpha_{\rm L}$ is equal to $-\hbar
G^{-1}_{\Delta}(0,{\bf 0})$, while $\gamma_{\rm L}=-\partial
\hbar G^{-1}_{\Delta}(0,{\bf 0})/\partial {\bf k}^2=0$ at the
Lifshitz point. By applying this condition we find that the Lifshitz point occurs for a majority of heavy $^{40}$K atoms, which can be qualitatively explained in the following way. The ideal case for pairing is an equal amount of particles with the same mass, so that the two Fermi surfaces fully overlap. In the case of a mass difference, the Fermi energy of the lighter particles is larger than that of the heavier particles, while the Fermi wavevectors stay the same. If consequently the number of heavy particles is increased, then the difference in Fermi energies is reduced, enhancing the tendency for pairing. Since now the Fermi wavevectors are different, the pairing is expected to occur at nonzero momentum. If the number of light particles is increased, then the difference in Fermi energies is  further enhanced resulting ultimately in phase separation. There is a critical mass ratio for which the phase diagram changes its form from having two tricritical points (like for the mass-balanced case) to a tricritical and a Lifshitz point (like in Fig.\  \ref{fig:lik}). At unitarity, it is given by $r=4.2$ according to mean-field theory. At this multicritical point we thus have that $\alpha_{\rm L} = \beta_{\rm L}=\gamma_{\rm L} = 0$. A continuously tunable mass difference might be achievable with two different atomic species in an optical lattice.

What precisely happens below the Lifshitz point is
an open question for further research  \cite{Casalb04a}. In Fig.~\ref{fig:lik}, we have
sketched a simple scenario, where there is a second-order
transition from the normal to the supersolid phase, for which the
condition is $G^{-1}_{\Delta}(0,{\bf k}_{\rm s})=0$ with ${\bf
k}_{\rm s}$ the wavevector of the supersolid. However, this scenario might not be sufficient since the
transition could in principle also be of first order, where the realized
supersolid periodicity can contain more than one wavevector
\cite{Mora05a}. The transition from supersolidity to the
homogeneous superfluid phase is also expected to be of first
order. The calculation for the stability regions of all possible
supersolid lattices is beyond the scope of this review, so that
the dashed lines in Fig.~\ref{fig:lik} are merely guides to the eye. A more elaborate discussion on various possible supersolid phases is given in Ref. \cite{Casalb04a}.

To include the presence of an external trapping potential, the local-density approach can be used for the (local) superfluid phases, like the BCS phase or the Sarma phase. This gives rise to shell structures for the phases in the trap, just like for the mass-balanced case in Section \ref{par:pdtrap}. For the mass-imbalanced case such an approach was pursued in Ref. \cite{Iskin07a}. However, to treat (non-local) supersolid phases in the trap we need theories that go beyond LDA. The Bogoliubov-de Gennes approach, which will be treated in Section \ref{par:bdg}, is a well-established approach to treat non-local  superfluid phases and it has been applied to the mass-imbalanced case at zero temperature in Ref. \cite{Iskin08a}. Another approach that not only goes beyond LDA, but also beyond mean-field theory is the Monte-Carlo approach, which was applied to few-particle systems by Blume \cite{Blume08a}. This method is restricted so far to zero temperature.

\section{Diagrammatic approaches}\label{sec:diag}

In Section \ref{sec:mf}, we theoretically studied the strongly interacting Fermi gas with a population imbalance using the mean-field BCS thermodynamic potential. In this section, we start from the microscopic Hamiltonian that describes the interacting atomic Fermi gas and derive diagrammatic approaches that go beyond mean-field theory. We mainly use the functional formalism, in which the central object is the fermionic action that belongs to the microscopic Hamiltonian. The diagrammatic approaches in the functional formalism give rise to a versatile theoretical toolbox. This allows us to understand the main shortcomings of the mean-field approximation, which already was adequate to qualitatively describe several aspects of the experiments in the strongly interacting regime. In this section, we try to improve the theoretical calculations so that also quantitative agreement with experiments is reached.

The starting point is the Hamiltonian for interacting
fermions in two different spin states, which we label with a
spin index $\sigma = \pm$. As mentioned before, for the atomic population-imbalanced Fermi gas the spin label represents two different internal hyperfine states of the atomic ground state. Since the resulting hyperfine
space is only two dimensional, it is often also referred to as
a (pseudo)spin-$1/2$ space. We consider utracold atoms that interact via $s$-wave
interactions, which is the dominant scattering mechanism at
the very low energies and momenta of interest, as
explained in Section \ref{sec:scat}. There, it was also shown that
$s$-wave scattering only occurs between fermions in different spin
states. Considering the point interaction $V({\bf r}-{\bf r}')=V_0 \delta ({\bf r}-{\bf r}')$ to model the short range interactions between the atoms, we can write down the following second-quantized Hamiltonian in the grand-canonical ensemble,
\begin{eqnarray}\label{eq:sqham}
\hat{H} \nonumber  &=& \sum_{\sigma=\pm}
 \int d{\bf r}~\hat{ \phi}^{\dagger}_{\sigma}({\bf
r})\left\{ -
\frac{\hbar^2 \boldsymbol{\nabla}^2}{2m_{\sigma}} -
\mu_{\sigma} \right\} \hat{\phi}_{\sigma}({\bf r}) \nonumber \\
&&+ \int d{\bf r}~   V_{0}
\hat{ \phi}^{\dagger}_{+}({\bf
r})\hat{ \phi}^{\dagger}_{-}({\bf r})
\hat{\phi}_{-}({\bf r}) \hat{\phi}_{+}({\bf r}) ~.
\end{eqnarray}
Here, the creation operator $\hat{ \phi}^{\dagger}_{\sigma}({\bf
r})$ (annihilation operator $ \hat{ \phi}_{\sigma}({\bf
r})$)  creates (annihilates) a particle with
mass $m_{\sigma}$ in spin state $\sigma$ at position ${\bf r}$, where \ $\mu_{\sigma}$ is the corresponding chemical potential. In this section, we discuss interacting particles with identical mass so $m_{+} = m_{-}=m$. However, the two chemical potentials can be different in order to account for a difference in the population of the two spin states.

To the second-quantized Hamiltonian of Eq.\ (\ref{eq:sqham}) corresponds a microscopic action $S[\phi]$, which is obtained from a derivation that generalizes Feynman's path integral approach to many-body quantum physics, see e.g. Ref.\ \cite{Negele98a}. The big advantage of the functional or path-integral approach is that the operators $\hat{ \phi}^{\dagger}_{\sigma}({\bf r})$ and $ \hat{ \phi}^{\dagger}_{\sigma}({\bf r})$ make place for fields  $\phi^{*}_{\sigma}(\tau,{\bf r})$ and $\phi_{\sigma}(\tau,{\bf r})$, which are often easier to deal with in calculations. The variable $\tau$ is then an (imaginary) time variable, that can be used to incorporate either the dynamics, or, as in our case, the equilibrium quantum statistics of the system. The position vector is given by ${\bf r}$. The fields are ordinary complex numbers in the case of bosons and anticommuting Grassmann numbers in the case of fermions. More precisely, the fields $\phi_{\sigma}(\tau, {\bf r})$ correspond to eigenvalues of the annihilation operators, whose eigenstates are called coherent states. Working with the eigenvalues of the operators is often convenient, because it is typically easier to manipulate numbers than operators. Following the derivation in Ref.~\cite{Stoof09a}, we have that the microscopic action in the functional formalism becomes
\begin{eqnarray}\label{eq:acfer}
S[\phi] \nonumber  &=& - \sum_{\sigma = \pm}
\int_0^{\hbar\beta} d\tau~d\tau'\int d {\bf r}~d {\bf r}'~\phi^*_{\sigma}(\tau,{\bf r})\hbar G^{ -1}_{0,\sigma}(\tau,{\bf r};\tau',{\bf r}') \phi_{\sigma}(\tau',{\bf r}') \nonumber \\
&& + V_{0} \int_0^{\hbar\beta} d\tau \int d {\bf r}~
\phi^*_{+}(\tau,{\bf r}) \phi^*_{-}(\tau,{\bf r})
\phi_{-}(\tau,{\bf r}) \phi_{+}(\tau,{\bf r}),
\end{eqnarray}
where we have introduced the so-called non-interacting (inverse) Green's function $\hbar G^{-1}_{0,\sigma}$, given by
\begin{eqnarray}\label{eq:greenf}
&&\hbar G^{ -1}_{0,\sigma}(\tau, {\bf r};\tau', {\bf r}') =- \left\{ \hbar \frac{\partial}{\partial\tau} -
\frac{\hbar^2 \boldsymbol{\nabla}^2}{2m} -
\mu_{\sigma} \right\} \delta({\bf r}-{\bf r}')\delta(\tau-\tau').
\end{eqnarray}
The Fourier transform of the inverse Green's function yields
\begin{equation}\label{eqgreenfk}
\hbar G^{ -1}_{0,\sigma}(i\omega_n,{\bf k})=i\hbar\omega_{n}-\epsilon_{\bf k}+\mu_{\sigma},
\end{equation}
where $\epsilon_{\bf k}=\hbar^2{\bf k}^2/2m$ is the kinetic energy, and $\omega_{n}$ are the odd fermionic Matsubara frequencies, as seen for example in Refs.\ \cite{Bruus04a,Stoof09a}.

\subsection{Hubbard-Stratonovich transformation}\label{par:hstrans}

As mentioned in Section \ref{sec:mf}, BCS theory can be physically interpreted as the Bose-Einstein condensation of Cooper pairs. This follows from the observation that the
order parameter for the BCS transition is proportional to the expectation
value of the pair annihilation operator $\langle \hat{\phi}_{-}({\bf r})
\hat{\phi}_{+}({\bf r}) \rangle$, which means that the Cooper pairs are in a coherent state analogous to a condensate of point-like bosons.
For the
transition to the paired condensate, we require that the two-body
interaction potential is attractive, because otherwise the
formation of pairs would not be energetically favorable. From now
on, we therefore consider an attractive potential. Note that this
does not necessarily mean that the corresponding scattering length
$a$ of the interaction is negative. In Section \ref{sec:scat}, we namely saw that
attractive potentials can also give rise to a positive scattering
length when there is a two-body bound state present in the
potential. To introduce the BCS order parameter exactly into the many-body
theory, we use the so-called Hubbard-Stratonovich (HS) transformation \cite{Strato58a,Hubbar59a,Kleine78a}.

To perform this transformation, we start from the expression of the partition sum $Z$ in the functional formalism, whose derivation can be found for example in Ref. \cite{Negele98a}. We have that
\begin{equation}\label{eq:parsum}
Z = \int \mathcal{D}\phi~e^{-S[\phi]/\hbar},
\end{equation}
which represents a functional integral over all fermion
fields $\phi$ and $\phi^*$ that are antiperiodic on the imaginary time interval
$[0,\hbar\beta]$. In the functional integral of Eq.\ (\ref{eq:parsum}), we can insert the following identity
\begin{eqnarray}\label{eq:hstrans}
1 &=& \int  \frac{\mathcal{D}\Delta}{\mathcal{N}}\exp \left\{\int_0^{\hbar\beta} d\tau \int d {\bf r} \left[\Delta^* (\tau,{\bf r})-V_{0} \phi^*_{+}(\tau,{\bf r})\phi^*_{-}(\tau,{\bf r}) \right] \right. \nonumber \\
&&  \left. {\phantom \int}\times V_{\bf 0}^{-1}  \left[\Delta (\tau,{\bf r})-V_{0} \phi^{\phantom *}_{-}(\tau,{\bf r})\phi^{\phantom *}_{+}(\tau,{\bf r})\right]/\hbar \right\},\qquad
\end{eqnarray}
which represents a functional integral over all bosonic
fields $\Delta$ and $\Delta^*$ that are periodic on the imaginary time interval
$[0,\hbar\beta]$. Here, $\mathcal{N}$ is a normalization constant that cancels the outcome of the Gaussian functional integral. Since this constant does not depend on any thermodynamic variable, it turns out to be irrelevant for our purposes. For notational conveniency,  we will absorb this numerical constant in the measure of Eq.\ (\ref{eq:hstrans}). Moreover, from Eq.\ (\ref{eq:hstrans}) it is also clear that on average the complex pairing field $\Delta(\tau,{\bf r})$ satisfies
\begin{eqnarray}\label{eq:avgap}
\langle \Delta (\tau,{\bf r}) \rangle = V_{\bf 0} \langle
\phi_{-} (\tau,{\bf r})  \phi_{+} (\tau,{\bf r})
\rangle.
\end{eqnarray} More details and a more general discussion about the HS transformation is given in Refs. \cite{Kleine78a,Stoof09a}.

By combining Eqs. (\ref{eq:parsum}) and (\ref{eq:hstrans}), we obtain
\begin{equation}\label{eq:parsum2}
Z = \int \mathcal{D}\phi~\mathcal{D}\Delta~e^{-S[\phi,\Delta]/\hbar},
\end{equation}
where the new action is given by
\begin{eqnarray}\label{eq:acbosfer}
S[\phi, \Delta] &=& -\int_0^{\hbar\beta} d\tau \int d {\bf r}
\frac{|\Delta(\tau,{\bf r})|^2}{V_{\bf 0}} \\
&-&\hbar \int_0^{\hbar\beta} d\tau~d\tau'\int d {\bf r}~d {\bf r}'~ \boldsymbol{\Phi}^{\dagger}(\tau,{\bf r})\cdot {\bf G}_{\rm BCS}^{-1}(\tau,{\bf r};\tau',{\bf r}')\cdot \boldsymbol{\Phi}(\tau',{\bf r}').\nonumber
\end{eqnarray}
Here, we have introduced $\boldsymbol{\Phi}^{\dagger}(\tau,{\bf r})= [ \phi^*_{+}(\tau,{\bf r}) , \phi_{-}(\tau,{\bf r})]$ and the $2\times 2$ inverse Green's function matrix ${\bf G}_{\rm BCS}^{-1}$ in spin space, which is given by
\begin{equation}\label{eq:gfbcs}
{\bf G}_{\rm BCS}^{-1}(\tau,{\bf r};\tau',{\bf r}') = {\bf G}_{0}^{-1}(\tau,{\bf r};\tau',{\bf r}') - \boldsymbol{\Sigma}_{\rm BCS}(\tau,{\bf r};\tau',{\bf r}'),
\end{equation}
with the noninteracting part
\begin{eqnarray}\label{eq:g0m}
{\bf G}_0^{-1}(\tau,{\bf r};\tau',{\bf r}') =
\left[\begin{array}{cc}
  G^{-1}_{0,+}(\tau,{\bf r};\tau',{\bf r}') &
 0 \\
0 & - G^{-1}_{0,-}(\tau,{\bf r};\tau',{\bf r}')
\end{array}\right].~
\end{eqnarray}
and the self-energy part
\begin{eqnarray}\label{eq:sebcs}
\hbar\boldsymbol{\Sigma}_{\rm BCS}(\tau,{\bf r};\tau',{\bf r}') =
\left[\begin{array}{cc}
 0 & \Delta(\tau,{\bf r}) \\
\Delta^*(\tau,{\bf r}) & 0
\end{array}\right]\delta(\tau-\tau')\delta({\bf r}-{\bf r}').~
\end{eqnarray}
The minus sign in the matrix element $G^{-1}_{0,22}$ of Eq.~(\ref{eq:g0m}) comes from reversing the order of the two corresponding fermionic Grassmann fields in Eq.~(\ref{eq:acbosfer}). The pairing field $\Delta$ couples to two fermionic creation fields in Eq.~(\ref{eq:acbosfer}), showing that a pair can decay into two fermions of opposite spin, while $\Delta^*$ couples to two fermionic annihilation fields showing that two fermions can form a Cooper pair. The action of Eq.~(\ref{eq:acbosfer}) can be physically interpreted as an interacting Bose-Fermi mixture. We see that by performing the exact HS transformation, we have eliminated the fourth-order interaction term in Eq.~(\ref{eq:acfer}), so that the resulting action of Eq.\ (\ref{eq:acbosfer}) is only quadratic in the fermion fields. However, this goes at the cost of introducing an additional functional integral over the pairing field $\Delta$.

Since the action of Eq.~(\ref{eq:acbosfer}) is quadratic in the fermionic fields, the corresponding functional integral is Gaussian and can be performed exactly. Using the standard formula for sermonic Gaussian functional integrals, see for example Ref.~\cite{Stoof09a}, we obtain
\begin{equation}\label{eq:parsum3}
Z = \int \mathcal{D}\Delta ~e^{-S[\Delta]/\hbar},
\end{equation}
with the action given by
\begin{eqnarray}\label{eq:acbos}
&&S[\Delta]   =  -\int_0^{\hbar\beta} d\tau \int d {\bf r}
\frac{|\Delta(\tau,{\bf r})|^2}{V_{\bf 0} } -\hbar{\rm Tr}[ \log (-{\bf G}_{\rm BCS}^{-1})].\quad
\end{eqnarray}
Here, the trace is to be taken over spin space, imaginary time and position space, or, after Fourier transformation, over spin space, frequencies and momenta.

We thus see that with the HS transformation, we have been able to derive two actions that are equivalent to Eq.~(\ref{eq:acfer}), and that thus exactly introduce the order parameter field into the theory. Since none of the three actions are exactly solvable, we have to invoke approximations in order to get actual results. However, having three actions at our disposal greatly enhances the scope of possible approximation schemes. Moreover, it often depends on the physical quantity of interest which action is most suitable as a starting point. For example, if we are interested in the fermionic self-energy in the normal state, we might start from Eq.~(\ref{eq:acfer}), while if we are interested in the properties of the Cooper pairs, we could start from Eq.~(\ref{eq:acbos}).

By expanding the logarithm in Eq.~(\ref{eq:acbos}) in powers of $\Delta$, we obtain an infinite series. As just mentioned, this prohibits an exact solution to the problem, and approximations have to be made in order to proceed. The part in Eq.~(\ref{eq:acbos}) that is independent of $\Delta$ gives rise to the partition sum of the ideal Fermi gas in Eq.~(\ref{eq:parsum3}). By performing the mean-field, or saddle-point, approximation, the full path integral over the bosonic field $\Delta(\tau,{\bf r})$ is simply approximated by the value of the integrand associated with the global minimum of the action. This approximation results in the BCS grand potential of Eq.~(\ref{eq:tpdBCS}). For a detailed derivation, see for example Ref. \cite{Stoof09a}.

\subsection{Pair fluctuations} \label{par:nsr}

Although mean-field BCS theory provides an elegant description of superconducting electrons in metals or superfluidity of weakly interacting atomic Fermi gases, it has some major shortcomings in the strongly interacting regime. In the normal state, when $\langle \Delta \rangle$ = 0, the BCS grand-potential density of Eq.~(\ref{eq:tpdBCS}) reduces to the grand-potential density of the ideal gas. This is because the BCS theory only takes the interaction of (quasi)particles with the condensate of Cooper pairs
non-pertubatively into account, but not the interaction between the
(quasi)particles themselves. Once the
condensate has vanished in the normal state, we thus have that $\omega_{\rm BCS}(0;T,\mu_{\sigma})=\omega_{\rm ig}(T,\mu_{\sigma})$ with
\begin{eqnarray}
\omega_{\rm ig}[\Delta;T,\mu_{\sigma}]&=& -\frac{1}{\beta \mathcal{V}}\sum_{{\bf k},\sigma}\ln(1+e^{-\beta
(\epsilon_{\mathbf{k}}-\mu_{\sigma})}).
\end{eqnarray}
As a result, BCS mean-field theory does not lead to an accurate description of the normal
state in the unitarity limit, because the particles are then actually very strongly
interacting, as follows from the diverging scattering length.

Another issue is that BCS theory describes only formation of pairs with zero center-of-mass momentum, i.e., it describes only a condensate of Cooper pairs. The critical temperature following from BCS theory then corresponds to the typical temperature at which the pairs are broken. This is in particular a bad approximation in the BEC limit of the crossover that was discussed in Section \ref{par:becbcs}. In the BEC limit the superfluidity is namely lost at the critical temperature due to thermal occupation of non-zero momentum states by the (tighly-bound) pairs, that are still intact.

Both problems of the BCS theory can already be overcome to a significant extent by considering also the Gaussian fluctuations of the order parameter around its mean-field value. Since Gaussian functional integrals can be performed analytically, the contribution of these pair fluctuations can thus be studied exactly, resulting in a theory that was pioneered by Nozi\`{e}res and Schmitt-Rink \cite{Nozier85a}. Diagrammatically this theory corresponds to summing the so-called ladder diagram, or ring diagram contributions to the grand potential.  Summarizing their analysis, we start in the normal state and expand the logarithm in Eq. (\ref{eq:acbos}) to second order in the pairing field, as also more elaborately explained in Ref.~\cite{Stoof09a}. The Cooper pair Green's function $G_{\Delta}^{-1}$ is then defined through this quadratic part of the Cooper pair action, and is found to be given by
\begin{eqnarray}
\hbar G_{\Delta}^{-1}(\tau,{\bf r};\tau',{\bf r}')&=&\frac{1}{V_0}\delta(\tau-\tau')\delta({\bf r}-{\bf r}')
\nonumber\\
&&+ \frac{1}{\hbar}G_{0,+}(\tau,{\bf r};\tau',{\bf r}')
G_{0,-}(\tau,{\bf r};\tau',{\bf r}'),
\end{eqnarray}
or, after Fourier transformation, by
\begin{eqnarray}\label{eq:cpgf}
&& \hbar G_{\Delta}^{-1}(i\omega_n, {\bf k})=\nonumber\\
&&\frac{m}{4 \pi \hbar^2 a}+ \frac{1}{\mathcal{V}}\sum_{\bf k'} \left\{\frac{1-f_+(\epsilon_{\bf k'})-f_{-}(\epsilon_{\bf k-k'})}{-i\hbar\omega_n+\epsilon_{\bf k'}+\epsilon_{\bf k-k'}-2\mu}-\frac{1}{2\epsilon_{\bf k'}}\right\},
\end{eqnarray}
where we eliminated the microscopic interaction strength $V_0$ in favour of the two-body scattering length $a$ using Eq.~(\ref{eq:t2b-v0}).

The resulting functional integral after substituting the quadratic part of the Cooper pair action in Eq.\ (\ref{eq:parsum3}) can be evaluated exactly by using the formula for bosonic Gaussian functional integrals, see e.g. Ref.\ \cite{Negele98a}. The partition sum $Z$ up to quadratic order is found to be given by a part that is independent of the non-condensed Cooper pairs, $Z_0$, and a part coming from the Gaussian integration over the Cooper pair propagator, $Z_{\Delta}$, which can thus be physically interpreted to describe a noninteracting gas of Cooper pairs. The latter is given by
\begin{eqnarray}
 Z_{\Delta} = e^{-{\rm Tr}\{\log[-G^{-1}_{\Delta}(i\omega_n,{\bf k})] \}},
\end{eqnarray}
while $Z_0$ is given by the ideal gas partition sum for fermions \cite{Nozier85a,Stoof09a}.  By using $Z=Z_0 Z_{\Delta}$ and the thermodynamic relation $Z\equiv e^{-\beta \mathcal{V} \omega_{\rm gc} (T,\mu_{\sigma})}$ with $\omega_{\rm gc} (T,\mu_{\sigma})$ the grand-canonical thermodynamic potential density, we thus conclude that
\begin{eqnarray}\label{eq:tptot}
&&  \omega_{\rm gc}(T,\mu_{\sigma}) = \omega_{\rm ig}(T,\mu_{\sigma}) + \omega_{\Delta}(T,\mu_{\sigma}),
\end{eqnarray}
where the contribution due to the Cooper pairs $\omega_{\Delta}(T,\mu_{\sigma})$ is given by
\begin{eqnarray}\label{eq:tpnsr}
\omega_{\Delta}(T,\mu_{\sigma}) &=& \frac{1}{\beta \mathcal{V}}\sum_{n,{\bf k}} \log[-G^{-1}_{\Delta}(i\omega_n,{\bf k})] \\
&=&\frac{\hbar}{\pi \mathcal{V}} \sum_{\bf k}\int d\omega ~ b(\omega){\rm Im}(\log[-G^{-1}_{\Delta}(\omega^+,{\bf k})]),\nonumber
\end{eqnarray}
with $b(\epsilon)=1/(e^{\beta\epsilon}-1)$ the Bose distribution function and $\omega^+=\omega+ \eta i$, where $\eta$ is infinitesimally small and positive. In the second step of Eq.~(\ref{eq:tpnsr}), we have used contour integration in order to write the sum over Matsubara frequencies as an integral along the real axis \cite{Nozier85a}. This is convenient because the resulting frequency integral converges very fast. Moreover, the imaginary part of $G^{-1}_{\Delta}(\omega^+,{\bf k})$ along the real axis can be determined analytically and yields
\begin{eqnarray}\label{eq:imcpgf}
&&{\rm Im}(G^{-1}_{\Delta}(\omega^+,{\bf k}))=\sum_{\sigma=\pm 1}\frac{\theta(\hbar\omega+2\mu-\epsilon_{\bf k}/2)m^2}{4\pi \beta |{\bf k}|\hbar^4}\times\\
&&\log\left[\frac{\cosh\left[\frac{\beta}{2}\left(\frac{\hbar\omega}{2}+\sqrt{(\hbar\omega+2\mu-\frac{\epsilon_{\bf k}}{2})\frac{\epsilon_{\bf k}}{2}}-\sigma h\right)\right]}{\cosh\left[\frac{\beta}{2}\left(\frac{\hbar\omega}{2}-\sqrt{(\hbar\omega+2\mu-\frac{\epsilon_{\bf k}}{2})\frac{\epsilon_{\bf k}}{2}}-\sigma h\right)\right]}\right],\nonumber
\end{eqnarray}
where $\theta(x)$ is the Heaviside step function.
The real part of $G^{-1}_{\Delta}(\omega^+,{\bf k})$ can then be determined by a Kramers-Kronig transform. Note that to this end, we have to subtract the analytically known high-frequency tail of the imaginary part, which is given by $(\hbar\omega+2\mu-\epsilon_{\bf k}/2)^{1/2}m^{3/2}/4\pi\hbar^3$. Only then, the remaining part decays fast enough to be transformed.

For the spin-balanced Fermi gas, the Nozi\`{e}res-Schmitt-Rink approach has been applied succesfully to the whole crossover from the BCS regime to the BEC regime \cite{Sademe93a}. The theory gives rise to a critical temperature curve that smoothly interpolates between the result from BCS theory in the weakly-interacting limit and the result for Bose-Einstein condensation of noninteracting bosons in the extremely strongly interacting limit, where a two-body bound state is present. The critical condition for superfluidity within this theory is that $G^{-1}_{\Delta}(0,{\bf 0})=0$, which can physically be interpreted as the chemical potential for the Cooper pairs being equal to zero. This thus corresponds to the well-known criterium for condensation of bosons. By comparing Eq. (\ref{eq:cpgf}) with Eq.~(\ref{eq:alphaL}), we see that the critical condition has not changed from mean-field theory, resulting in the unitarity limit in $k_{\rm B}T_{\rm c} = 0.66 \mu_{\rm c}$ for the critical temperature. This is rather large compared to most recent measurements by Ku {et al.}, that give rise to $k_{\rm B}T_{\rm c} = 0.40 \mu_{\rm c}$ \cite{Ku12a}, or to Monte-Carlo results, that predict $k_{\rm B}T_{\rm c} = 0.31 \mu_{\rm c}$ \cite{Burovs06a}. But since the equation of state has changed due to the inclusion of the non-condensed Cooper pairs in Eq.~(\ref{eq:tptot}), we find by calculating the total particle density at the critical temperature with $n(T,\mu) =-\partial \omega_{\rm gc}(T,\mu) / \partial \mu$ that the NSR theory leads to $T_{\rm c} = 0.23 T_{\rm F}$, which is a significant decrease from mean-field theory. The latter prediction is rather close to the results from various more sophisticated strong-coupling theories, that typically predict critical temperatures in the range $T_{\rm c} = 0.15 \pm 0.05 T_{\rm F}$ \cite{Burovs06a,Haussm07a,Nishid07a,Nikoli07a,Bulgac08b}, while the most recent experimental value is  $T_{\rm c} = 0.167 \pm 0.013 T_{\rm F}$ \cite{Ku12a}.

\begin{figure}
\begin{center}
\includegraphics[width=0.75\columnwidth]{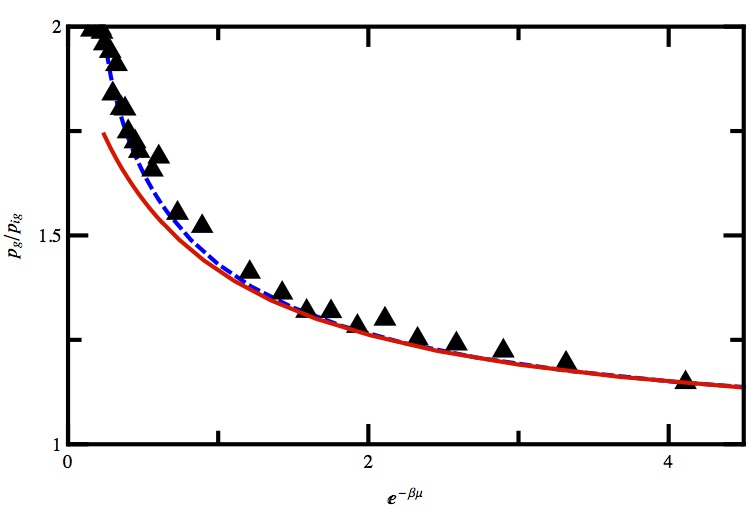}
\caption{\label{fig:press} (Color online) Equation of state for the normal phase of a strongly interacting balanced Fermi gas at unitarity in the grand-canonical ensemble. The pressure $p_{\rm g}$ of the gas is calculated as a function of the inverse fugacity $e^{-\beta \mu}$ with the renormalization group approach of Section \ref{par:cprg} (full line) and the Nozi\`{e}res-Schmitt-Rink approach (dashed line). The pressure of the ideal two-component Fermi gas is given by $p_{\rm ig}$. The triangles  are the experimental results of Nascimb\`{e}ne {\it et al.} \cite{Nascim10a}.}
\end{center}
\end{figure}

Using Eqs.~(\ref{eq:tptot}) and (\ref{eq:tpnsr}) we can calculate the pressure of a strongly interacting Fermi gas, given by $p_{\rm g}(T,\mu_{\sigma})=- \omega_{\rm gc}(T,\mu_{\sigma})$. In the unitarity limit, this pressure was measured for a balanced Fermi gas by Nascimb\`{e}ne {\it et al.} \cite{Nascim10a}, Horikoshi {\it et al.} \cite{Horiko10a}, and by Ku {\it et al.} \cite{Ku12a}. We show the results of Nascimb\`{e}ne {\it et al.} and of the prediction by the NSR theory  in Fig.\ \ref{fig:press}, where  the agreement is seen to be very good. We note that the experimental results of Nascimb\`{e}ne {\it et al.} and Ku {\it et al.} agree reasonably well with each other for the temperature range of Fig.\  \ref{fig:press}, while the results of Horikoshi {\it et al.} \cite{Horiko10a} do not match the high-temperature virial expansion \cite{Liu09a}. At lower temperatures the results of Nascimb\`{e}ne {\it et al.} and Ku {\it et al.} differ quite significantly, where the results of Ku {\it et al.}  seem to  agree better with most recent Monte-Carlo calculations \cite{Houcke12a}. It is possible to generalize the NSR theory to the superfluid state by making in Eq.~(\ref{eq:acbosfer}) the substitution $\Delta(\tau,{\bf r})=\Delta_0+\Delta'(\tau,{\bf r})$, expanding Eq.~(\ref{eq:acbos}) up to second order in the fluctuations $\Delta'$, and performing the resulting Gaussian functional integral \cite{Romans05a,Hu07a,Diener08a}. In the superfluid state, the NSR theory gives rise to good agreement with earlier experiments \cite{Luo07a,Hu10a}, while the most recent experimental data of Ku {et al.} agrees best with diagrammatic Monte-Carlo calculations  \cite{Houcke12a}. A
self-consistent fluctuation theory that takes fermionic self-energy effects into account also leads to rather good agreement with these recent experiments \cite{Haussm94a,Haussm07a}. From this discussion, we conclude that current high-precision measurements allow for a very sensitive test of many-body theories.  Although the NSR theory is not exact, and therefore is expected to give rise to deviations from experiments in the unitarity regime, it is seen from Fig.\  \ref{fig:press} to give a big step forward compared to mean-field theory.

The fact that the NSR theory of pair fluctuations is not perfect is most clearly seen at nonzero polarizations, where it gives rise to unphysical results. Namely, for small imbalances the NSR theory predicts near the critical temperature a negative polarization $p=(n_{+}-n_{-})/(n_{+}+n_{-})$ for a positive chemical potential difference $(\mu_{+}-\mu_{-})$, which corresponds to a compressibility matrix $-\partial^2 \omega_{\rm gc}(T,\mu_{\sigma})/\partial \mu_{\sigma} \partial \mu_{\sigma'}$ that is not positive definite \cite{Parish07b}. It is a quite unsatisfactory situation that the NSR theory, which gives such good results for the balanced Fermi gas, already gives unphysical results for very small population imbalances. In Section \ref{par:cprg}, we discuss a way to overcome this problem.

\subsection{Fermionic self-energy }

So far, we have seen that the pair fluctuation effects in a strongly-interacting Fermi gas lead to an accurate description of the thermodynamics for a balanced Fermi gas, but fail to describe to imbalanced Fermi gas. An extreme example of an imbalanced Fermi gas is a single spin-down particle interacting with a Fermi sea of spin-up particles, also called the Fermi polaron \cite{Lobo06a,Chevy06b}. The self-energy $\hbar \Sigma_{-}$ of the Fermi polaron has been calculated with various methods including Monte-Carlo calculations \cite{Lobo06a,Prokof08a,Pilati08a}, variational methods \cite{Chevy06b,Combes08a,Punk09a}, the many-body $T$ matrix in the ladder approximation \cite{Combes07a}, and RG methods \cite{Gubbel08a}. The various theories are in good agreement with each other, and with experiment \cite{Nascim09a,Schiro09a}, leading to about $\hbar \Sigma_{-} = - 0.6 \mu_{+}$ at zero temperature. We note that this large self-energy effect is not incorporated in the pair-fluctuation theory of the previous section.

\begin{figure}[t]
\begin{center}
\includegraphics[width=0.7\columnwidth]{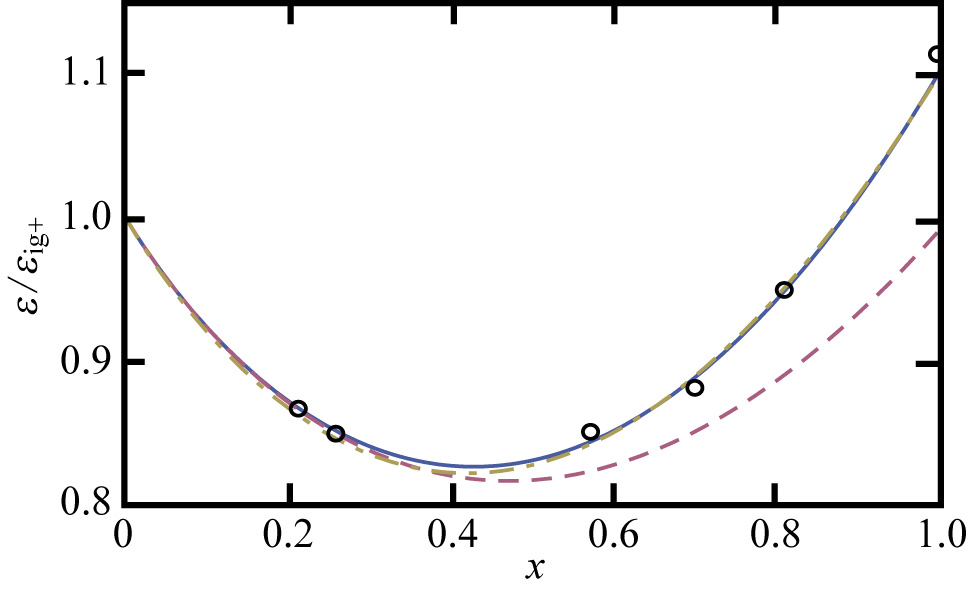}
\caption{\label{fig:eosmc} Equation of state for the
zero-temperature normal state of the spin-imbalanced Fermi mixture
with unitary interactions. The Monte Carlo results by Lobo {\it et
al}. \cite{Lobo06a} are shown by circles, while $x=n_-/n_+$. The full
line represents the best fit of Lobo {\it et
al}. to their data. The dashed line gives
the result for considering only the self-energy of the minority
particles. Upon taking also the self-energy of the majority
particles into account through a symmetric ansatz, the
dashed-dotted line is obtained. The latter is hard to discern from
the best fit. The energy density $\epsilon$ is scaled with the
ideal gas energy for the spin-up particles $\epsilon_{\rm ig+}=
3\epsilon_{{\rm F}+} n_{+}/5$ with $\epsilon_{{\rm
F}\sigma}=\hbar^2(6 \pi^2 n_{\sigma})^{2/3}/2m$.}
\end{center}
\end{figure}

The goal of the present section is to discuss in more detail the strongly-interacting normal state of the imbalanced Fermi gas at zero temperature as described by the equation of state from the pioneering Monte-Carlo study of Lobo {\it et al.} \cite{Lobo06a}.
The results of their calculations are shown in Fig.~\ref{fig:eosmc}. The circles present the Monte-Carlo data,
while the full line is the best fit to this data. The dashed line
represents the following Ansatz by Lobo {\it et al.} for the total energy density
$\epsilon$ of the interacting system,
\begin{eqnarray}\label{eq:eoslobo}
\epsilon &=&\frac{3}{5}n_{+} \epsilon_{\rm F+}+\frac{3}
{5}\frac{m}{m^*}n_{-} \epsilon_{\rm F-}-\frac{3}{5}A
\epsilon_{\rm F+} n_{-}\nonumber\\
&=&\frac{3}{5}n_{+} \epsilon_{\rm F+}\left(1+
\frac{m}{m^*}x^{5/3}-A x\right),
\end{eqnarray}
which was constructed to treat the strongly polarized regime. This Ansatz physically describes a few spin-down atoms in a Fermi sea of spin-up atoms, where the minority atoms have acquired an effective
mass of $m^*=1.04 m$ and a self-energy of $\hbar \Sigma_{-}=-3 A \epsilon_{\rm
F+}/5$ with $A=0.97$. Both effects are the result of the strong
attractive interactions with the spin-up Fermi sea, which
remains unaffected. The Ansatz of Eq.~(\ref{eq:eoslobo})
is less suitable to describe the balanced case, $x=1$, where it
causes for example the two chemical potentials $\mu_{\sigma}$ to
be unequal.

To have an Ansatz that is suitable for all polarizations, we may
write the equation of state from the beginning in a form that is
symmetric upon interchange of $n_{+}$ and $n_{-}$. This reflects
the fact that it does not matter for the energy density which of
the two spin states is the majority state. We try a self-energy of the form
\begin{equation}\label{eq:selfn}
\hbar\Sigma_{\sigma}(n_{+},n_{-})= - \frac{3 A (6
\pi^{2})^{2/3}\hbar^2}{10m}\frac{ n_{-\sigma}}{(n^{\alpha}_+ +
n^{\alpha}_-)^{1/3\alpha}},
\end{equation}
where we still have to specify the power $\alpha$. If we use the self-energy of Eq.~(\ref{eq:selfn})
to obtain a total energy density similar to Eq.~(\ref{eq:eoslobo}),
we find
\begin{eqnarray}
\epsilon&=&\frac{3}{5}n_{+} \epsilon_{\rm F+}+\frac{3}{5}n_{-}
\epsilon_{\rm F-}-\frac{3 A (6 \pi^{2})^{2/3}\hbar^2}{10m}
\frac{n_{+} n_{-}}{(n^{\alpha}_+
+ n^{\alpha}_-)^{1/3\alpha}}\nonumber\\
&=&\frac{3}{5}n_{+} \epsilon_{\rm F+}\left(1+x^{5/3}-A \frac{x}
{(1+x^{\alpha})^{1/3\alpha}}\right),
\end{eqnarray}
which agrees for small $x$ with Eq.~(\ref{eq:eoslobo}). If we wish to
obtain agreement with the full Monte-Carlo equation of state, we
may use $A=1.01$ and $\alpha=2$, which results in the
dashed-dotted line shown in Fig.~\ref{fig:eosmc}. It is hardly
discernible from the full line that gives the best fit of Lobo {\it et al}. The value
of $A$ is calculated in the next section, while there does not
seem to be a direct microscopic argument why $\alpha$ should be 2.
It can thus be interpreted as a single fit parameter.

\subsubsection{The extremely imbalanced case}\label{par:polaron}

We have just seen that the equation of state for the normal state
at zero temperature is to a large extent determined by self-energy
effects of the minority (spin-down) atoms in the Fermi sea of
majority (spin-up) atoms. In this section, we calculate the
self-energy of a single spin-down atom in a Fermi sea of spin-up atoms analytically in the
so-called ladder or many-body $T$ matrix approximation as was done by Combescot {\it et al.} \cite{Combes07a}. The effect of this
self-energy $\hbar \Sigma_{\sigma}(\mu_{\sigma})$ is to renormalize the chemical potential according to  $\mu'_{\sigma}(\mu_{\sigma})= \mu_{\sigma} - \hbar
\Sigma_{\sigma}(\mu_{\sigma})$. We may let the renormalized chemical
potentials $\mu'_{\sigma}$ consequently enter the grand potential density
of the normal gas at zero temperature $\omega_{{\rm n}}$ in the
following way
\begin{eqnarray}\label{eq:tpifgr}
\omega_{{\rm n}}(\mu_{\sigma})&=&\omega_{{\rm ig}}(\mu'_{\sigma}(T=0,\mu_{\sigma})) \\
&=& - \frac{(2m)^{3/2}}{15
\pi^2 \hbar^3}\left(\mu'_+(\mu_{\sigma})^{5/2}+\mu'_-
(\mu_{\sigma})^{5/2}\right).\nonumber
\end{eqnarray}
The single-particle limit for spin state $\sigma$ is located right
at the transition from a nonzero particle density to a
zero-particle density. The latter takes place when the renormalized chemical potential $\mu'_{\sigma}$
goes to zero, i.e. when $\mu_{\sigma} = \hbar
\Sigma_{\sigma}(\mu_{\sigma})$. To be able to
solve this equation, we need to calculate the self-energy
$\hbar \Sigma_{\sigma}(\mu_{\sigma})$ in the extremely imbalanced
limit.

To perform the self-energy calculation for a single spin-down atom in a Fermi
sea of spin-up particles. The many-body $T$ matrix is treated in the so-called ladder approximation, which is for example described in Ref.\ \cite{Stoof09a}. For unitary interactions and at zero temperature, the many-body $T$ matrix in the extremely imbalanced
case is seen to have a particularly simple expression, because the
Fermi distributions become step functions and there is no
spin-down particle density, resulting in
\begin{eqnarray}\label{eq:timb}
T^{-1}({\omega,\bf k})&=&\frac{1}{T^{\rm 2b}}
+\frac{1}{\mathcal{V}}\sum_{\bf k'}\left\{ \frac{\theta(\epsilon_{\bf
k'}-\mu_{+})}{-\hbar\omega+\epsilon_{{\bf k'}}+\epsilon_{{\bf
k-k'}}-2\mu}-\frac{1}{2\epsilon_{\bf k'}}\right\},~
\end{eqnarray}
where $\theta(x)$ is the Heaviside step function. The $T$ matrix gives rise
to a self-energy for the minority particle, which at zero momentum
and frequency is given by \cite{Combes07a}
\begin{eqnarray}\label{eqselfpol}
\hbar \Sigma_-(\mu_{\sigma}) &=& \frac{1}{2\pi i \mathcal{V}}\sum_{\bf k'}\int_{-i\infty}^{i\infty}d\omega ~ T(\omega,{\bf k'}) G_{0,+}(\omega,{\bf k'}) \nonumber \\ &=&\frac{1}{\mathcal{V}}\sum_{\bf k'}
T(\epsilon_{\bf k'}-\mu_{+},{\bf k'})\theta(\mu_+-\epsilon_{\bf k'}),
\end{eqnarray}
where the substitution $\hbar\omega\rightarrow \epsilon_{\bf
k'}-\mu_{+}$ in the $T$ matrix comes from a contour integration over the frequency $\omega$. Then, to fulfill the initial assumption of
zero down particles, we still need to solve $\mu'_-=0$, which
leads to $\mu_{-}=-0.607\mu_+=-3 A\mu_+/5$ with $A=1.01$. This
result means that for a chemical potential lower than
$\mu_{-}=-0.6\mu_+$ there are no spin-down particles, whereas for
a higher chemical potential there is a nonzero density of
spin-down particles. Of course, the results would be the same for
a single spin-up particle in a spin-down sea. There have been
several Monte-Carlo calculations for a single fermion with spin
$\sigma$ in a Fermi sea of particles with spin $-\sigma$. The
results are $A=0.97$ \cite{Lobo06a}, $A = 0.99$ \cite{Pilati08a}
and $A = 1.03$ \cite{Prokof08a}, which all agree remarkably well
with the simple ladder calculation. In the following, we will keep using $A=1.01$.

\begin{figure}[t]
\begin{center}
\includegraphics[width=1.0\columnwidth]{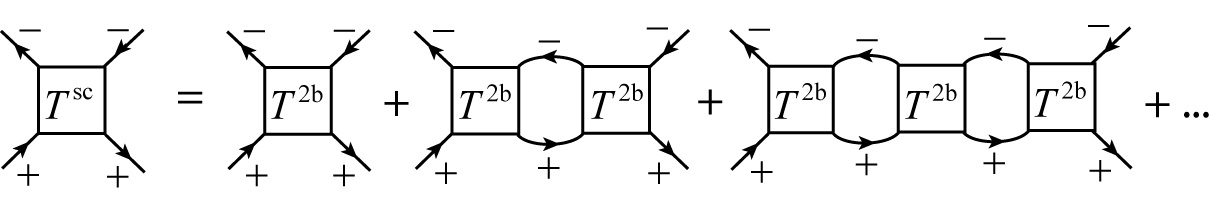}
\caption{\label{fig:bubble} Diagrammatic representation of the so-called bubble sum, leading to a screened two-body transition matrix $T^{\rm sc}$.}
\end{center}
\end{figure}

The fact that the $T$ matrix calculation in the ladder approximation agrees so well with Monte-Carlo results can be regarded to some extent as a coincidence \cite{Combes08a}. Namely, we could also try to perform a more selfconsistent calculation, in which the fermionic self-energy effects are also taken into account in the expression for the many-body $T$ matrix. Such a selfconsistent $T$ matrix approach was pioneered by Haussmann for the balanced Fermi gas \cite{Haussm94a}. In order to take fermionic self-energy effects into account, we could replace $\mu_{-}$ by $\mu'_- = 0$ in Eq. (\ref{eq:timb}). However, then we find that $\mu_{-}=\hbar\Sigma_{-}=-0.9\mu_+$, which does not agree with the Monte-Carlo results. To proceed further we could try to incorporate another effect, namely that of particle-hole excitations that tend to screen the interaction, as is well known for electrons in a metal \cite{Fetter71a}. We may account for this screening by replacing the two-body transition
matrix $T^{\rm 2b}$ with an effective transition matrix that
includes the particle-hole excitations through the so-called bubble sum, as is depicted in Fig.~\ref{fig:bubble}. We then have that $1/T^{\rm sc
}(\omega,{\bf k})= 1/T^{\rm 2b}-\hbar \Pi(\omega,{\bf k})$, where the expression for the polarization bubble diagram is given by, see e.g. \cite{Stoof09a},
\begin{eqnarray}\label{eq:bubble}
\hbar \Pi(\omega,{\bf k})&=&\frac{1}{V} \sum_{\bf k' }
\frac{f_+({\bf k}+{\bf k'})-f_-({\bf k'})}{\hbar\omega+2h' - \epsilon_{{\bf k}+{\bf k'}}+\epsilon_{{\bf k'}}}.\quad
\end{eqnarray}
If we now replace $1/T^{\rm 2b}$ with $1/T^{\rm sc}(0,{\bf 0})$ and $\mu_{-}$ by $\mu'_- = 0$ in Eq.~(\ref{eq:timb}), then we find $\mu_{-}=\hbar\Sigma_{-}=-0.5\mu_+$, which is again quite close to the mentioned Monte-Carlo results of $\mu_{-}=-0.6\mu_+$. Here, we have neglected the momentum and frequency dependence of the screened interaction. It actually turns out that nonzero external momenta reduce the effect of screening, where we note that a slight reduction of the screening effect would bring the present calculation even closer to the Monte-Carlo results and recent measurements \cite{Schiro09a}. We may conclude from this discussion that in order to accurately describe the normal state of the imbalanced Fermi gas, in particular the fermionic self-energy and the effect of screening seem to be of importance.

We then try to generalize the fermionic self-energy effect on the polaron to the full equation of state by using an Ansatz for the normal state that has the correct extremely imbalanced limits in it, but
actually leads to excellent results with Monte-Carlo calculations
for all polarizations. This is achieved by introducing the
following renormalized chemical potentials
\begin{equation}\label{eq:mup}
\mu'_{\sigma}=
\mu_{\sigma}+\frac{3}{5}A\frac{\mu'^2_{-\sigma}}{\mu'_{+}+\mu'_{-}}.
\end{equation}
We can now determine readily the
renormalized chemical potentials $\mu'_{\sigma}$ as a function of
the microscopic ones $\mu_{\sigma}$, and consequently use
$n_{\sigma}=-\partial\omega_{\rm n}/\partial \mu_{\sigma}$ to
determine the densities. The two extremely imbalanced solutions to
Eq.\ (\ref{eq:mup}) are such that the chemical potential of the
majority species ($-\sigma$) is not renormalized, $\mu'_{-\sigma}=
\mu_{-\sigma}$, while the renormalized chemical potential of the
minority species $\sigma$ is zero, $\mu'_{\sigma}= \mu_{\sigma}+3
A \mu_{-\sigma}/5=0$. It is easy to invert the quadratic relations
in Eq.\ (\ref{eq:mup}) to obtain $\mu'_{\sigma}(\mu_{\sigma})$,
which motivates the choice for a quadratic Ansatz. However, a
better validation of the approach is obtained by comparing with
the Monte-Carlo equation of state. We determine the energy density
using $\epsilon=\omega_{\rm n}+\mu_+n_++\mu_-n_- $, for which the
result is shown in Fig.~\ref{fig:eosmc}. Here, both the result of
the present procedure (dashed line) and the Monte-Carlo data of Lobo {\it et al.}
\cite{Lobo06a} (squares) are shown, giving very good agreement. Moreover, we also applied the same procedure to the case of the strongly-interacting $^6$Li-$^{40}$K mixture \cite{Gubbel09a}, where we compared with the Monte-Carlo data of Gezerlis {\it et al}.
\cite{Gezerl09a}. Again, the agreement is excellent.

\begin{figure}
\begin{center}
\includegraphics[width=0.6\columnwidth]{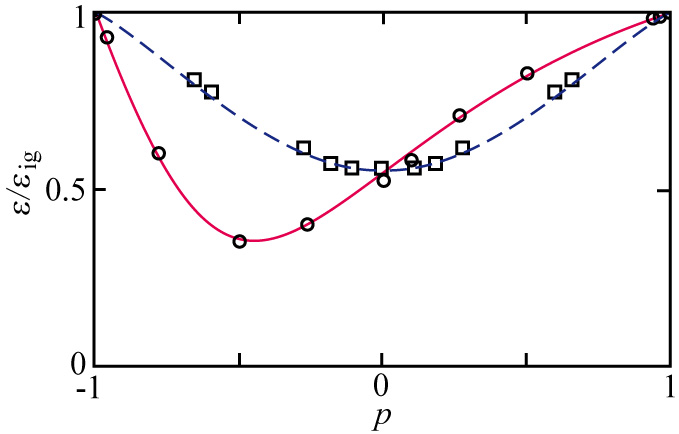}
\caption{\label{fig:eoslik} (Color online) Equations of state for the
zero-temperature normal state of the ${}^6$Li-${}^{40}$K mixture
(full line) and the mass-balanced mixture (dashed line) in the
unitarity limit. The Monte Carlo results for the
${}^6$Li-${}^{40}$K mixture by Gezerlis {\it et al}.
\cite{Gezerl09a} are shown by circles, while the Monte Carlo results
for the mass-balanced mixture by Lobo {\it et al}. \cite{Lobo06a} are
shown by squares. The energy density $\epsilon$ is scaled with the
ideal gas result $\epsilon_{\rm ig}= 3(\epsilon_{{\rm F}+} n_{+} +
\epsilon_{{\rm F}-} n_{-})/5$ with $\epsilon_{{\rm
F}\sigma}=\hbar^2(6 \pi^2 n_{\sigma})^{2/3}/2m_{\sigma}$. The
deviations from one thus show the strong interaction effects.}
\end{center}
\end{figure}

The approach is also expected to be appropriate at small nonzero
temperatures, since the coefficient $A$ is not expected to be
strongly temperature dependent at low temperatures in the normal phase. This is because the Fermi sea of
majority particles remains unaffected for $T \ll T_{\rm F}$, where
the Fermi temperature typically represents a large energy scale in
the system. It is given by $k_{\rm B}T_{\rm F}=\epsilon_{\rm F}
=\hbar^2(3 \pi^2 n)^{2/3}/2m$ with $n=n_{+}+n_{-}$ the total
atomic density. We can check this assumption by comparing with the
Monte-Carlo calculation at nonzero temperature of Burovski {\it et al.}. In
this calculation, the chemical potential at the critical temperature in the unitarity limit was determined to be $\mu(T_{\rm c} = 0.15 T_{\rm F})=0.49 \epsilon_{\rm F}$
\cite{Burovs06a}. We can compare the last result with the direct
extension of the present approach to nonzero temperature, which follows
from using $\omega_{{\rm n}}(T,\mu_{\sigma})=\omega_{{\rm
BCS}}(0;T,\mu'_{\sigma})$. By applying $n_{\sigma}=
\partial\omega_{{\rm n}}(0.15 T_{\rm
F},\mu_{\sigma})/\partial\mu_{\sigma}$, we obtain $\mu(0.15 T_{\rm
F})=0.53 \epsilon_{\rm F}$ in quite good agreement with the Monte-Carlo
calculation. However, we note that both results differ somewhat from the most recent measurements for the chemical potential as a function of temperature, which had a maximum at $\mu(0.17 T_{\rm
F})=0.42 \epsilon_{\rm F}$ \cite{Ku12a}.

\subsubsection{Density profiles}\label{par:denprof}

Having obtained the grand potential density for a
homogeneous Fermi gas in the strongly interacting normal
state, we can now use it to also study the normal trapped gas
by means of the local-density approximation, which was discussed in Section \ref{par:lda}. Namely,
since we have that the local chemical potential for spin state
$\sigma$ is given by $\mu_{\sigma}({\bf r})= \mu_{\sigma} - V^{\rm
ex}({\bf r})$, where $\mu_{\sigma}$ is the chemical potential in
the trap center, we immediately obtain the particle density
at each point in the trap by using $n_{\sigma}({\bf r})=-
\partial\omega_{{\rm n}}(T,\mu_{\sigma}({\bf r}))/\partial
\mu_{\sigma}$. In the case that the gas cloud does not
have a superfluid core, this procedure gives rise to the complete
density profile for the trapped normal phase in the unitarity
limit.

\begin{figure}[t]
\begin{center}
\includegraphics[width=1.0\columnwidth]{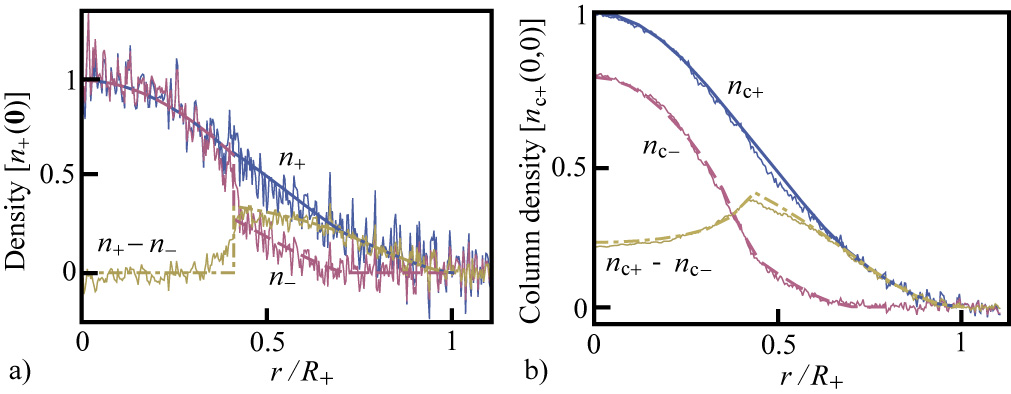}
\caption{\label{fig:mitprof} Observation of phase separation in a
trapped Fermi mixture by Shin {\it et al.} \cite{Shin08a}. The
measurement was performed at their lowest temperature of about
$T=0.02 T_{{\rm F}+}({\bf 0})$ and a total polarization of $P=0.44$. Their
data are shown by the noisy curves. a) Calculated density profile
for the spin-up particles $n_{+}$ (full line), the spin-down
particles $n_{-}$ (dashed line), and their difference
(dashed-dotted line). The densities are scaled by the central
density of the spin-up particles and $R_+ = (2\mu_+/m\bar{\omega}^2)^{1/2}$. b) Exactly the same, but now for
the column densities, which means that an additional integration
along the $y$ axis is performed. Even though the calculations are
performed at zero temperature, the agreement is very good.}
\end{center}
\end{figure}

However, if the spin imbalance is not too large, then the core of
the trapped unitary Fermi mixture is superfluid. In Section
\ref{sec:mf}, we discussed the mean-field BCS theory for a superfluid Fermi gas. At zero temperature, BCS theory predicts for the equal-density superfluid equation of state at unitarity that  $\mu=(1+\beta_{\rm BCS})\epsilon_{\rm F}$ with $\beta_{\rm BCS}=-0.41$ \cite{Haque06a}. If Gaussian fluctuations are included in the superfluid state, we have that $\beta_{\rm NSR}=-0.60$ \cite{Hu06a}, while Monte-Carlo calculations give $\beta_{\rm MC}=-0.58$ \cite{Astrak04a}. Measurements give results ranging from $\beta_{\rm exp}=-0.49$ to $-0.68$ \cite{Barten04a, Bourde04a, Kinast05a, Partri06a}. In the following we use the Monte-Carlo value of $\beta_{\rm MC}=-0.58$. From the equation of state, we can
immediately calculate the energy density of the superfluid.
Namely, from the relation $\mu=\partial\epsilon_{\rm sf}/\partial
n$, we obtain upon integration that $\epsilon_{\rm sf}=
3(1+\beta_{\rm MC})\epsilon_{\rm F}n/5$. As a result, we find for
the grand potential density in the superfluid phase at zero temperature that
\begin{eqnarray}\label{eq:tpsfmc}
\omega_{\rm sf}(\mu_{\sigma})&=&\epsilon_{\rm sf}-\mu n
=-\frac{2}{5}(1+\beta_{\rm MC})\epsilon_{\rm F}n \\ &=&
-(1+\beta_{\rm MC}) \frac{2(2m)^{3/2}}{15 \pi^2
\hbar^3}\left(\frac{\mu}{1+\beta_{\rm MC}}\right)^{5/2}.\nonumber
\end{eqnarray}
When we have that $\omega_{\rm sf}(\mu_{\sigma})=\omega_{\rm
n}(\mu_{\sigma},0)$, then a first-order phase transition between
the equal density superfluid and the polarized normal phase
occurs. From Eqs.\ (\ref{eq:tpifgr}), (\ref{eq:mup}) and
(\ref{eq:tpsfmc}), it follows that the critical difference in the
chemical potentials is given by $h_{\rm c} = 0.93 \mu$. At this
critical value, the (local) polarization jumps from zero in the
superfluid phase to $p_{\rm c}=(n_+-n_-)/(n_++n_-)= 0.38$ in the normal
phase as found by Lobo {\it et al.} \cite{Lobo06a}. Note that this result is very different from
the mean-field result of Section \ref{par:hompd}, where we
obtained $p_{\rm c}=0.93$ for the critical polarization. The difference
shows the crucial quantitative effect of the strong interactions
in the normal state. In the trap, the equation of state of Lobo {\it et al.} in combination with the local-density approximation gives rise to $P_{\rm c} = 0.78$ \cite{Lobo06a}, as opposed to $P_{\rm c} = 0.998$ that follows from mean-field theory, see Fig.\ \ref{fig:pdtrap}.

By comparing $\omega_{\rm sf}$ with $\omega_{\rm n}$ everywhere in
the trap, we can determine which of the two phases is locally most
favorable and calculate the particle densities from the
corresponding equation of state \cite{Recati08a}. For the superfluid phase, the
local particle densities are obtained from $\mu({\bf
r})=(1+\beta_{\rm MC})\epsilon_{\rm F}({\bf r})$. We are now thus
in the position to determine the density profiles for the trapped
phase-separated state, which has a superfluid core and a
first-order transition to the normal state as a function of
position in the trap. This discontinuous transition has been
observed in the density profiles measured by Shin {\it et al.}
\cite{Shin08a}, as shown in Fig.\ \ref{fig:mitprof}. In panel (a), the
density profiles are shown for the spin-up particles, the
spin-down particles and their difference. The noisy curves are the
data, which compare well with the calculated lines. The
experimental profiles were obtained at a temperature of about
$T=0.02 T_{{\rm F}+}({\bf 0})$, where $T_{{\rm F}+}({\bf 0})$ is
the local Fermi temperature of the spin-up particles in the center
of the trap. The total polarization was about $P=0.44$ with a
total number of spin-up particles given by $N_+ = 6\cdot 10^6$.
Moreover, we specified the trapping potential $V^{\rm ex}(\rho,z)=
m (\omega^2_\perp (x^2+y^2)+\omega^2_z z^2)/2$ for the MIT
experiments in Section \ref{par:lda}.  From Fig.\ \ref{fig:mitprof}(a),
we see that the measured densities are equal in the superfluid
core, but upon going outwards from the center of the trap there is
a sudden rise in the density difference, indicating a first-order
transition. In the figure, we used the radial coordinate $r$ given
by $\bar{\omega}^2r^2=\omega^2_\perp (x^2+y^2)+\omega^2_z z^2$ with
$\bar{\omega}=(\omega_\perp^2\omega_z)^{1/3}$. The calculated profiles
are shown by the full line for $n_+ (r)$, by the dashed line for
$n_- (r)$, and by the dashed-dotted line for their difference. The
densities are all scaled by the central density of the spin-up
particles $n_{+}({\bf 0})$. The radial distances are scaled in the
figure by the radial cloud size of the spin-up atoms, given by
$R_{+}=(2\mu_+/m\bar{\omega}^2)^{1/2}$.

In Fig.\ \ref{fig:mitprof}(b), also the column densities $n_{\rm
c\sigma}$ are shown, which follow from
\begin{equation}
n_{\rm c\sigma}(x,z)=\int^{\infty}_{-\infty} dy~
n_{\sigma}(x,y,z).
\end{equation}
The column densities are less noisy, because they are directly
probed by the in-situ imaging that is performed in the experiment,
while the densities have to be reconstructed. The radial
coordinate for the column density is now given by
$\bar{\omega}^2r^2=\omega^2_\perp x^2+\omega^2_z z^2$, while the plotted
column densities are all scaled by the central column density of
the spin-up particles $n_{{\rm c}+ }(0,0)$. In general, the
agreement between the experimental curves and the theoretical
curves is very good. The small differences could be due to the
fact that the calculation was performed at zero temperature, while
the experiment is at very small nonzero temperature. Moreover, the
local-density approximation is expected to fail close to the
interface, because a true jump in the order parameter would cost
an infinite amount of gradient energy. If gradient terms
are taken into account, then a smooth order-parameter profile is
calculated, as we will see in more detail in Section \ref{sec:inh}.

Note that throughout the calculation in this section, we have
assumed that more exotic superfluid phases, like the FF and LO
phases mentioned in the introduction, or induced $p$-wave superfluidity \cite{Patton11a}, do not play a role at
zero temperature in the unitarity limit. To put it differently, we
have assumed that the mean-field phase diagrams of Figs.\
\ref{fig:hpdmf} and \ref{fig:pdtrap} are topologically correct. Seen the good agreement between experiments and theory, this assumption seems to be justified so far.

\section{The renormalization-group approach}\label{sec:rg}

In the following section, we discuss in some detail the renormalization-group (RG)
approach to interacting quantum gases in order to dicuss in a different way the importance of fermionic self-energy and screening effects. Renormalization
group theory is not only a powerful technique for studying quantitatively
strongly-interacting gases, but it also gives an elegant
conceptual framework for understanding many-body physics and phase transitions in general. The latter comes about because we are often
interested in determining the physics of a many-body system at the
macroscopic level, i.e., at long wavelengths or at low momenta. As
a result, we then need to determine the effect of
microscopic degrees of freedom with high momenta to arrive at an
effective quantum field theory for the long-wavelength physics. This is achieved elegantly by the momentum-space renormalization group approach as formulated originally by Wilson \cite{Wilson71a,Wilson74a}. In this section, we study his approach because of its simplicity both conceptually and technically. However, nowadays many formulations of the renormalization-group equations are available \cite{Wegner73a,Polchi84a,Wetter93a,Berges02a}, having each their advantages and disadvantages. The renormalization group procedure and related ideas have been also applied to the unitary Fermi gas, where early examples include the use of the $\epsilon$ expansion \cite{Nishid06a}, the $1/N$ expansion \cite{Nikoli07a,Veille07a}, and the
functional renormalization group approach \cite{Birse05a,Diehl07a}.

The goal of Wilsonian renormalization is to construct a
transformation that maps an action, characterized by a certain set
of coupling constants, to a new action for longer wavelengths, where the values of the
coupling constants have changed. This is achieved by performing
two steps. First, an integration over high-momentum degrees of
freedom is carried out, and the effect of this integration is
absorbed in the coupling constants of the action that are therefore said
to flow. Second, a re-scaling of all momenta and fields is
performed to bring the relevant momenta for the action back to
their original domain. By repeating these two steps over and over
again, it is possible to arrive at highly nonperturbative
approximations to the exact effective action.

At a continuous phase transition, the correlation length diverges
which implies that critical fluctuations are dominant at each
length scale and that the system becomes scale invariant. This
critical behavior is elegantly captured by the
renormalization-group approach, in which a critical system is
described by a fixed point of the above two-step transformation.
By studying the properties of these fixed points, it is possible
to obtain accurate predictions for the critical exponents that
characterize the nonanalytic behavior of various thermodynamic
quantities near the critical point. Moreover, the renormalization-group approach also explains universality, which is the observation that very
different microscopic actions give rise to exactly the same
critical exponents. It turns out that these different microscopic
actions actually flow to the same fixed point, which is to a large
extent solely determined by the dimensionality and the symmetries
of the underlying theory. As a result, critical phenomena can be
categorized in classes of models that share the same critical
behavior.

In this review, however, we do not wish to calculate universal critical exponents, since these have already been determined with great accuracy for the $XY$ universality class, which is the class of the transition to the superfluid state in three spatial dimensions. Rather, we want to calculate quantities like the critical temperature of a unitary interacting Fermi gases, that have only been measured in recent years. Such quantities are not universal from a renormalization group point of view, and depend on the microscopic details of the Hamiltonian. Therefore, in this section we use the renormalization group only as a nonperturbative method to iteratively solve a many-body
problem, rather than as a map between actions with the same high-momentum cutoff from which critical scaling relations can be derived. As a result, the renormalization step in which the fields and coupling constants are rescaled is not particularly useful for our goals and we leave it away.

\subsection{The first step of Wilsonian renormalization}\label{par:wilrg}

Thus, we only perform the first step of the renormalization method, which is to
evaluate the Feynman diagrams that renormalize the coupling
constants of interest, while keeping the integration over the
internal momenta restricted to the considered high-momentum shell. Only one-loop and tree diagrams contribute to the
flow, because we consider the width of the momentum shell to be infinitesimally small, and each loop introduces an additional factor proportional to the infinitesimal width. Although the one-loop structure of the infinitesimal Wilsonian RG is exact, it does not mean that it is easy to also obtain exact results, since this would require the consideration of an infinite number of coupling constants. Since the latter is usually not possible in practice, the
RG distinguishes between the relevance of the coupling constants, so that a small set of them may already lead to useful non-perturbative results.

To start the derivation of the Wilsonian renormalization group equations, we consider the action in momentum space, $S =S_0 +S_I$, where the noninteracting part is given by
\begin{eqnarray}\label{eq:s0}
S_0[\phi]=-\sum_{{\bf k},n,\sigma}\phi^*_{\sigma,n,{\bf
k}}\hbar G^{-1}_{0,\sigma}(i \omega_n,{\bf k})\phi_{\sigma,n,{\bf k}},
\end{eqnarray}
with the noninteracting Green's function of Eq.~(\ref{eq:greenf}). The interacting part is given by
\begin{eqnarray}\label{eq:si}
S_I[\phi] =\frac{V'_{0}}{\hbar\beta V}\sum_{\substack{{\bf k},{\bf k'},{\bf
q}\\{n},{n'},{m}}} \phi^*_{+,n',{\bf k'}}\phi^*_{-,m-n',{\bf
q-k'}}\phi_{-,m-n,{\bf q-k}}\phi_{+,n,{\bf
k}}~,
\end{eqnarray}
where the notation with the prime, namely $V'_{0}$, indicates that the interaction strength is a renormalizing quantity in our treatment. In the same way, we will use the notation $\mu_{\sigma}'$ for the renormalizing chemical potentials. To arrive at the one-loop corrections to this action arising from the integration over fluctuations in an infinitesimal high-momentum shell, it is particularly convenient to use the following procedure, where all one-loop corrections can be obtained by performing a single Gaussian functional integral. Namely, by splitting the functional integral of Eq.~(\ref{eq:parsum}) into a part that contains the integration over low-momentum modes  $\phi^{<}_{n,{\bf k}}$ with small wavenumbers $|{\bf k}|<\Lambda$, and a part that contains the integration over high-momentum modes $\phi^{>}_{n,{\bf k}}$ with large wavenumbers in an infinitesimal shell $\Lambda-d\Lambda<|{\bf k}|<\Lambda$, we obtain
\begin{eqnarray}\label{eq:derrggf}
Z &=&\int \mathcal{D}\phi^{<} e^{-S_{0}[\phi^{<}]/\hbar} \int
\mathcal{D}\phi^{>}e^{-S_{0}[\phi^{>}]/\hbar}e^{-S_{\rm I}[ \phi^{<},
\phi^{>}]/\hbar}\nonumber \\
&=& \int \mathcal{D}\phi^{<} e^{-S[\phi^{<}]/\hbar}\nonumber \\
& \times &\int
\mathcal{D}\phi^{>} \exp\left\{
\sum_{n',{\bf k}'} \boldsymbol{\Phi}^{>\dagger}_{n',{\bf k}'}\cdot {\bf G}_>^{-1}(i \omega_{n'},{\bf k}')\cdot \boldsymbol{\Phi}^{>}_{n',{\bf k}'}\right\}\nonumber\\
&=&\int  \mathcal{D}\phi_{<} e^{-S[\phi_{<}]/\hbar} \exp\left\{-{\rm Tr}[ \log(-{\bf G}_>^{-1})]\right\},
\end{eqnarray}
where we introduced the notation $\boldsymbol{\Phi}^{>\dagger}_{n',{\bf k}'}=\left[ \phi_{+,n',{\bf k}'}^{>*},\phi_{-,-n',-{\bf k}'}^{>} \right]$ for the fast modes in the high-momentum shell. In the second line of Eq.~(\ref{eq:derrggf}), we moved all terms that only depend on long-wavelength modes ($\phi_{n,{\bf k}}^{<}$) to the first integral, while in the exponent of the second integral on the third line we only kept the terms that are quadratic in the short-wavelength modes ($\phi_{n',{\bf k}'}^{>}$). We note that other terms, like of linear order or higher orders in $\phi_{>}$, would lead after expansion of the exponent and integration over $\phi_{>}$ either to tree diagrams or to higher-loop Feynman diagrams. The tree diagrams that are allowed by momentum conservation give rise to effective three-body interactions or higher, which we ignore in the present discussion. Moreover, we argued that higher-loop diagrams vanish because we work in the limit of infinitesimal shell width. The inverse Green's function matrix
${\bf G}_>^{-1}$ of Eq.~(\ref{eq:derrggf}) is given by ${\bf G}_>^{-1}={\bf G}_{0}^{-1}-\boldsymbol{\Sigma}_>$, where ${\bf G}_{0}^{-1}$ is the Fourier transform of Eq.~(\ref{eq:g0m}), while we have for $\boldsymbol{\Sigma}_>$ that
\begin{eqnarray}
\boldsymbol{\Sigma}_>=  \frac{V'_0}{\hbar^2\beta V}\sum_{n,{\bf k}} \left[
\begin{array}{cc}
\phi^{<*}_{-,n,{\bf k}}\phi^<_{-,n,{\bf k}}  & \phi^<_{-,-n,-{\bf k}}\phi^<_{+,n,{\bf k}} \\
\phi^{<*}_{+,n,{\bf k}}\phi^{<*}_{-,-n,-{\bf k}}  & \phi^<_{+,n,{\bf k}}\phi^{<*}_{+,n,{\bf k}}
\end{array} \right],
\end{eqnarray}
which follows directly from Eq.~(\ref{eq:si}), Eq.~(\ref{eq:derrggf}) and assuming that the interaction stays local during the RG flow.

In the last step of Eq.~(\ref{eq:derrggf}), we have performed an exact integration over the fast modes, resulting in the well-known expression for a Gaussian functional integral. The trace is to be taken over the 2$\times$2 spin space structure of ${\bf G}^{-1}$, over all Matsubara frequencies, and over the momenta in the considered high-momentum shell.
To expand ${\rm Tr}[ \log(-{\bf G}_>^{-1})]$ in powers of $\phi_{<}$, we use
\begin{equation}
\log(-{\bf G}_>^{-1}) = \log(-{\bf G}_0^{-1})+ \log({\bf 1}-{\bf G}_0\boldsymbol{\Sigma}_>).
\end{equation}
For example, expanding the logarithm to first order in $\boldsymbol{\Sigma}_>$, we find, after taking the trace over the 2$\times$2 Nambu space, the contributions proportional to $\phi^{<*}_{\sigma}\phi^{<}_{\sigma}$. Namely, we have that
\begin{eqnarray}\label{eq:shell}
&&\frac{1}{\hbar}\sum_{{\bf k},n}\phi^{<*}_{-,n,{\bf k}}\phi^{<}_{-,n,{\bf k}}V'_0\sum_{{\bf k}',n'} \frac{G_{0,+}({\bf k}',n')}{\hbar\beta V} =\nonumber\\
&&\frac{1}{\hbar}\sum_{{\bf k},n}\phi^{<*}_{-,n,{\bf k}}\phi^{<}_{-,n,{\bf k}}d\Lambda\frac{\Lambda^2}{2
\pi^2}V'_0 f_+(\Lambda),
\end{eqnarray}
where in the second line we first summed the Green's function over the fermionic Matsubara frequencies resulting in the Fermi distribution $f_{\sigma}({\bf k})=1/\{\exp[\beta
(\epsilon_{{\bf k}}-\mu'_{\sigma})]+1\}$ (see e.g. Ref \cite{Bruus04a} for more details), after which we converted the sum over ${\bf k}'$ into an integral. Since we integrate over a shell of infinitesimal width, the outcome is simply the value of the integrand at the momentum $\hbar \Lambda$ times the width of the shell $d\Lambda$.  Note that Eq. (\ref{eq:shell}) is valid for integration from $\Lambda$ to $\Lambda + d\Lambda$, i.e., integrating out fluctuations from lower to higher momenta, where the cutoff $\Lambda$ sets the scale up to which fluctuations have already been integated out. As a result, we would have an additional minus sign for integration of fluctations from higher to lower momenta, since then $d\Lambda$ becomes negative.

Eq.~(\ref{eq:shell}) can be interpreted as a self-energy contribution, which can be absorbed in the 'renormalized' chemical potential of the minority particles, $\mu'_{-}$. We thus find that
\begin{equation}\label{eq:dmu}
d\mu'_{\sigma}=-\frac{\Lambda^2}{2 \pi^2}
V'_{0}f_{-\sigma}(\Lambda)d\Lambda~,
\end{equation}
which is a differential equation for $\mu'_{\sigma}$. The change in the chemical potential is seen to be proportional to the (renormalized) interaction strength $V'_{\bf 0}$ and the (renormalized) average density of particles, as expected for a self-energy contribution. By expanding ${\rm Tr}[ \log(-{\bf G}_>^{-1})]$ to second order in $\Sigma_{>}$, we can also find the renormalization of the interaction strength, resulting ultimately in
\begin{equation}\label{eq:renint}
dV'_{0} =-{V'_{0}}^2\frac{\Lambda^2}{2
\pi^2}
\left[\frac{1-f_{+}(\Lambda)-f_{-}(\Lambda)}{2(\epsilon_{\Lambda}-\mu')}-\frac{f_{+}(\Lambda)-f_{-}(\Lambda)}{2
h'}\right]d\Lambda,
\end{equation}
with $\mu'=(\mu'_{+}+\mu'_{-})/2$ and $h=(\mu'_{+}-\mu'_{-})/2$. The first term in Eq. (\ref{eq:renint}) corresponds to
the so-called ladder diagram and describes the scattering between particles.
The second term corresponds to the so-called bubble diagram and describes
screening of the interaction by particle-hole excitations. Also
note that due to the coupling of the differential equations for
$\mu'_{\sigma}$ and ${V'_{0}}^{-1}$, we automatically
generate an infinite number of Feynman diagrams, showing the
nonperturbative nature of the RG. The expressions in Eqs.~(\ref{eq:dmu}) and (\ref{eq:renint}) are
diagrammatically represented in Fig.~\ref{fig:rgfd}. The only difference here compared to the usual Feynman rules is that the internal momenta in the one-loop diagrams are restricted to stay in the infinitesimal high-momentum shell.

\subsection{Renormalization theory for fermions}

\begin{figure}
\includegraphics[width=0.8\columnwidth]{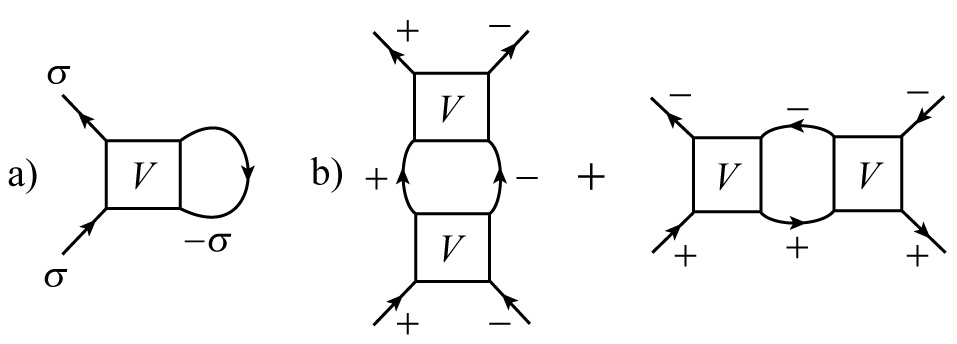}
\caption{\label{fig:rgfd} Feynman diagrams renormalizing a) the
chemical potentials and b) the interatomic interaction.}
\end{figure}

From the previous discussion, we conclude that if we consider the interaction vertex to remain frequency and momentum independent during the RG procedure, and if we only take into account the renormalization of the
three coupling constants $\mu'_{\sigma}$ and $V'_{{\bf 0}}$,
we find the following coupled RG equations \cite{Gubbel08a}
\begin{eqnarray}
\frac{d{V'_{0}}^{-1}}{d\Lambda} &=&\frac{\Lambda^2}{2
\pi^2}
\left[\frac{1-f_{+}(\Lambda)-f_{-}(\Lambda)}{2(\epsilon_{\Lambda}-\mu')}-\frac{f_{+}(\Lambda)-f_{-}(\Lambda)}{2
h'}\right], \label{eq:RGI}\\
\frac{d\mu'_{\sigma}}{d\Lambda}&=&-\frac{\Lambda^2}{2 \pi^2}
\frac{f_{-\sigma}(\Lambda)}{{V'_{0}}^{-1}}~, \label{eq:RGC}
\end{eqnarray}
with the Fermi distribution $f_{\sigma}(\Lambda)=1/\{\exp[\beta (\epsilon_{\Lambda}-\mu'_{\sigma})]+1\}$. To be able to solve these coupled differential equations numerically, we need to specify the initial conditions. We start the RG flow at a high-momentum cutoff $\hbar\Lambda_0$, where the chemical potential is not yet renormalized, so that we have $\mu'_{\sigma}(\Lambda_0)= \mu_{\sigma}$ with $\mu_{\sigma}$ the (not renormalized or bare) chemical potential, which is a given thermodynamic variable when working in the grand-canonical ensemble. As an initial condition for the interaction, we use $V'_{0}(\Lambda_0)=4 \pi^2 a \hbar^2/m(\pi-2 a\Lambda_0)$ with $a$ the $s$-wave scattering
length. We encountered this expression already before in Eq.~(\ref{eq:intvscut}). There, this expression for the two-body case was found to be the exact relation between the microscopic (not renormalized or bare) interaction parameter $V_{\bf 0}$ and the effective interaction strength, namely the two-body transition matrix $T^{\rm 2b}$ with scattering length $a$. When we take the two-body limit of Eq.~(\ref{eq:RGI}), i.e. $f_{\sigma} \rightarrow 0$ and $\mu'_{\sigma} \rightarrow 0$, we find the differential form of Eq.~(\ref{eq:t2b-v0}). This implies that the two-body limit of the RG equations gives after integration rise to a renormalized interaction strength of $T^{\rm 2b}=4\pi a \hbar^2/m$. The initial condition for the interaction is thus chosen to
make sure that we automatically incorporate the correct two-body
scattering length into the RG theory. Moreover, this initial condition also ensures that at the end of the RG calculation no dependence on the arbitrary high-momentum
cut-off $\hbar\Lambda_0$ remains. The results depend on the thermodynamic variables $\mu_{\sigma}$, $T$ and the scattering length $a$. In the unitarity limit, when $a\rightarrow\infty$, we have that ${V'_{0}}^{-1}(\Lambda_0)=-m \Lambda_0/2\pi^2\hbar^2$, so that then also the dependence on $a$ disappears.

Although we have specified the initial conditions, the outcome of the differential equations is not yet completely fixed. Because the differential equations are coupled, this final outcome depends on the specific way we perform the intermediate steps, i.e. it depends on the way in which we flow. Ultimately, we want to take the effect of the fluctuations in all momentum shells into account, and to this end we have to specify which momentum shells to consider first and which to consider last. If we would be able to calculate all (infinitely many) coupling constants generated by the RG procedure exactly, then the order in which we integrate out the fluctuations would not matter anymore, because all effects of the fluctuations in one momentum shell on the other shells would be fully included. However, for our current simplest nontrivial RG equations, this is not the case. If the RG is formulated for classical systems or bosonic systems, then usually the most relevant excitations of lowest energy have a momentum near zero. Therefore, the RG equations are then taken to start at the high-momentum cutoff $\Lambda_0$, and the effects of an increasing amount of fluctuations on the low-momentum modes is considered by using for example $\Lambda(l)=\Lambda_0 e^{-l}$. If we insert $\Lambda(l)$ and $d\Lambda(l)/dl$ into Eqs.~(\ref{eq:RGI}) and (\ref{eq:RGC}), we get a set of coupled differential equations in $l$ which we can numerically evolve from $l = 0$ to $l=\infty$.
Note that then an additional minus sign is needed, because we flow from high to low momenta.

\begin{figure}
\begin{center}
\includegraphics[width=1.0\columnwidth]{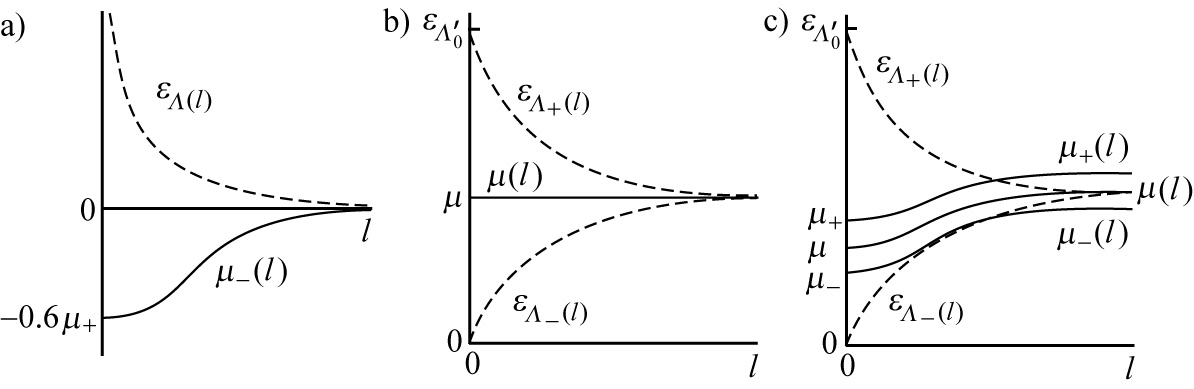}
\caption{\label{fig:flow} a) Position of the momentum shells
(dashed lines) and flow of the chemical potentials (solid lines)
for a) the strongly interacting extremely imbalanced case, b) the
weakly interacting balanced case, and c) the strongly interacting
imbalanced case. }
\end{center}
\end{figure}

For a fermionic system the relevant excitations of lowest energy have a momentum near the Fermi energy, which is therefore a natural endpoint for a RG flow \cite{Shankar94a}. In an imbalanced system, actually three Fermi levels are relevant, namely one for each species and the average Fermi level. Moreover, these Fermi levels, given by the renormalized chemical potentials $\mu'_{\sigma}(l)$, are shifting during the flow. If we would want to flow automatically to the renormalized average Fermi level, we could use the following procedure. First, we start at a
high momentum cutoff $\hbar\Lambda_0$ and flow to an intermediate cutoff
$\hbar\Lambda'_0$ to integrate out high-momentum (two-body) physics. Then, we start integrating out the rest of the
momentum shells approximately symmetrically with respect to the flowing average
Fermi level to treat the many-body physics. This is achieved by using \cite{Gubbel08a}
\begin{equation}\label{eq:flowmom1}
\Lambda_{+}(l)=
\left(\Lambda'_0- \sqrt{\frac{2m\mu}{\hbar^2}}
\right)e^{-l}+\sqrt{\frac{2m\mu'(l)}{\hbar^2}}
\end{equation}
and by
\begin{equation}\label{eq:flowmom2}
\Lambda_{-}(l)= - \sqrt{\frac{2m\mu}{\hbar^2}}e^{-l}+
\sqrt{\frac{2m\mu'(l)}{\hbar^2}}.
\end{equation}
Note that, as desired, we have that $\Lambda_{+}(l)$ starts at
$\Lambda'_0$ and automatically flows from above to $\sqrt{2
m\mu'(\infty)}/\hbar$, whereas $\Lambda_{-}(l)$ starts at 0 and
automatically flows from below to $\sqrt{2 m\mu'(\infty)}/\hbar$. This behavior is illustrated in Fig.~\ref{fig:flow}.
By substituting $\Lambda_{+}(l)$, $\Lambda_{-}(l)$ and their
derivatives in Eqs.\ (\ref{eq:RGI}) and (\ref{eq:RGC}), we obtain a set of coupled differential equations in
$l$ that again can be solved numerically. Note that this procedure leads to different results than from using $\Lambda(l)=\Lambda_0 e^{-l}$. A similar procedure can also be used for the imbalanced case to flow automatically to either one of the Fermi levels $\mu'_{\sigma}(l)$, or even to both Fermi levels simultaneously.

As a somewhat technical side comment we note that the two functions $\hbar\Lambda_{\sigma}(l)$ are not exactly symmetric with respect to $\sqrt{2 m\mu'(l)}$. This can be a problem when there are true poles in the RG equations, as happens for example to Eq.~(\ref{eq:RGI}) at zero temperature. The RG equation for the interaction can then be interpreted as the calculation of the principal value for a shifting pole. In order to get finite results the divergent parts on each side of the pole have to exactly cancel each other, which can be somewhat cumbersome to achieve, since also the derivative of the shifting pole plays a role. This can be seen from Eqs.~(\ref{eq:RGI}), (\ref{eq:flowmom1}) and (\ref{eq:flowmom2})  and noting that $d\Lambda_{\pm}/dl$ enter the differential equation on both sides of the pole. However, for our present purpose this problem does not play a major role, since we are interested primarily in nonzero temperatures and then Eq.~(\ref{eq:RGI}) has a well-defined right-hand side for all $\Lambda(l)$.

Unfortunately, close to the critical temperature, Eq.~(\ref{eq:RGI}) gives rise to another problem. Namely,
when the Fermi mixture is critical, the inverse many-body
vertex $V^{-1}_{{\bf 0}}$ flows to zero according to the
Thouless criterion \cite{Nozier85a,Thoule60a} and the chemical potentials in Eq.~(\ref{eq:RGC}) consequently
diverge, which is unphysical. The most simple way to solve this problem and calculate critical properties more realistically, is to take the momentum dependence of the renormalized interaction vertex into account. Namely, considering the ladder diagram in Fig.~(\ref{fig:rgfd}), we note that in general this diagram depends on the external
center-of-mass momentum ${\bf q} $ and  $\omega_m$ of Eq.~(\ref{eq:si}). As a result, this diagram leads in general to the contribution
\begin{eqnarray}\label{eq:ladderrgf}
&&d\Xi(q^2,i \omega_m)=\int_{d\Lambda}\frac{d{\bf k
}}{(2\pi)^3}\frac{1-f_{+}(\epsilon_{\bf
k})-f_{-}(\epsilon_{\bf
q-k})}{-i\hbar\omega_m+\epsilon_{\bf k}+\epsilon_{\bf
q-k}-2\mu'}~,
\end{eqnarray}
where during integration over high-momentum modes both ${\bf k}$ and ${\bf q-k}$ have to
remain in the infinitesimal shell of width $d\Lambda$. We note that the expression on the right-hand side of Eq.\ ({\ref{eq:ladderrgf}}) depends only on $q^2$, rather than on ${\bf q}$.

We thus have that the renormalization procedure can lead to a momentum-dependent interaction, even though the microscopic interaction is to a very good approximation momentum independent. If we consider only ladder diagrams, which is a commonly used approximation for dilute quantum gases because ladder diagrams do not vanish in the two-body limit, then the renormalized interaction will depend in first instance only on ${\bf q} $ and  $\omega_m$. In order to simplify things even further, we only take the
center-of-mass momentum into account by expanding the (inverse)
interaction in the following way: ${V'_{{\bf
q}}}^{-1}={V'_{0}}^{-1}-{Z'_{q}}^{-1}q^2$. The flow
equation for the additional coupling constant ${Z'_{q}}^{-1}$ is then obtained by \cite{Gubbel08a}
\begin{eqnarray}\label{eq:zqrg}
 d{Z'_{q}}^{-1}=\left.\frac{\partial d\Xi(q^2,0)}{\partial q^2}\right|_{q=0}.
\end{eqnarray}

\subsection{The extremely imbalanced case}

To apply the discussed RG formalism, we consider first one spin-down particle in a Fermi sea of spin-up particles at zero temperature in the unitarity limit, also called the Fermi polaron, which was discussed in Section \ref{par:polaron}. When we
calculate the self-energy of the polaron with the present renormalization group approach, the RG equations are simplified, because $f_{-}(\Lambda)$ can be set to zero and thus $\mu_{+}$ is not renormalized. In Section \ref{par:polaron}, it was mentioned that the contour integration to determine the one-loop Matsubara sum in the diagram of Fig.~\ref{fig:rgfd}(a), leads to the frequency substitution $i\hbar \omega_m\rightarrow \epsilon_{\bf q}-\mu'_{+}$ in Eq.\
(\ref{eq:ladderrgf}). Using this and accounting for the external momentum dependence with the coupling
${Z'_{q}}^{-1}$, we obtain \cite{Gubbel08a}
\begin{eqnarray}
\frac{d {V'_{0}}^{-1}}{d\Lambda} &=&\frac{\Lambda^2}{2 \pi^2}
\left(\frac{1-f_{+}(\Lambda)}{2\epsilon_{\Lambda}-
\mu'_{-}}-\frac{f_{+}(\Lambda)}{2
h'}\right)~,\label{eq:RGei1}\\
\frac{d\mu'_{-}}{d\Lambda}&=&\frac{\Lambda^2}{2 \pi^2}
\frac{f_{+}(\Lambda)}{-{V'_{0}}^{-1}+{Z'_{q}}^{-1}\Lambda^2}~,\label{eq:RGei2}\\
\frac{d{Z'_{q}}^{-1}}{d\Lambda} &=& - \frac{\hbar^4\Lambda^4}{6
\pi^2m^2}
\frac{1-f_{+}(\Lambda)}{(2\epsilon_{\Lambda}-\mu'_{-})^3}~.\label{eq:RGei3}
\end{eqnarray}
Eq.~(\ref{eq:RGei1}) is similar to Eq.~(\ref{eq:RGI}), where the main difference is the denominator of the ladder diagram contribution caused by the just described substitution that takes into account the frequency-dependence of the interaction. We note that in the above equations we have not considered the effect of external momenta and frequencies on the bubble diagram contribution, which is an approximation.
Eq.~(\ref{eq:RGei2}) is similar to Eq.~(\ref{eq:RGC}), but the denominator has changed because we include the effect of the momentum-dependent interaction $V_{\bf q}$ in the self-energy contribution. As a result, we have that the interaction in Eq.~(\ref{eq:shell}) would now be given by $V_{{\bf k}+{\bf k}'}$ rather than $V_{{\bf 0}}$. This also means that the resulting renormalization contribution to the self-energy depends on ${\bf k}$, giving rise to a renormalization of the effective mass of the fermions. However, in our present treatment we only consider the renormalization of the chemical potential, i.e., the momentum-independent part of the self-energy, and we put ${\bf k}=0$. Eq.~(\ref{eq:RGei3}) follows from Eqs.~(\ref{eq:ladderrgf}) and (\ref{eq:zqrg}), where we set $f_-$ to zero and perform the mentioned frequency substitution. The initial conditions for the above coupled differential equations are ${V'_{0}}^{-1}(\Lambda_0)=-m \Lambda_0/2\pi^2\hbar^2$, $\mu'_{-}(0)=\mu_{-}$ and
${Z'_{q}}^{-1}(0)=0$, since the interaction starts out as being
momentum independent.

Note that for the case to our interest we have that
$\mu'_{-}$ is initially negative and increases during
the flow due to the strong attractive interactions. The quantum phase transition from a zero density to a nonzero density of spin-down particles occurs for the initial
value $\mu_{-}$ that at the end of the flow precisely leads to
$\mu'_{-}(\infty)=0$. We find $\mu_{-}=-0.52 \mu_{+}$, where we used $\Lambda(l)=\Lambda_0 e^{-l}$ to flow to the final value of the renormalized minority Fermi surface. Note that we thus again have to include the additional minus sign for going from high to low momenta. This initial value of $\mu'_{-}$ is therefore also the self-energy of a strongly interacting
spin-down particle in a sea of spin-up particles \cite{Combes07a}. Our value for the self-energy is somewhat less than the results obtained by Monte-Carlo calculations \cite{Lobo06a,Prokof08a,Pilati08a}, giving rise to $\mu_{-}=-0.6 \mu_{+}$, while the experiment of Schirotzek {\it et al.} gives $- 0.64(7)$  $\mu_{+}$ \cite{Schiro09a}. The difference is probably mainly caused by an
overestimation of the screening of the interaction, as also explained in Section \ref{par:polaron}. We come back to this point in the next section. In Fig.~\ref{fig:flow}(a) the flow $\mu'_{-}(l)$ of
the spin-down chemical potential is depicted as a function of $l$.

\subsection{Homogeneous phase diagram}

Next, we determine the nonuniversal critical properties of the
strongly interacting Fermi mixture at nonzero temperatures with the RG approach \cite{Gubbel08a}, and in
particular the location of the tricritical point in the
homogeneous phase diagram. We use
Eq.\ (\ref{eq:RGI}) for the flow of ${V'_{0}}^{-1}$, while the
expression for the flow of $\mu'_{\sigma}$ is given by
\begin{eqnarray}
\frac{d\mu'_{\sigma}}{d\Lambda}&=&\frac{\Lambda^2}{2 \pi^2}
\frac{f_{-\sigma}(\Lambda)}{-{V'_{0}}^{-1}+{Z'_{q}}^{-1}\Lambda^2}~, \label{eq:RGCI}
\end{eqnarray}
similar to Eq.~(\ref{eq:RGei2}).
The flow equation for ${Z'_{q}}^{-1}$
can be obtained analytically from Eqs.~(\ref{eq:ladderrgf}) and (\ref{eq:zqrg}), but is somewhat cumbersome to write down explicitly. Note that now we do not use the substitution $i\hbar \omega_m\rightarrow \epsilon_{\bf q}-\mu_{+}$ anymore in Eq.~(\ref{eq:ladderrgf}), since this substitution is only correct for the extremely imbalanced case at zero temperature. The initial conditions are
the same as before with in addition
$\mu'_{+}(0)=\mu_{+}$. As
mentioned previously, the critical condition is the Thouless criterion
that the fully renormalized vertex ${V'_{0}}^{-1}(\infty)$
flows to zero. From Eq.\ (\ref{eq:RGCI}), we see that incorporating
the coupling constant ${{Z'_{q}}}^{-1}$
solves the problem of the diverging chemical
potential.

We first apply the above procedure to study the equal density
case (i.e., $h\rightarrow 0$) as a function of the negative
scattering length $a$. The scattering length enters the
calculation through the initial condition of ${V'_{0}}^{-1}$.
To be able to express our results in terms of the Fermi energy
$\epsilon_{\rm F}= \epsilon_{{\rm F}\sigma}$, we calculate the (renormalized)
densities of the atoms from
\begin{equation}\label{eq:densrg}
n_{\sigma}= \int \frac{d{\bf k}}{(2\pi)^3}\frac{1}
{e^{\beta(\epsilon_{\bf k}-\mu'_{\sigma}(\infty))}+1}.
\end{equation}
In the weak-coupling limit, $a \rightarrow 0^{-}$, the chemical
potentials hardly renormalize, so that only Eq.\ (\ref{eq:RGI}) is
relevant. The critical temperature becomes exponentially small,
which allows us to integrate Eq.\ (\ref{eq:RGI}) exactly with the
result $k_{\rm B}T_{\rm c}=8\epsilon_{\rm
F}e^{\gamma-3}\exp\{-\pi/2k_{\rm F} |a|\}/\pi$ and $\gamma$
Euler's constant. Compared to the standard BCS-result we have an
extra factor of $1/e$, coming from the screening effect of the
bubble diagram that is not present in BCS theory. It is to be
compared with the so-called Gor'kov correction \cite{Gorkov61a},
which is known to reduce the critical temperature by a factor of
2.2 in the weak-coupling BCS-limit \cite{Heisel00a}. The
difference with our present result is that we have only allowed
for a nonzero center-of-mass momentum, whereas to get precisely
the Gor'kov correction we would also need to include the relative
momentum. We see that due to our approximation of neglecting the
relative momenta in the bubble diagram, we overestimate the
screening effect on the critical temperature by 20\%. Note that in
the previous section, we also already concluded that screening
effects are slightly overestimated by the present simplest nontrivial RG. At larger values of $|a|$, the flow of the chemical potential
becomes important and we obtain higher critical temperatures. In
the unitarity limit, when $a$ diverges, we obtain that $T_{\rm c}=0.14 T_{\rm F}$ and
$\mu (T_{\rm c}) = 0.59 \epsilon_{\rm F}$.

\begin{figure}[t]
\begin{center}
\includegraphics[width=0.7\columnwidth]{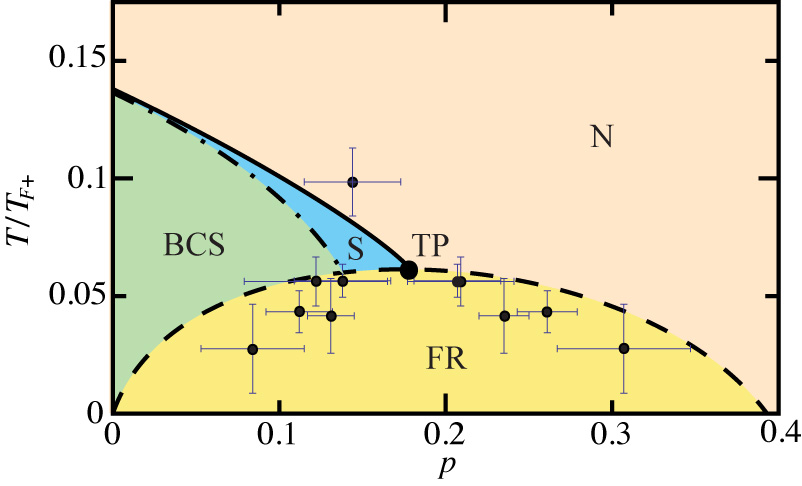}
\caption{\label{fig:pdrg} The phase diagram of the homogeneous
two-component Fermi mixture in the unitarity limit, containing the
superfluid Sarma (S) and BCS phases, the normal phase (N) and a
forbidden region (FR). The solid black line is the result of the
RG calculations. The dots with error bars are experimental data along the phase boundaries as determined by Shin {\it et al.} \cite{Shin08a}. The dashed and
dashed-dotted lines are only guides to the eye.}
\end{center}
\end{figure}

We can also calculate the critical
temperature as a function of polarization $p$ with the RG approach \cite{Gubbel08a} and compare with the
recent experiment of Shin {\it et al.} \cite{Shin08a}. The result
is shown in Fig.\ \ref{fig:pdrg}. The location of the tricritical point (TP) is determined from the fourth-order
coefficient in the Landau theory for the superfluid phase
transition. If this coefficient changes sign, then the nature of the phase transition
changes from second order to first order. This was explained more
elaborately in Section \ref{par:hompd}, where the right-hand side
of Eq.\ (\ref{eq:betaL}) is the mathematical expression
for the Feynman diagram in Fig.~\ref{fig:fdcp}(b), which is the diagrammatic representation of this fourth-order coefficient. In our RG
approach, we calculate this one-loop diagram, where during the flow the integration over internal momenta is confined to the infinitesimal shell that is integrated out.
Moreover, we have that during the RG flow the chemical potentials
are renormalizing, so that self-energy corrections to the Feynman
diagram of Fig.~\ref{fig:fdcp}(b) are automatically taken into account. With this procedure we obtain for the tricritical point that $P_{\rm c3}=0.18$ and $T_{\rm c3} =0.063$ $T_{{\rm
F}+}$.

The results of Fig.~\ref{fig:pdrg} with the RG approach should be compared with those of Fig.~\ref{fig:hpdmf} with the mean-field approach. We see that the location of the tricritical point obtained from RG theory has shifted to much lower temperatures and polarizations than the one from mean-field theory. As a result, the RG calculation is in much better agreement with experiments. We believe that the RG captures two main shortcomings of the mean-field theory, namely it takes into account fermionic self-energy effects and screening effects. Actually, the level of agreement with experiment is rather remarkable considering the simplicity of our RG. To some extent this is a coincidence, since there are many couplings whose renormalization we have ignored here although they could have a quantitative influence, such as e.g. the effective mass of the fermions. In Ref.~\cite{Gubbel08a}, we for example also included the center-of-mass frequency dependence of the interaction and found $P_{\rm c3}=0.24$ and $T_{\rm c3} =0.063$ $T_{{\rm F}+}$. Moreover, the results of the RG are also sensitive to the precise way in which we flow, so that the results depend for example on the intermediate cut-off $\hbar\Lambda'_0$. We pick $\Lambda'_0$ such that the high-energy two-body physics has been integrated out to a large extent, but the many-body physics not yet. This means that we take $\Lambda'_0$ to be a few times the Fermi wavevector. However, this procedure has some arbitrariness, and in an exact treatment the results should be fully independent of $\Lambda'_0$. We note that in Fig.~\ref{fig:pdrg}, the dashed and dashed-dotted lines have the same meaning as in the homogeneous phase diagram of Fig.~\ref{fig:hpdmf}. However, with our current RG for the normal phase these lines cannot be calculated, since for this a treatment of the superfluid phase would be required. Finally, we mention that at zero temperature, the Monte-Carlo treatment of Lobo {\it et al.} predicts a quantum phase transition from the equal-density superfluid to the polarized normal phase
at a critical imbalance of $p=0.38$, as was discussed in Section~\ref{par:denprof} \cite{Lobo06a}. This value seems to be in reasonably good agreement with experiments as seen from Fig.~\ref{fig:pdrg}.

\subsection{Renormalization theory for Cooper pairs}
\label{par:cprg}

In the previous paragraphs we studied renormalization effects using the fermionic action of Eq.~(\ref{eq:acfer}).  In Section~\ref{par:hstrans} we showed that the Hubbard-Stratonovich transformation can be used to derive two more actions that are equivalent to Eq.~(\ref{eq:acfer}). These two can also be used for a renormalization-group study of the strongly-interacting Fermi gas. In Refs.~\cite{Birse05a,Diehl07a,Bartos09a,Diehl10a}, the Bose-Fermi action of Eq.~(\ref{eq:acbosfer}) was used as a starting point, while next we study renormalization effects for the Cooper-pair action of Eq.~(\ref{eq:acbos}). The Cooper-pair action is an exact microscopically derived action for the order parameter to our interest, which is the expectation value of the BCS pairing field $\Delta(\tau,{\bf r})$.  In Section \ref{sec:mf}, we extensively studied the saddle-point or mean-field approximation to the Cooper-pair action. This simple approximation was seen to give a very satisfactory qualitative description of the experimentally observed phase transitions in the strongly-interacting regime. In Section \ref{par:nsr}, we also studied the Gaussian fluctuations of the pairing field in the normal phase for the balanced Fermi gas, called the NSR theory. This theory resulted in good quantitative agreement with thermodynamic measurements on the equation of state. Moreover, it was shown in Ref.~\cite{Hu10a} that the agreement between the Gaussian fluctuation theory and experiments is also good in the superfluid phase.

Unfortunately, as soon as we consider even the smallest population imbalances, the Gaussian fluctuation theory gives rise to unphysical results \cite{Parish07b}. Namely, near the critical temperature the Nozi\`{e}res-Schmitt-Rink theory predicts a negative polarization $p=(n_{+}-n_{-})/(n_{+}+n_{-})$ for a positive chemical potential difference $(\mu_{+}-\mu_{-})$, which corresponds to a compressibility matrix $-\partial^2 \omega_{\rm gc}/\partial \mu_{\sigma} \partial \mu_{\sigma'}$ with $\omega_{\rm gc}$ the grand-potential density that is not positive definite. This is a very unsatisfactory situation, especially considering the success of the NSR theory for the balanced case. In this section we try to solve this fundamental problem of the Gaussian fluctuation theory with the renormalization-group approach. The NSR theory for the balanced Fermi gas \cite{Nozier85a} takes into account only a noninteracting gas of non-condensed Cooper pairs. We can therefore improve the theory for the Cooper-pairs by taking also into account the effect of the interactions between the pairs. Namely, as seen from Eq.~(\ref{eq:acbos}), the pair action does not only contain a noninteracting part, but also two-pair interactions, three-pair interactions, and all higher-order interactions. A Popov theory for the bosonic pairs that includes the pair interaction effects was formulated by Pieri and Strinati, leading in the BEC regime to Popov's results for point-like bosons \cite{Pieri00a,Pieri05a}. In this regime the theory of Pieri and Strinati gives rise to a scattering length for the pair interaction of $0.7 a$ with $a$ the fermionic scattering length. This is rather close to the later obtained exact result of $0.6 a$ \cite{Petrov04a}. Below the critical temperature, also the Bogoliubov theory for interacting Cooper pairs was studied \cite{Romans05a, Hu06a, Diener08a}. Other strong-coupling approaches that go beyond the NSR theory include so-called self-consistent ladder approximations  \cite{Haussm94a, Haussm07a} and Monte-Carlo calculations \cite{Carlso03a,Astrak04a,Burovs06a, Bulgac08b, Burovs08a}.

\begin{figure}
\begin{center}
\includegraphics[width=0.7\columnwidth]{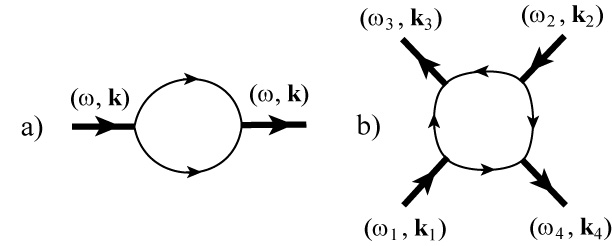}
\caption{\label{fig:fdcp} Diagrammatic representation of a) the bare Cooper-pair propagator and b) the bare Cooper-pair interaction. The Cooper pairs are represented by thick lines, while the thin lines correspond to fermionic propagators. }
\end{center}
\end{figure}

In this section, we apply the renormalization-group techniques developed in the previous section to study non-condensed interacting Cooper pairs in the unitarity limit. To this end, we have to generalize the Wilsonian RG theory for point-like bosons to the more complicated case of Cooper pairs \cite{Gubbel11a}. The inverse propagator for the non-condensed Cooper pairs $G_{\Delta}^{-1}$ that follows from the quadratic part in the pairing field of Eq.~(\ref{eq:acbos}) is given by Eq.~(\ref{eq:cpgf}). Here, we note that $G_{\Delta}^{-1}(i\omega_n, k)$ only depends on the magnitude of ${\bf k}$. The Feynman diagram that corresponds to the Cooper-pair propagator is shown in Fig.~\ref{fig:fdcp}(a). We call $G_{\Delta}^{-1}$ the bare or microscopic propagator, indicating that no Cooper-pair interaction effects have been taken into account yet. Note that the bare propagator is exact, in the sense that it follows from an exact transformation of the fermionic action.  With the RG approach we can consequently systematically include Cooper-pair interaction effects that lead to self-energy corrections to the bare propagator.

The Cooper-pair interaction $V_{\Delta}$ follows from the quartic part in the pairing field of Eq.~(\ref{eq:acbos}) and is diagrammatically represented in Fig.~\ref{fig:fdcp}(b) \cite{Pieri00a,Pieri05a}. Here, we do not take the full frequency and momentum dependence of the Cooper-pair interaction vertex into account, but we consider only two external frequencies and momenta to be nonzero, namely either $\omega_1=-\omega_2$ and ${\bf k}_1=-{\bf k}_2$, or $\omega_3=-\omega_4$ and ${\bf k}_3=-{\bf k}_4$, where the labeling is given in Fig.~\ref{fig:fdcp}b. This specific choice corresponds physically to considering only zero center-of-mass frequencies and momenta, which is motivated later on. The resulting expression is given by
\begin{eqnarray}\label{eq:intcp}
 V_{\Delta}(i\omega_n, k) &\equiv& V_{\Delta}G_{V}(i\omega_n, k)  \nonumber\\
 &=&\frac{1}{\hbar^3\beta \mathcal{V}}\sum_{n',{\bf k'}}G_{0,-}(n',{\bf k'}) G_{0,+}(-n',-{\bf k'})\nonumber\\
&&\times G_{0,-}(n'+n,{\bf k'}+{\bf k}) G_{0,+}(-n'-n,-{\bf k'}-{\bf k}),
\end{eqnarray}
where we defined $V_{\Delta}\equiv V_{\Delta}(0,0)$, so that $G_V$ encapsulates the considered (relative) momentum and frequency dependence of the Cooper-pair interaction. The function $G_V$ is thus equal to the second and third line of Eq.\ \ref{eq:intcp} divided by $V_{\Delta}$. The Matsubara sum over odd fermionic frequencies $n'$ in Eq.~(\ref{eq:intcp}) can be performed analytically. Performing the sum for zero external momentum and frequency, we find for example the expression of Eq.~(\ref{eq:betaL}). Indeed, the fourth-order coefficient $\beta_{\rm L}$ in the Landau theory can physically be interpreted as an interaction strength for the Cooper pairs. We call $V_{\Delta}(i\omega_n,k)$ the bare or microscopic interaction, in order to make the distinction with the effective or renormalized Cooper-pair interaction $V'_{\Delta}$, which includes the effect of Cooper-pair fluctuations and that is calculated next during the RG flow.

\begin{figure}
\begin{center}
\includegraphics[width=0.7\columnwidth]{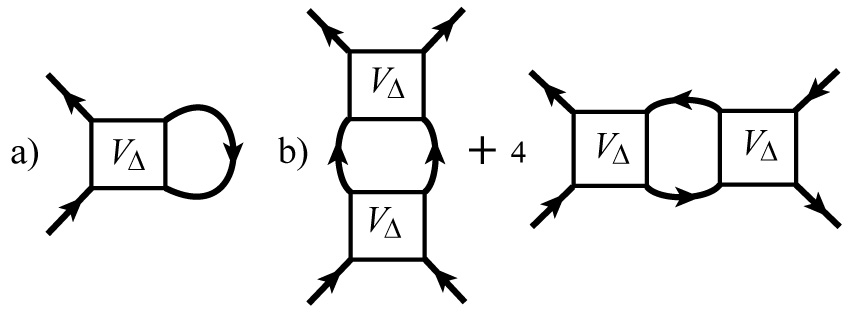}
\caption{\label{fig:rgcp} Diagrammatic representation of the `$\beta$ functions'. a) Feynman diagram determing the self-energy of the Cooper pairs. b) Feynman diagrams renormalizing the Cooper-pair interaction. The middle diagram is also called the ladder diagram, the right diagram is called the bubble diagram. Note that the lines are thick and correspond to the Cooper-pair propagator.}
\end{center}
\end{figure}

Due to the repulsive interaction between the Cooper pairs, they acquire a self-energy $\Sigma_{\Delta}$. In the simplest approximation, the self-energy is momentum- and frequency independent, so that the full renormalized propagator becomes ${G'_{\Delta}(i\omega_n,k)}^{-1}=G_{\Delta}^{-1}(i\omega_n,k)-\Sigma_{\Delta}$. We define the bare Cooper-pair chemical potential as $\mu_{\Delta}\equiv \hbar G^{-1}_{\Delta}(0,0)$, while the renormalized chemical potential is given by $\mu'_{\Delta}=\mu_{\Delta}-\hbar\Sigma_{\Delta}$. The renormalized chemical potential thus includes pair self-energy effects. As a result, the full Cooper-pair propagator $G'_{\Delta}(i\omega_n,k,\mu'_{\Delta})$ depends on the renormalized chemical potential $\mu'_{\Delta}$.
The simplest RG calculation that gives nontrivial results treats both the renormalization of the chemical potential $\mu'_{\Delta}$ and the interaction strength $V'_{\Delta}$. It ignores the three-pair interactions and higher. The flow equations for these two coupling constants can be derived along exactly the same lines as for point-like fermions, as we did in Section \ref{par:wilrg}. The corresponding one-loop Feynman diagrams are diagrammatically represented in Fig.~\ref{fig:rgcp}. Note that a thick line denotes a Cooper pair propagator, whereas the thin lines in Fig.~\ref{fig:fdcp} denote fermion propagators. Since we are now performing a RG study solely for the Cooper pairs, only the momenta of the thick lines are restricted to the considered high-momentum shell. The bare coupling constants for the pair action are derived from the Hubbard-Stratonovich transformation and integrating over all fermionic fields. As a result, the integration over momenta in Fig.~\ref{fig:fdcp} for the thin lines is thus in the present case over all momenta.

When doing the derivation of Section \ref{par:wilrg} also for the Cooper pair RG equations, a few things should be kept in mind. The Cooper pairs do not carry spin as the fermions, so that the Cooper pair interaction is between equal pairs, in contrast to the interaction for fermions between opposite spin. As a result, the Cooper pair interaction parameter is divided by the usual factor of two in the pair action to avoid double counting problems. Another difference with the fermionic action is that the frequency dependence of the Cooper-pair propagator is more complicated. As a result, the Matsubara sums of the one-loop Feynman diagrams cannot be performed analytically anymore, but have to be evaluated numerically within each momentum shell. More detailed accounts for deriving the RG equations for bosons can be found in Refs. \cite{Bijlsm96a,Stoof09a}.

The RG flow of the Cooper-pair interaction strength and chemical potential are found to be determined by the following set of coupled differential equations
\begin{eqnarray}\label{eq:rgcp}
\frac{d\mu'_{\Delta}}{d l} = \beta_{\mu}(l,\mu'_{\Delta},V'_{\Delta}),\quad \frac{dV'_{\Delta}}{d l} = \beta_{V}(l,\mu'_{\Delta},V'_{\Delta})~,~
\end{eqnarray}
where the `$\beta$-functions' are given by \cite{Gubbel11a}
\begin{eqnarray}\label{eq:betamu}
\beta_{\mu} &=& -\dot{\Lambda}_l\frac{\Lambda_l^2V'_{\Delta}}{\pi^2\beta}\sum_n \hbar G'_{\Delta}(i\omega_n,\Lambda_l,\mu'_{\Delta}),\\
\beta_{V} &=& \dot{\Lambda}_l\frac{\Lambda_l^2{V'_{\Delta}}^2}{2\pi^2} \left\{\Xi(\Lambda_l,\mu'_{\Delta})+4\Pi (\Lambda_l,\mu'_{\Delta}) \right\}.~\label{eqbetav}
\end{eqnarray}
Here, $\Pi$ and $\Xi$ are the so-called `bubble' and `ladder' contributions to the effective Cooper-pair interaction, which we also encountered in the RG theory for the fermionic action. Moreover, $\Lambda_l$ denotes the wavevector of the Cooper pairs in the shell of infinitesimal width. This wavevector is parametrized by the flow parameter $l$, and we start the RG flow at the high-momentum cutoff $\hbar\Lambda_0$ and decrease as $\Lambda_l=\Lambda_0 e^{-l}$. In addition, $\dot{\Lambda}_l$ is the derivative of $\Lambda_l$ with respect to $l$. Solving Eq.~(\ref{eq:rgcp}) for increasing $l$ means that we are including the effect of pair fluctuations with lower and lower momenta, while due to the coupling of the differential equations we automatically generate an infinite number of Feynman diagrams, showing the nonperturbative nature of the RG. The initial conditions for  Eq.~(\ref{eq:rgcp}) are $\mu'_{\Delta}(l=0)=\mu_{\Delta}=\hbar G^{-1}_{\Delta}(0, 0)$ and $V'_{\Delta}(l=0)=V_{\Delta}$, which are calculated from Eqs.~(\ref{eq:cpgf}) and (\ref{eq:intcp}).

The one-loop expression for the renormalization of the chemical potential in Eq.~(\ref{eq:betamu}) that determines the self-energy of the Cooper pairs, has a clear physical meaning, since it is seen to be proportional to the renormalized pair interaction strength and to the density of Cooper pairs. The `bubble' diagram $ \Pi(k,\mu_{\Delta}')$ describes the effect of `particle-hole' excitations on the effective Cooper-pair interaction, where these `particles' are now actually Cooper pairs. It is given by
\begin{eqnarray}\label{eq:bubblecp}
\Pi(k,\mu'_{\Delta}) = \frac{\hbar^2}{\beta}\sum_n G'_{\Delta}(i\omega_n,k,\mu'_{\Delta})^2.
\end{eqnarray}
The `ladder diagram' describes the Bose-enhanced scattering of the bosonic Cooper pairs, given by
\begin{eqnarray}\label{eq:ladderrgb}
\Xi(k,\mu'_{\Delta}) = \frac{\hbar^2}{\beta}\sum_n  G_{V}(i\omega_n,k)^2| G'_{\Delta}(i\omega_n,k,\mu'_{\Delta})|^2,
\end{eqnarray}
where the momentum and frequency dependence of the interaction in Eq.~(\ref{eq:intcp}), i.e. $G_V$, is seen to enter. This frequency and momentum dependence is important, since otherwise Eq.~(\ref{eq:ladderrgb}) ultimately would lead to an ultraviolet divergence. The divergence physically arises from approximating the pair interaction as a point interaction, which is therefore insufficient. We also note that the self-energy diagram of Fig.~\ref{fig:rgcp}(a) and Eq.~(\ref{eq:betamu}), and the bubble diagram of Fig.~\ref{fig:rgcp}(b)  and Eq.~(\ref{eq:bubblecp}) do not lead to divergencies. As a result, our present scheme for including the Cooper-pair interactions is the minimal choice for obtaining divergence-free, or equivalently, cutoff independent results.

\subsection{Results}

In Section \ref{par:nsr} it was explained that the NSR theory of the strongly interacting normal state gives rise to two contributions to the grand-canonical thermodynamic potential, namely a contribution describing an ideal gas of fermions and a contribution describing an ideal gas of noncondensed Cooper pairs \cite{Nozier85a}. The contribution to the grand potential density due to the Cooper pairs is given in our RG approach by the one-loop expression
\begin{eqnarray}\label{eq:tpcp}
\frac{d\omega_{\Delta}}{d l} = -\dot{\Lambda}_l\frac{\Lambda_l^2}{2\pi^2\beta}\sum_n \log[-{G'_{\Delta}}^{-1}(i\omega_n,\Lambda_l,\mu'_{\Delta})] ,
\end{eqnarray}
where the first minus sign on the right-hand side is only present when $\Lambda_l$ is a decreasing function. Note that this last expression reduces to the differential form of the NSR contribution to the grand potential density in Eq. (\ref{eq:tpnsr}) when the Cooper-pair chemical potential is not renormalized ($\mu'_{\Delta}\equiv\mu_{\Delta}$), i.e., when we consider the Cooper pairs to be noninteracting. If the exact Cooper-pair propagator is inserted in Eq.~(\ref{eq:tpcp}), then the exact grand potential density is obtained. However, this would require the treatment of all $n$-body interactions, which is presently out of reach.

To be able to evaluate Eqs.~(\ref{eq:rgcp}) and (\ref{eq:tpcp}) numerically, it is convenient to perform contour integration. The Green's function of the Cooper pairs from Eq.~(\ref{eq:cpgf}) can be written in the spectral form \cite{Perali02a}
\begin{equation}\label{eq:spec}
G'_{\Delta}(i\omega_n,k,\mu'_{\Delta})=\frac{1}{\pi}\int d\omega \frac{\ {\rm Im}[G'_{\Delta}(\omega^{+},k,\mu'_{\Delta})]}{\omega-i\omega_n},
\end{equation}
where $\omega^{+} = \omega + i \eta$ with $\eta \downarrow 0$. The imaginary part of the Green's function can be obtained analytically and was given in Eq. (\ref{eq:imcpgf}). With the spectral representation, we can rewrite Matsubara sums over the pair Green's function as frequency integrals that are convenient for numerical evaluation. For example, we have
\begin{eqnarray}\label{eq:gfint}
&&\frac{1}{\hbar\beta}\sum_n G'_{\Delta}(i\omega_n,k,\mu'_{\Delta})=\frac{1}{\pi}\int d\omega\, b(\hbar\omega) {\rm Im}[G'_{\Delta}(\omega^{+},k,\mu'_{\Delta})],
\end{eqnarray}
where $n$ is even and $b(\epsilon)=1/(e^{\beta \epsilon}-1)$ is the bosonic distribution function. Moreover, the pair bubble diagram from Eq.~(\ref{eq:bubblecp}) becomes
\begin{eqnarray}\label{eq:bubint}
&&  \Pi(k,\mu_{\Delta}) =\frac{1}{\hbar\beta}\sum_n   G'_{\Delta}(i\omega_n,k,\mu'_{\Delta})^2=\\
&& \frac{2}{\pi}\int d\omega ~ b(\omega){\rm Im}[G'_{\Delta}(\omega^+,k,\mu'_{\Delta})]{\rm Re}[G'_{\Delta}(\omega^+,k,\mu'_{\Delta})],  \nonumber
\end{eqnarray}
where we used Eq.~(\ref{eq:spec}) and the Kramers-Kronig relation to relate the real and imaginary part of the Cooper-pair Green's function. For the grand potential density from Eq.~(\ref{eq:tpcp}), the frequency integral form is given in Eq.~(\ref{eq:tpnsr}).

\begin{figure}
\begin{center}
\includegraphics[width=0.8\columnwidth]{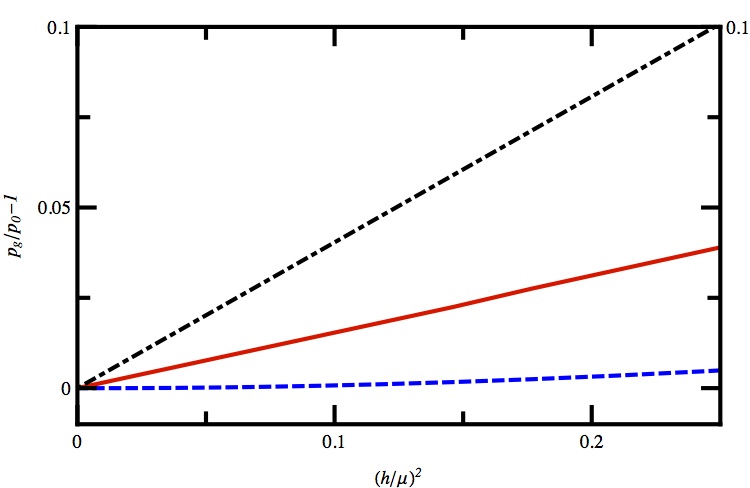}
\caption{\label{fig:eoscp} Equation of state for the normal phase of a strongly interacting imbalanced Fermi gas at unitarity. The pressure $p_{\rm g}=-\omega_{\rm gc}(T,\mu,h)$ of the gas is calculated at temperature $T =0.75\mu$ as a function of $h/\mu=(\mu_+ - \mu_-)/(\mu_+ + \mu_-)$ with the renormalization group approach (full line), the Nozi\`{e}res-Schmitt-Rink approach (dashed line), and for the ideal Fermi gas (dashed-dotted line). For each curve, the pressure of the imbalanced gas is normalized to the corresponding pressure of the balanced gas $p_{0} = -\omega_{\rm gc}(T,\mu,0)$. We note that $p_0$ is different for the three different methods, as can be seen from Fig.\ \ref{fig:press}. }
\end{center}
\end{figure}

After having performed the RG calculations, we have that the total thermodynamic potential density is given by
\begin{equation}
\omega_{\rm gc}(T,\mu_{\sigma})= \frac{\Omega (T,\mu_{\sigma})}{\mathcal{V}}=\omega_{\rm ig}(T,\mu_{\sigma})+\omega_{\Delta, \infty}(T,\mu_{\sigma}),
\end{equation}
with $\Omega$ the grand potential, $\mathcal{V}$ the volume, and $\omega_{\Delta, \infty}(T,\mu_{\sigma})= \omega_{\Delta}(l\rightarrow \infty)$. From $\omega_{\rm gc}$, all other thermodynamic quantities of interest can be obtained by the standard thermodynamic relations. For the balanced case, the results of the renormalization procedure are shown in Fig.\ \ref{fig:press}. Here, it is seen that due to the inclusion of the repulsive Cooper-pair interactions the equation of state shifts slightly down towards the noninteracting result. However, the effect is not strong and the agreement with the experiment is still reasonable. We note that a more elaborate treatment of fermionic self-energy effects on the Cooper pair propagator would increase the pressure again \cite{Haussm07a,Hu10a}. The results of the present procedure for the imbalanced case are shown in Fig.~\ref{fig:press}.
Here, we have calculated the equation of state for the imbalanced Fermi gas. In Fig.~\ref{fig:eoscp}, we show the pressure $p_{\rm g}$ as a function of $h/\mu$ for the temperature $T/\mu = 0.75$, where $h=(\mu_+-\mu_-)/2$. At this temperature, the NSR approximation predicts a negative polarization for positive $h$, which is an unphysical result. The present RG approach that treats the Cooper pair interaction effects beyond NSR theory does not have this problem. We see that, as a result, the present RG theory, the NSR theory and the mean-field theory give very different results for the pressure of the imbalanced Fermi gas in Fig.~\ref{fig:eoscp}. This pressure has recently been measured at zero temperature, where good agreement with Monte-Carlo calculations was obtained \cite{Nascim11a}.

Finally, we discuss the effect of the Cooper-pair interactions on the critical temperature for the balanced Fermi gas. In the unitarity regime, the effective two-pair interaction is repulsive in the normal state. Taking into account the repulsive two-pair interaction makes the ratios $k_{\rm B}T_{\rm c}/\mu$ and $\epsilon_{\rm F}/\mu$ smaller, because the repulsion lowers the effective chemical potential of the non-condensed Cooper pairs. As a result, the density of non-condensed Cooper pairs becomes lower, which decreases the total density and the critical temperature for condensation. More precisely, when $G^{-1}_{\Delta} (0,0) = \mu_{\Delta} =0$, then both mean-field theory and NSR theory predict a transition to the superfluid state, i.e., the condensation of Cooper pairs at $k_{\rm B}T^{\rm c}_{\rm mf} =0.66 \mu $. Upon lowering the temperature from $T^{\rm c}_{\rm mf}$, $\mu_{\Delta}$ becomes positive, but due to the repulsive interactions the renormalized chemical potential goes down, and at the end of the flow $\mu'_{\Delta, \infty}$ might still be negative. Therefore, the critical condition in the presence of Cooper-pair interactions becomes $\mu'_{\Delta, \infty}= 0$. With our present approach, this results in $k_{\rm B}T^{\rm c}_{\rm rg} =0.43 \mu $, which is much closer to the most recently measured value of  $k_{\rm B}T^{\rm c}_{\rm exp} =0.40 \mu $ \cite{Ku12a}. Note that this result could be further improved by performing a RG calculation for the superfluid state. Namely, close to $T^{\rm c}$ we then initially have $\mu'_{\Delta} (l) > 0$, and the RG flow starts out in a superfluid state. Due to the repulsive interactions, the effective chemical potential is then again lowered, and the system can flow into the normal state \cite{Bijlsm96a}.

 Having discussed in this section in more detail the effect of interactions on strongly-interacting Fermi mixtures for the homogeneous case, we look next more accurately at the effects of the trapping potential, which we have so far only treated in the local density approximation.

\section{Inhomogeneous Fermi gases}
\label{sec:inh}

Ultracold atom experiments are always performed in a trap to avoid contact of the atoms with
material walls that would heat up the cloud. Due to this
trapping potential the atomic cloud is never truly homogeneous.
However, typically the trapping frequency corresponds to a
small energy scale in the system, so that the inhomogeneity is not very
prominent. In this case, we may use the so-called local-density
approximation (LDA), which was discussed in Section \ref{par:lda}. It physically implies that the gas is
considered to be locally homogeneous everywhere in the trap.
The density profile of the gas is then fully determined by the
condition of chemical equilibrium, which causes the shape of the
cloud to follow an equipotential surface of the trap.

But even if the trap frequency is small, LDA may still break
down in a few specific cases. A first example is near the edge of the cloud, where the local-density approximation predicts at zero temperature that the particle density goes strictly to zero beyond the radius where the chemical potential equals the potential energy in the trap. This is not fully correct, because quantum-mechanically there is an exponentially  suppressed probability that the particles reside in the (semi-)classically forbidden areas. Another example occurs when an interface is present
in the trap due to a first-order phase transition. For a
resonantly interacting Fermi mixture with a population
imbalance in its two spin states \cite{Zwierl06a,Partri06a}, such interfaces were
encountered in the experiments by Partridge {\it et al}.~\cite{Partri06a} and by Shin {\it et al}.~\cite{Shin08a} at sufficiently low temperatures. Here, the application of LDA leads to a discontinuity in the density profiles of the two spin states, which would cost an infinite amount
of energy when gradient terms are taken into account.
Experimental profiles are therefore never truly discontinuous,
but are always smeared out.

The presence of a superfluid-normal interface also can have further
consequences. Namely, in a very elongated trap Partridge {\it
et al}. observed a strong deformation of the minority cloud at
their lowest temperatures. At higher temperatures the shape of
the atomic clouds still followed the equipotential surfaces of
the trap \cite{Partri06b}. A possible interpretation of
these results is that for temperatures sufficiently far below
the tricritical point \cite{Parish07b,Gubbel06a},
the gas shows a phase separation between a balanced superfluid
in the center of the trap and a fully polarized normal shell
around this core. Because the trap shape is ellipsoidal, the superfluid core could consequently be deformed from this shape due to the surface tension of the interface
between the two phases, which prefers shapes with the least possible area \cite{Haque07a}. This deformation then
causes an even more dramatic break down of LDA. We note here that to understand the large surface tension of the Rice experiment that causes rather extreme deformations in their case, it is probably needed to take non-equilibrium effects into account \cite{Parish09a, Liao11a}.

In order to further investigate the inhomogeneity of trapped Fermi gases, we thus need to go beyond the local density approximation. There are several approaches to incorporate the inhomogeneity
of the system beyond LDA, of which we discuss two frequently used methods. First, we discuss an approach based on Landau-Ginzburg (LG) theory, which includes a gradient energy penalty for a spatially varying order parameter. We try to incorporate the knowledge from normal and superfluid Monte-Carlo equations-of-state into the LG functional and calculate the surface tension of the superfluid-normal interface in the unitarity limit. We also compare the LG approach with the interface observed at the MIT experiment \cite{Shin08a}. Next, we discuss the Bogoliubov-de Gennes (BdG) equations, which take the single-particle states of the trapping potential into account. As a result, the BdG approach gives the exact result for the noninteracting trapped Fermi gas, so that also the edge of the noninteracting cloud is correctly described. The interactions between the particles in the BdG approach are treated on the mean-field level, so that the results from these equations can be trusted at best qualitatively. We will see that the BdG equations give rise to oscillations in the superfluid order parameter and the particle densities near the superfluid-normal interface. These oscillations are interpreted in terms of the so-called proximity effect.

\subsection{Landau-Ginzburg approach}

When using Landau-Ginzburg theory, we write the grand-canonical thermodynamic potential as an expansion in the order parameter, where usually only the first few local terms and gradient terms are kept, as was for example done in Eq. (\ref{eq:omlan}). In most cases, it is not possible to find exact expressions for the coefficients in the expansion, so either mean-field theory, or other approximations have to be invoked in order to determine these parameters. As we discussed in Section \ref{sec:mf}, the BCS mean-field theory seems to give a satisfactory description of the qualitative phase diagram for the imbalanced Fermi gas, as determined by recent experiments \cite{Shin08a}.  We therefore use the BCS theory as a starting point for our discussion of the inhomogeneous Fermi gas. In Section \ref{par:pdtrap}, we discussed the phase diagram for the trapped Fermi gas, where we used the local-density approximation to account for the trapping potential. This means that we applied the homogeneous theory with spatially varying chemical potentials, $\mu_{\sigma}({\bf r })=\mu_{\sigma}-V^{\rm ex}({\bf
r})=\mu_{\sigma}-m\bar{\omega}^2r^2/2$. Due to the fact that a first-order transition could occur as a function of the chemical potentials, the LDA approximation predicted a jump in the superfluid order parameter and the corresponding particle densities.

As just mentioned, a true jump in the order parameter is unphysical, and the easiest way to go beyond LDA for a more realistic description of the superfluid-normal interface is to take also into account the gradient energy of the order parameter in the grand potential functional by
\begin{equation}\label{eq:inhtp}
\Omega[\Delta;\mu,h] = \int d {\bf r}~\left\{\gamma|\nabla\Delta({\bf r})|^2 + \omega_{\rm gc}[\Delta({\bf r});\mu({\bf r}),h]\right\}.
\end{equation}
with  $\omega_{\rm gc}[\Delta;\mu,h]$ the local grand potential density. We encountered the coefficient $\gamma$ already in Section \ref{par:mi}, where $\gamma$ was interpreted as being inversely proportional to the effective Cooper pair mass. Note that in Eq.\ (\ref{eq:inhtp}) we neglect the dependence of $\gamma$ on $\Delta$  and we also ignore all terms that are of higher order in the gradients, resulting in the simplest extension beyond the local-density approximation. In Section \ref{par:mi}, we showed how to calculate $\gamma$ from the normal state using the Cooper-pair propagator, diagrammatically represented in Fig.\ \ref{fig:fdcp}. A second way to compute a value for $\gamma$ is to use the fact that this
coefficient can be related to the superfluid density $\rho_{\text{s}}$. Namely, it is a standard result from Landau-Ginzburg theory that $\gamma =\hbar^2\rho_{\text{s}}/8m^2|\langle\Delta\rangle|^2$ \cite{Legget06a}. Using the knowledge from the superfluid Monte-Carlo equation of state, we get the zero-temperature result
\begin{equation}\label{eq:gammamc}
\gamma_{\rm MC}=\frac{\sqrt{m}}{6\pi^2\hbar \zeta_{\rm MC}^2 (1+\beta_{\rm MC})^{3/2} \sqrt{2\mu}},
\end{equation}
where we used that at zero temperature the superfluid density is equal to the total density. We have that $\beta_{\rm MC}= -0.58$ \cite{Carlso03a,Astrak04a,Carlso05a}, which was already encountered in Section \ref{par:denprof}, and that $\zeta_{\rm MC}=1.07$, which gives the Monte-Carlo value \cite{Carlso08a} for the universal relation between the pairing gap and the chemical potential for the unitary Fermi gas, $\langle\Delta\rangle=\zeta \mu$. Note that by using BCS theory, we would have obtained $\gamma_{\rm BCS}=\sqrt{m}/6\pi^2\hbar\zeta_{\rm BCS}^2(1+\beta_{\rm BCS})^{3/2}\sqrt{2\mu}$, where $\beta_{\rm BCS} = -0.41$ and $\zeta_{\rm BCS}=1.16$ are the mean-field results for the same universal constants.

\subsubsection{Surface tension}

Since we are studying the superfluid-normal
interface beyond the LDA, we can
determine the surface tension of the calculated interface \cite{Silva06a}. The surface tension is the work per unit area that has to be done to increase the area of the interface. Since the LDA assigns no energy to the interface, we can calculate the surface tension from the difference in the grand potential between
the LDA result with a discontinuous step in
$\langle\Delta\rangle({\bf r})$ and the Landau-Ginzburg result with a
smooth profile for the order parameter $\langle\Delta\rangle({\bf r})$.
We will calculate the surface tension by considering a flat interface in a homogeneous box, which has an interface around $z=0$ with a normal phase at the bottom of the box ($z<0$) and a superfluid phase on top ($z>0$). We do not consider any gravitational effects on the gas. The superfluid-normal interface occurs when the imbalance is critical, i.e., when $h=h_{\rm c}(\mu)=\kappa\mu$, where in Section \ref{par:hompd} we found $\kappa_{\rm BCS} = 0.81$ using mean-field theory, while in Section \ref{par:denprof} we found $\kappa_{\rm MC} = 0.93$ using the results from the Monte-Carlo equation of state \cite{Lobo06a}. Since at the critical imbalance the grand potential of the
normal state minimum is exactly equal to the grand
potential at the superfluid state minimum, we have that the surface tension
is given by the difference between a
system that stays in one minimum and a system that goes near the
interface smoothly from one minimum to the other.

To determine the resulting shape of the interface in the box, we use the grand potential
\begin{equation}\label{eq:inhtp2}
\Omega[\Delta;\mu,h_{\rm c}] = \mathcal{A}\int d z~\left\{\gamma\left(\frac{d\Delta(z)}{d z}\right)^2 + \omega_{\rm gc}[\Delta(z);\mu,h_{\rm c}]\right\}.
\end{equation}
where we condider a real-valued order-parameter profile $\Delta(z)$, and $\mathcal{A}$ is the area of the box at $z=0$. Extremizing this functional results in the Euler-Lagrange equation for the interface
\begin{eqnarray}\label{eq:eomgap}
 \frac{\partial\omega_{\rm gc}[\Delta;\mu,h_{\rm c}]}{\partial\Delta}
     - 2\gamma \frac{d^2 \Delta}{d z^2} =
     0\;.
\end{eqnarray}
which can be readily
solved numerically. This results in a smooth monotonic function
$\langle\Delta\rangle(z)$ that on the normal side of the interface
approaches zero and on the superfluid side approaches the
equilibrium value of the superfluid minimum $\langle \Delta \rangle$. Inserting this profile in Eq. (\ref{eq:inhtp}) and computing the difference with $\Omega[0;\mu,h_{\rm c}]$ then results after division by the area in the surface tension.

There is also a more elegant way to compute the surface tension $\sigma$, which does not explicitly require the shape of the interface \cite{Dieder12a}. Namely, we have that
\begin{eqnarray}
    \sigma = \int_{-\infty}^{\infty} d z \;
     \left(\omega_{\rm gc}[\langle \Delta \rangle(z);\mu,h_{\rm c}] - \omega_{\rm gc}[0;\mu,h_{\rm c}]\right),
\end{eqnarray}
which can be rewritten as an integral over a new integration variable $\Delta'$,
using that $\langle \Delta \rangle(z)$ is a monotonically increasing
function between zero and $\langle \Delta \rangle$. Doing so, we obtain
\begin{eqnarray}\label{eq:tension}
    \sigma = \int_0^{\langle \Delta \rangle} d
    \Delta' \frac{dz}{d
    \Delta'} \left(\omega_{\rm gc}[\Delta';\mu,h_{\rm c}] - \omega_{\rm gc}[0;\mu,h_{\rm c}]\right),
\end{eqnarray}
which still requires the determination of $dz/d\Delta'$. This can be done by considering
\begin{eqnarray}
  \frac{d}{dz}\left(\frac{d \Delta}{d z}\right)^2=2  \frac{d \Delta}{dz}\frac{d^2 \Delta}{d z^2}=\frac{1}{\gamma}\frac{d \Delta}{dz} \frac{\partial\omega_{\rm gc}}{\partial\Delta},
\end{eqnarray}
where we used Eq.\ (\ref{eq:eomgap}). After integration over $z$ this gives
\begin{eqnarray}\label{eq:dergap}
 \left(\frac{d \Delta}{d z}\right)^2=\frac{1}{\gamma} \left(\omega_{\rm gc}[\Delta(z);\mu,h_{\rm c}] - \omega_{\rm gc}[0;\mu,h_{\rm c}]\right),
\end{eqnarray}
where we used that $(d \Delta/d z)^2=0$ at $z=-\infty$. Upon taking the square root of Eq.\ (\ref{eq:dergap}), inverting the result and insertion in Eq.\ (\ref{eq:tension}), we obtain the final result for the surface tension
\begin{eqnarray}\label{eq:tension2}
    \sigma = \int_0^{\langle\Delta\rangle} d
    \Delta' \sqrt{\gamma\left(\omega_{\rm gc}[\Delta';\mu,h_{\rm c}] - \omega_{\rm gc}[0;\mu,h_{\rm c}]\right)}\;,
\end{eqnarray}
which does not explicitly depend on the solution $\langle \Delta \rangle(z)$ anymore, but only on $\gamma$ and the shape of the barrier in the grand potential density $\omega_{\rm gc}$.
Writing the surface tension as $\sigma = \eta \mu^2 m/\hbar^2$, we have that the dimensionless number
$\eta$ only depends on the temperature. So in order to calculate the surface tension we need an expression for the grand potential as a function of the order parameter $\Delta$. Using the BCS mean-field thermodynamic potential from Eq.\ (\ref{eq:tpdBCS}), we find that $\eta_{\rm BCS}=8.6 \times 10^{-3}$.

\subsubsection{Monte-Carlo improved grand potential}

In Section \ref{sec:diag}, we discussed the results of Monte-Carlo calculations for the superfluid equation of state and the normal equation of state in the unitarity limit. There we saw that the BCS mean-field theory gives in particular a rather poor description of the strongly-interacting normal state. In Section \ref{par:polaron}, we also proposed a simple way in which the Monte-Carlo equation of state for the normal state could be included in the grand potential by using the Ansatz for the `renormalized' chemical potentials in Eq.\ (\ref{eq:mup}). Here, we apply a similar procedure to the superfluid state. The first step is to insert the normal state renormalized chemical potentials in the BCS grand potential density, giving $\omega_{\rm imp}(\Delta;\mu_{\sigma}) =\ \omega_{\rm BCS}(\Delta;\mu_{\sigma}'(\mu_{\sigma}))$, so that the strongly-interacting normal equation-of-state is automatically included when $\Delta =0 $.

\begin{figure}[t]
\begin{center}
\includegraphics[width=0.7\columnwidth]{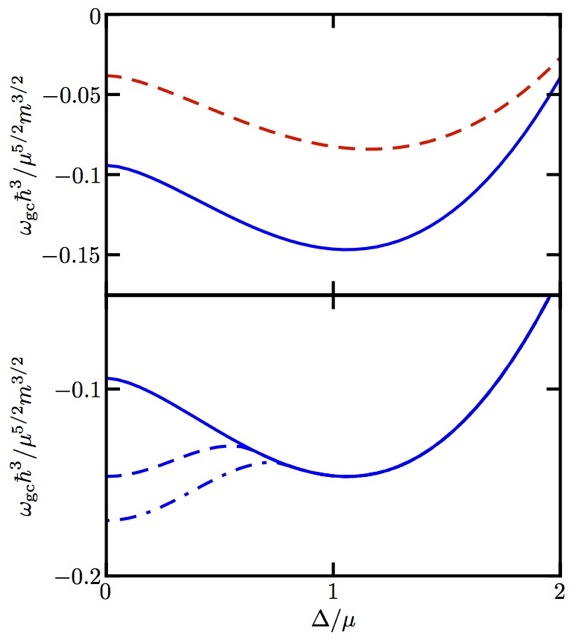}
\caption{\label{fig:inhtp} Upper panel: thermodynamic potential density $\omega$ for the unitary Fermi gas as a function of the superfluid order parameter $\Delta$. The dashed line gives the result for BCS mean-field theory, while the full line effectively includes the superfluid and normal equations of state from Monte-Carlo calculations. Lower panel: the 'Monte-Carlo' improved thermodynamic potential density for the balanced case, $h=0$ (full line),  for the critical chemical potential difference, $h_{\rm c}=0.99\mu$ (dashed line), and for $h=1.2 h_{\rm c}$ (dash-dotted line).}
\end{center}
\end{figure}

Although the Monte-Carlo results for the normal state are now included, the grand potential needs to be further modified in order to also give agreement with the Monte-Carlo superfluid equation of state \cite{Carlso05a}, for which we have at zero temperature that $\mu= 0.42 \epsilon_{\rm F}$ and $\langle\Delta\rangle = 0.45 \epsilon_{\rm F} = 1.07 \mu$ \cite{Carlso08a}. One issue with the construction so far is that $\partial \omega_{\rm imp}(\langle\Delta\rangle;\mu,h)/\partial h$ is nonzero in the superfluid state, because $(\partial \omega_{\rm BCS}(\langle\Delta\rangle;\mu',h')/\partial \mu')(\partial \mu' / \partial h) $ is nonzero. The reason for this behavior is that the self-energy corrections captured by $\mu'$ depend on $h$, which is correct for the normal state, but not for the superfluid state, where the experiments of Section \ref{par:denprof} show an equal-density superfluid core in the unitarity limit, so that the grand potential should become independent of $h$. This behavior is correctly captured by the BCS potential itself without the `renormalization' of the chemical potentials, which indeed becomes independent of $h$ for $\Delta > h$. From this discussion, we conclude that $\mu'$ should become independent of $h$ as a function of $\Delta$. Another issue is that the parameter $A$ in Eq.\ (\ref{eq:mup}), which describes the interaction effects in the normal state, is too big for the superfluid state, since in the latter case the BCS mean-field theory already takes into account a large part of the interaction effects through the presence of the Cooper-pair condensate $\Delta$. We thus expect an improved result if we let the self-energy parameter $A$ decrease as a function of $|\Delta|^2/\mu^2$. Assuming that there is no $h$ dependence of $\mu'$ in the superfluid state, we find that by using $A'=A - B |\Delta|^2/\mu^2$ we simultaneously satisfy $\mu= 0.41 \epsilon_{\rm F}$ and $\langle\Delta \rangle = 1.06 \mu$, if we take $B=0.135$, while $A=1.01$ is the earlier obtained value for the normal state.

However, to have a complete grand potential we still must specify how $\mu'$ evolves as a function of $\Delta$ in order to lose its dependence on $h$ for $\Delta > h'$. To ensure a smooth interpolation between the normal state and the superfluid state, we use the switching function
\begin{equation}
f_{s}(\Delta,h')=\left( \frac{1}{2}+\frac{1}{2}\cos\left(\pi \frac{|\Delta|}{h'}\right)\right)\theta(h'-|\Delta|)
\end{equation}
which is a function of $|\Delta|^2$ that equals 1 for $\Delta=0$ and which goes to zero with vanishing derivative at $|\Delta|=h'$, the point where the modified BCS thermodynamic potential loses its dependence on $h'$. The switching function stays zero at larger values of $|\Delta|$ due to the presence of the stepfunction $\theta$. If we include the switching function in the expression for $\mu'$ as a function of $\mu$ and $h$, obtained from Eq. (\ref{eq:mup}), we get
\begin{eqnarray}\label{eq:mupsf}
\mu'&=& \frac{5\mu}{10-3A'}\left(1+\sqrt{1+\frac{3(10-3A') A'}{(5+3 A')^2}\left(\frac{h}{\mu}\right)^2 f_{s}\left(\Delta,h'\right)}\right),\\
h'&=&\frac{5 h}{5+3A'},\label{eq:hpsf}
\end{eqnarray}
where we have also replaced $A$ with $A'$. The latter is a function of $|\Delta|/\mu$ in the way mentioned above. The expressions in Eqs. (\ref{eq:mupsf}) and (\ref{eq:hpsf}) are valid for $\mu>h(1-3A/5)/(1+3A/5)$, while for smaller values of $\mu$ the system is in a fully polarized phase described by an ideal gas of majority particles. When $\Delta=0$, we have $A'=A$ and $f_s = 1$, so that then Eqs.\ (\ref{eq:mupsf}) and (\ref{eq:hpsf}) contain exactly the same information as Eq.\ (\ref{eq:mup}). Note that the chosen switching function is arbitrary in the sense that it is solely motivated by its smooth interpolation properties between 0 and 1. However, by including it we have obtained a grand potential that not only has the Monte-Carlo equation-of-state for the normal state and the equilibrium superfluid state incorporated, but also smoothly interpolates between these two regimes. We will use this zero-temperature `Monte-Carlo improved' grand potential density $\omega_{\rm imp}$ to compare with the results from BCS mean-field theory and experiments. It is plotted in Fig.\ \ref{fig:inhtp}, where in the upper panel $\omega_{\rm imp}$ and $\omega_{\rm BCS}$ are compared. The `Monte-Carlo improved' potential is seen to give rise to a much larger pressure of the gas $p_{\rm g}$ ($p_{\rm g}=-\omega_{\rm gc}$), which is due to the effective inclusion of the self-energy effects. In the lower panel, we show $\omega_{\rm imp}$ for different values of $h$, showing that at zero temperature the critical chemical potential difference is determined by $h_c=0.99 \mu$, which is very close to the Monte-Carlo result of $h_{\rm c}=0.96 \mu$ \cite{Lobo06a}.

\begin{figure}
\begin{center}
\includegraphics[width=0.7\columnwidth]{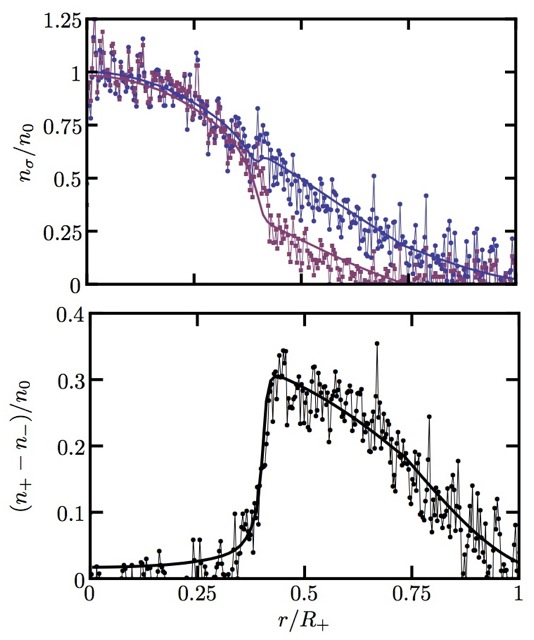}
\caption{\label{fig:lgprof} The density profiles of a unitary Fermi mixture in a harmonic trap with a spin polarization of $P=0.44$. To realize this polarization, the chemical potentials $\mu_+ = 81 \hbar\bar{\omega}$, $\mu_- = 19 \hbar\bar{\omega}$, and temperature $k_{\rm B}T=10 \hbar\bar{\omega}$ were used with $\bar{\omega}$ the (effective) trapping frequency. The upper figure shows the majority (upper blue curve) and minority (lower red curve) densities as a
function of the position in the trap. The lower figure shows the density difference. The horizontal axes are scaled by $R_+ =\sqrt{2\mu_+/m\bar{\omega}^2}$, while the vertical axis is scaled by the central spin-up density $n_0 = n_+({\bf 0})$. The experimental data (dots) are from Shin {\it et al}. \cite{Shin08a}.
}
\end{center}
\end{figure}

\subsubsection{Results}

We calculate the surface tension with the use of Eq.\ (\ref{eq:tension2}), where we insert for the grand potential density $\omega_{\rm imp}(\Delta;\mu,h)=\omega_{\rm BCS}(\Delta;\mu',h')$ with $\omega_{\rm BCS}$, $\mu'$ and $h'$ given by Eqs.\ (\ref{eq:tpdBCS}), (\ref{eq:mupsf}) and (\ref{eq:hpsf}), respectively. We find that at zero temperature $\eta_{\rm imp}= 1.6 \times 10^{-2}$. This result is to be compared with the value of $\eta_{\rm BCS} = 8.6 \times 10^{-3}$ from mean-field theory, and $\eta_{\rm fit} = 0.6$, which was fitted to the large deformations of the superfluid core observed in the Rice experiment \cite{Haque07a}. The experiment at MIT of Shin {\it et al}. does not show any deformation, putting an upper bound on $\eta$ of about
$0.1$ \cite{Haque07a}. This is in agreement with the above value of $\eta_{\rm imp}$ at zero temperature. For higher temperatures the surface tension would further decrease due to a lowering of the superfluid barrier, and ultimately vanish at the tricritical point. Understanding the large value for the surface tension found in the early Rice experiments is not possible within the present approach, since in that case non-equilibrium effects seem to be important \cite{Liao11a,Parish09a}. The experimental results of Shin {\it et al}. were shown already in Fig.\ \ref{fig:mitprof}, where they were compared with the theoretical results from the Monte-Carlo equations-of-state in combination with the LDA. In Fig.\ \ref{fig:lgprof}, the same experimental data is shown again, but now compared with the present LG approach. For simplicity, we have assumed here a spherically symmetric trapping potential, although the actual experimental trapping potential was elongated in one direction. The trapping frequencies for the MIT experiments were given in Section \ref{par:lda}. As explained in the same section, the elongation can be scaled away by a coordinate transformation as long as the different directions in the trap are not coupled, as was done for example in applying the local-density approximation. Then, the elongation plays no physical role and by transforming the coordinates a spherically symmetric potential is obtained. However, in the Landau-Ginzburg functional of Eq.\ (\ref{eq:inhtp})  the directions in the trap are coupled by the gradient term, so that the mentioned coordinate transformation then actually transfers the effect of the trap anisotropy to the gradient term. It was shown in Ref.\ \cite{Dieder12a} that the effect of this anisotropy is actually small, which also agrees with the experimental results of Shin {\it et al}. that find a gas cloud which follow the equipotential lines of the trap very closely. To get the order-parameter profile for $\langle\Delta\rangle({r})$, we solve the Euler-Lagrange equation for the spherically symmetric case with $\gamma$ determined by Eq.\ (\ref{eq:gammamc}), where we used for $\mu$ the local value in the trap at the interface.

To obtain the density profiles, we use the local thermodynamic relation
$n_{\sigma}(r)=-\partial\omega_{\rm gc}(T,\mu_{\sigma}(r))/\partial\mu_{\sigma}$, by which we mean that we differentiate the used grand potential density $\omega_{\rm gc}=\omega_{\rm imp}$ with respect to the chemical potentials $\mu_{\sigma}$ and evaluate the result at the local value for the chemical potential $\mu_{\sigma}(r)=\mu_{\sigma} - V^{\rm ex}(r)$ with $V^{\rm ex}(r)=m\bar{\omega}^2r^2/2$ and the order parameter $\langle\Delta\rangle(r)$.
This allows for a full comparison
between theory and experiment as shown in Fig.~\ref{fig:lgprof}. The upper panel shows the majority and minority densities as a function of the radial position in the trap, while the lower panel shows the density difference. The agreement is seen to be very good, although the data is somewhat too noisy for an accurate study of the interface. Although not very clearly visible in Fig.~\ref{fig:lgprof}, we find that there is a small kink in the majority density profile. This kink signals the transition
to the gapless Sarma phase. At the transition, the order parameter $\langle\Delta\rangle$ becomes locally smaller than the renormalized chemical potential difference
$h'$ and the unitarity limited attraction is no longer able to
fully overcome the frustration induced by the imbalance. As a
result the gas becomes a polarized gapless Sarma superfluid. We found in Section \ref{par:hompd} that for the homogeneous mass-balanced case at zero temperature, the Sarma phase was unstable. However, because of the
inhomogeneity of the trapped gas, the order parameter is forced to move from its equilibrium value at the center of the trap to zero at the edge of the cloud. Therefore, at the interface we unavoidably encounter locally a stabilized Sarma state, even at zero temperature. Notice that this is a feature of the smooth
behavior of the gap and
thus does not depend on the quantitative details of the grand potential
density functional.

\subsection{Bogoliubov-de Gennes approach}\label{par:bdg}

In the previous section we presented the simplest theory beyond the local-density approximation, namely a Landau-Ginzburg theory that takes into account the gradient energy of the order parameter. Although this resulted in a nonlocal treatment of the order-parameter profile, the underlying fermionic wavefunctions were still treated in a local-density approximation as follows from the use of the homogeneous thermodynamic potential density with a spatially varying chemical potential in Eq.\ (\ref{eq:inhtp}). To overcome this local approximation for the fermions, we now use a different approach to study the inhomogeneous system, namely the Bogoliubov-de Gennes (BdG) equations \cite{DeGenn89a}. The BdG approach has been widely used in the superconductivity literature to study a wide range of inhomogeneous situations, for example involving interfaces between superfluids and metallic or magnetic phases. The method has also been applied by several groups to study the superfluid-normal interface in an imbalanced Fermi gas \cite{Castor05a,Mizush05a,Machid06a,Kinnun06a,Schaey07a,Liu07a}. Some of these studies found that near the interface the order parameter becomes oscillatory, which was suggested to be related to the Fulde-Ferrell or Larkin-Ovchinnikov (FF or LO) phases \cite{Castor05a,Kinnun06a,Machid06a} that were discussed in the introduction. In a follow-up study \cite{Jensen07a} the authors of Ref. \cite{Kinnun06a} showed that the oscillations are not a finite size effect \cite{Liu07a}, but rather caused by the so-called proximity effect \cite{Demler97a,Kontos01a,Buzdin05a,Schaey07a}. In this section, we briefly review the Bogoliubov-de Gennes method, where we start with a derivation, after which we present an efficient numerical scheme to solve the obtained coupled di fferential equations. We apply the method to the phase-separated imbalanced case and indeed find the mentioned oscillations near the interface. Our findings support the conclusion that they are caused by the proximity effect.

\subsubsection{Derivation}

To derive the Bogoliubov-de Gennes equations, we start with the Hubbard-Stratonovich transformed action of Eq.~(\ref{eq:acbosfer}). The action can be written in a $2\times 2$ matrix form, which we call spin or Nambu
space, as we saw explicitly in Section \ref{par:hstrans}. As a result, we have for the quadratic part in the fermionic fields of the action $S^{(2)}$ that
\begin{eqnarray}\label{eq:bdgac}
      S^{(2)}[\phi_{\sigma}] &=&\int_0^{\hbar\beta} d\tau \int d{\bf r}~(\phi^*_{+}(\tau,{\bf r}),\phi_{-}(\tau,{\bf r})) \\
      &\times& \begin{pmatrix}
      \hbar\partial_{\tau} - h + H_0({\bf r}) & \Delta ({\bf r})\\
      \Delta^* ({\bf r})  &\hbar\partial_{\tau} - h -H_0({\bf r}) \\
      \end{pmatrix}
      \begin{pmatrix}
      \phi_{+}(\tau,{\bf r}) \\
      \phi^{*}_{-}(\tau,{\bf r}) \\
      \end{pmatrix}, \nonumber
\end{eqnarray}
where we have made the approximation that the pairing field is independent of (imaginary) time. The diagonal part of the $2\times2$ matrix contains
\begin{equation}
H_0({\bf r}) = -\frac{\hbar^2}{2m}\nabla^2+
V^{\text{ex}}({\bf r})-\mu,
\end{equation}
where the external potential $V^{{\rm ex}}({\bf r})$ is added to describe the trapping potential. We consider a harmonic trapping potential, which approximates the experimental situation very well. Just like in the previous section, we use a spherically symmetric trapping potential
$V^{{\rm ex}}({\bf r})=m\bar{\omega}^2r^2/2$ with $\bar{\omega}$ the (effective) trap frequency.

Next, we perform a Bogoliubov transformation with the goal to diagonalize the position-dependent part of the $2\times2$ matrix in Eq.\ (\ref{eq:bdgac}). To this end, we apply the following unitary transformation \cite{DeGenn89a}
\begin{eqnarray}\label{eq:bdgtrans}
      \begin{pmatrix}
        \phi_{+}(\tau,{\bf r}) \\
        \phi^{*}_{-}(\tau,{\bf r}) \\
      \end{pmatrix}
       = \sum_{\bf n}
      \begin{pmatrix}
        u_{\bf n}({\bf r})&-v^*_{\bf n}({\bf r}) \\
        v_{\bf n}({\bf r})& u^*_{\bf n}({\bf r}) \\
      \end{pmatrix}
      \begin{pmatrix}
        \psi_{+,{\bf n}}(\tau) \\
        \psi^{*}_{-,-{\bf n}}(\tau) \\
      \end{pmatrix},
\end{eqnarray}
where $\pm {\bf n}=(n,\ell,\pm m_\ell)$ denotes the set of three quantum numbers required
to specify the single-particle eigenstates. We see that the unitary transformation indeed diagonalizes the spatial part of Eq.\ (\ref{eq:bdgac}), if the time-independent Bogoliubov-de Gennes equations are satisfied, which are given by
\begin{eqnarray}\label{eq:bdgeq}
      \begin{pmatrix}
        H_{0}(\bf r) & \Delta(\bf r) \\
        \Delta^{*}(\bf r) & - H_{0}(\bf r) \\
      \end{pmatrix}
      \begin{pmatrix}
        u_{\bf n}({\bf r})\\
        v_{\bf n}({\bf r}) \\
      \end{pmatrix}
      =E_{\bf n}
      \begin{pmatrix}
        u_{\bf n}(\bf r)\\
        v_{\bf n}(\bf r) \\
      \end{pmatrix}.
\end{eqnarray}
This is a set of coupled differential equations, for which we still need to specify the boundary conditions and the self-consistency relation for $\Delta(\bf r)$. The coherence factors $u_{\bf n}$ and $v_{\bf n}$ are normalized for each ${\bf n}$, namely
$\int d {\bf r}~(| u_{\bf n}|^2+| v_{\bf n}|^2)=1$. Notice that when $(u_{\bf n},v_{\bf n})$ is a solution to Eq.\ (\ref{eq:bdgeq}) for energy $E_{\bf n}$, then  $(-v_{\bf n}^*,u_{\bf n}^*)$ is a solution with $-E_{\bf n}$.
Since the considered trap is spherically symmetric, the gap $\Delta(r)$ is a function of the radius only. We therefore have that
\begin{eqnarray}
u_{\bf n}(r,\theta,\phi) = \frac{\bar{u}_{n\ell}(r)}{r}Y_{\ell m_\ell}(\theta,\phi),
\end{eqnarray}
where $Y_{\ell m_\ell}$ are the spherical harmonics, and we do the same
for $v$. The sum over ${\bf n}$ is now a sum over the
set $\{n,\ell,m_\ell\}$ with $\ell \ge 0$ and $m_\ell= -\ell, ..., \ell$. In the following, we will consider without loss of generality that $\bar{u}_{n\ell}(r)$, $\bar{v}_{n\ell}(r)$, and $\Delta({\bf r})$ are real-valued functions. As result, we find the following equation for
the functions $\bar{u}_{n\ell}(r)$ and $\bar{v}_{n\ell}(r)$, namely
\begin{align}\label{eq:bdgrad}
      \frac{d^2}{dr^2}
      \begin{pmatrix}
        \bar{u}_{n\ell}(r)\\
        \bar{v}_{n\ell}(r) \\
      \end{pmatrix}
      =-{\bf H}_{n\ell}(r)
      \begin{pmatrix}
        \bar{u}_{n\ell}(r)\\
        \bar{v}_{n\ell}(r) \\
      \end{pmatrix}\;.
\end{align}
The matrix ${\bf H}_{n\ell}$ is given by
\begin{equation}\label{eq:bdgham}
 {\bf H}_{n\ell} (r) = \  \frac{2 m}{\hbar^2}
    \begin{pmatrix}
        \mu-V^{\rm ex}_{\ell}(r)+E_{n\ell}& -\Delta(r)\\
        \Delta(r)  & \mu-V_\ell(r)-E_{n\ell} \\
    \end{pmatrix},
\end{equation}
with  $V^{\rm ex}_\ell(r)$
the effective external potential, i.e.,
\begin{equation}
    V^{\rm ex}_{\ell}(r) = \ m\bar{\omega}^2 \frac{r^2}{2}+\frac{\hbar^2}{2m}\frac{\ell(\ell+1)}{r^2}\;,
\end{equation}
which includes the effect of the centrifugal barrier for nonzero angular momentum $\ell$.
The boundary conditions for $\bar{u}_{n\ell}$ and $\bar{v}_{n\ell}$
are that they should be both zero in the origin and at infinity.

Before we start with the discussion of the numerical procedure to solve the BdG equations, we give the analytic solution for the non-interacting case, i.e., $\Delta(r) = 0$. Then, the differential equations are not coupled, so that they reduce to the 3-dimensional harmonic oscillator problem, whose solution can be found in many textbooks. The properly normalized noninteracting solutions $(u_{\bf n},
v_{\bf n})= (\phi_{n\ell m},0)$ are given by
\begin{eqnarray}
    \phi_{n\ell m}(r,\theta,\phi)&=&\mathcal{N}_{n\ell} e^{-\frac{r^2}{2 l^2}}\left(\frac{r}{l}\right)^\ell L_n^{\ell+\frac{1}{2}}\left(\frac{r^2}{l^2}\right)Y_{\ell m}(\theta,\phi),\\
    E_{n\ell}&=&\hbar\bar{\omega}\left(\frac{3}{2}+2n+\ell\right)-\mu\;,\\
    \mathcal{N}_{n\ell} &=& \sqrt{\frac{2^{n+\ell+2}n!}{{l^3(1+2\ell+2n)!!\sqrt{\pi}}}}\;,
\end{eqnarray}
with $L_n^\ell$ the associated Laguerre
polynomials, $\mathcal{N}_{n\ell}$ the corresponding normalization constants, and
$l=\sqrt{\hbar/m\bar{\omega}}$ the so-called harmonic trap length. Eq.\ (\ref{eq:bdgeq}) also gives rise to the solutions $(u_{\bf n}, v_{\bf n})= (0, \phi_{n\ell m})$ with energy $- E_{n\ell}$.

\subsubsection{Numerical methods}

There are multiple ways to solve the second-order matrix differential equation of Eq.\ (\ref{eq:bdgrad}). For example, it is possible to use the (renormalized) Numerov method, which is a well-established numerical approach for treating scattering and bound-state problems. In the (renormalized) Numerov method, the bound states in a certain potential are calculated one at a time. To this end, we start with a trial energy and for this energy we try to construct a smooth wavefunction that matches the boundary condition. If we fail, we need to have a criterion to improve our guess for the energy, so that a few times later we will succeed. We also need to be able to test if we have calculated all possible states, which is usually done by counting for each state the number of times that the wavefunction goes through zero. However, it turns out that for large values of the order parameter, the functions $\bar{u}_{n\ell}$ and $\bar{v}_{n\ell}$ can have very different shapes from the noninteracting solutions, so that counting zeroes leads to problems and states can be missed.

This problem is absent if we use a different approach. Namely, if we evaluate the matrix ${\bf H}_{n\ell}$ of Eq.\ (\ref{eq:bdgrad}) in a certain basis, then we can calculate all eigenstates and eigenenergies at once with matrix diagonalization. In Ref.\ \cite{Kinnun06a} the harmonic oscillator basis was used, which analytically solves the diagonal part of Eq.\ (\ref{eq:bdgeq}), while for the off-diagonal part integrals of the form $ \langle\phi_{n\ell m}|\Delta(r)|\phi_{n \ell m} \rangle$ have to be calculated. We use another method that is based on the discrete variable representation (DVR). Using a DVR based on sinc-functions (sinc-DVR) \cite{Groene:93}, we have that the matrix ${\bf H}_{n\ell}$ is expressed in a basis that results in simple analytic expressions both for the kinetic energy operator and for position-dependent functions.

We start with discussing the sinc-DVR method for the one-dimensional case, after which we generalize the method to the three-dimensional spherically symmetric case. More details can be found in Ref.\ \cite{Groene:93}. The basis set that is used is given by the following set of orthonormal functions
\begin{equation}\label{eq:dvrbas}
\chi_j (x) = \frac{1}{\sqrt{\varDelta x}} {\rm sinc}\left[\pi\left(\frac{x}{\varDelta x}-j\right)\right],
\end{equation}
where ${\rm sinc}(x)=\sin(x)/x$ and $j =-M , ..,M $ with $2 M + 1$ the number of basis functions. These basis functions are to be evaluated on an equidistant grid $x_i=i \varDelta x$ with  $i = -M , ..,M $ and grid spacing $\varDelta x$. Note that $\chi_j (x_j) =1/\sqrt{\varDelta x}$, while for all other grid points $\chi_j(x_i)=0$. For the sinc-DVR basis, we have that
\begin{eqnarray}
V_{ij} = \langle\chi_i|V|\chi_j\rangle \delta_{ij},
\end{eqnarray}
for an arbitrary potential $V(x)$, where the implied integral is to be evaluated using a quadrature rule on the discrete grid with constant quadrature weight $w_i=\varDelta x$. For the second derivative, we then have
\begin{eqnarray}
\langle\chi_i|\frac{\partial^2}{\partial x^2}|\chi_j\rangle =\left\{
\begin{array}{rl} -\frac{1}{3}\frac{\pi^2}{\varDelta x^2},  & \text{if} \quad i = j, \\
                 -\frac{2}{\varDelta x^2}\frac{(-1)^{i-j}}{(i-j)^2},  & \text{if} \quad i \neq  j.\\
\end{array}\right.
\end{eqnarray}
As a result, we find that position-dependent potentials are diagonal in the sinc-DVR basis on the specified grid, while the kinetic energy operator has a simple analytic form. To illustrate the performance of the sinc-DVR method we may apply it to the one-dimensional harmonic oscillator. Using a step size of half a harmonic trap length, $\varDelta x= 0.5 l$, and using $M=10$, we have to diagonalize a $21\times21$ hamiltonian matrix, for which standard numerical diagonalization routines can be used. Doing this, we get for the lowest twelve eigenvalues an error of less than 1 \% compared to the analytic result $ E_n =(n+1/2) \hbar\bar{\omega}$.

The three-dimensional harmonic oscillator in radial coordinates reduces for the spherically symmetric case to the corresponding one-dimensional problem after considering the functions $\bar{\phi}(r)=\phi(r)/r$. For the range $r \in [0,\infty]$, we can construct so-called wrapped sinc-functions \cite{Groene:93}, given by
\begin{equation}
\chi^{\pm}_j (r) = \chi_j (r) \pm \chi_j (-r).
\end{equation}
Here, we have $j =1, ..,M$ and $(r_i,w_i)=(i \varDelta x, \varDelta x)$ with $i = 1 , ..,M$ for the radial grid and the quadrature weight. Note that the antisymmetric basis functions $\chi^{-}_j (r)$ by construction fulfill the boundary condition $\chi^{-}_j (0)=0$, whereas the symmetric basis functions  $\chi^{+}_j (r)$ have a vanishing first derivative at $r=0$. For the wrapped sinc-function basis sets, an $r$-dependent potential function has again only diagonal matrix elements, while the second derivative gives rise to
\begin{eqnarray}
\langle\chi^{\pm}_i|\frac{d^2}{d r^2}|\chi^{\pm}_j\rangle =\left\{
\begin{array}{rl} -\frac{1}{3}\frac{\pi^2}{\varDelta x^2} \mp\frac{2}{\varDelta x^2}\frac{(-1)^{i+j}}{(i+j)^2},  & \text{if} \quad i = j, \\
      -\frac{2}{\varDelta x^2}\frac{(-1)^{i-j}}{(i-j)^2}\mp\frac{2}{\varDelta x^2}\frac{(-1)^{i+j}}{(i+j)^2},  & \text{if} \quad i \neq  j.\\
\end{array}\right.
\end{eqnarray}
Applying the sinc-DVR method to the three-dimensional harmonic oscillator, we find that for $\ell=0$ the antisymmetric basis set $\chi^{-}$ performs best, while for $\ell=1$ the symmetric basis set $\chi^{+}$ performs best. For higher angular momenta the difference between the two sets is very small. In the following, we choose the $\chi^{-}$ basis for $\ell$ even, and the $\chi^{+}$ basis for $\ell$ odd.

As a result, we are now in the position to evaluate the complete position-dependent $2 \times 2$ matrix ${\bf H}_{n\ell}$ of Eq. (\ref{eq:bdgrad}) in the sinc-DVR basis. Using $M$ grid points, we have for each angular momentum $\ell$ a $2M \times 2M$ matrix to diagonalize. Consider first the balanced case with $h=0$. In the presence of the Cooper-pair condensate $\Delta(r)$, we obtain positive eigenenergies $E_{n\ell}$ for the quasiparticles with spin $+$, and the same negative energies $-E_{n\ell}$ for the quasiparticles with spin $-$. The opposite sign is because we reversed the order of the fermionic fields in Eq.\ (\ref{eq:dvrbas}). However, after the diagonalization we wish to bring the fermionic quasiparticles back in their normal order, so that then all quasiparticles are seen to have positive energies $E_{n\ell}$. For the imbalanced case, we get after the normal ordering of the quasiparticle fields that the dispersions are given by $-\sigma h+E_{n\ell}$ with $\sigma = \pm$ the spin of the quasiparticle and $E_{n\ell}>0$. In the following, all summations are over positive energies $E_{n\ell}$.

For the gap profile $\Delta(r)$, we start with the profile given by the local-density approximation. To find a fully self-consistent solution to the BdG equations, the order parameter should satisfy
\begin{eqnarray}\label{eq:bdggapdiv}
 \frac{\Delta({\bf r})}{V_0} &=&  \langle \phi_{+}(\tau,{\bf r}) \phi_{-}(\tau,{\bf r}) \rangle  \\
&=&-{\sum_{n,\ell}}'\frac{2 \ell +1}{r^2} \bar{u}_{n\ell}(r) \bar{v}_{n\ell}(r) [1-f(E_{n\ell}+h)-f(E_{n\ell}-h)],\nonumber
\end{eqnarray}
where in the second step we used Eq.\ (\ref{eq:bdgtrans}) to write the expectation value in terms of the $\psi_{\sigma}$ fields. Since the action is diagonal in these quasiparticle fields, their correlation functions are readily calculated, resulting in the Fermi occupation numbers $f(\epsilon) = 1/(e^{\beta \epsilon}+1)$ in the second line of Eq.\ (\ref{eq:bdggapdiv}). The primed summation in Eq.\ (\ref{eq:bdggapdiv}) indicates that we sum over positive energies, while The factor $2 \ell+1$ arises from the summation over $m_\ell$. Eq.\ (\ref{eq:bdggapdiv}) contains a divergence due to the use of the contact potential, as was discussed in Section \ref{par:becbcs}. In order to obtain a divergence-free expression \cite{Grasso03a}, we combine Eqs.\ (\ref{eq:t2b-v0}) and (\ref{eq:bdggapdiv}) to
\begin{eqnarray}\label{eq:bdggapfor}
 \frac{\Delta({\bf r})}{T^{\rm 2b}(0)} =-{\sum_{\bf n}}' u_{\bf n}({\bf r}) v^*_{\bf n}({\bf r})[1-f(E_{\bf n}+h)-f(E_{\bf n}-h)]+\frac{1}{\mathcal{V}}\sum_{\bf k} \frac{\Delta({\bf r})}{ 2\epsilon_{\bf k}}.
\end{eqnarray}

In the unitarity limit, we have that the left-hand side of Eq. (\ref{eq:bdggapfor}) is zero. The right-hand side of Eq. (\ref{eq:bdggapfor}) is finite and we can find an analytic expression for the high-energy tail, which can be used to improve the convergence of the numerics. The analytic expression is proportional to $\Delta({\bf r})$ and the proportionality factor $T_{\Lambda}^{-1}({\bf r})$ is calculated from
\begin{eqnarray}\label{eq:bdgtail}
\frac{\Delta({\bf r})}{T_{\Lambda}({\bf r})}&\equiv&-\sum_{{\bf n} } u_{\bf n}({\bf r}) v^*_{\bf n}({\bf r})\theta(E_{\bf n}-E_{\Lambda})+\frac{1}{\mathcal{V}}\sum_{\bf k} \frac{\Delta({\bf r})}{ 2\epsilon_{\bf k}}\nonumber\\
&\approx& -\frac{1}{\mathcal{V}}\sum^{\infty}_{|{\bf k}|=k_{\Lambda}({\bf r})} \frac{\Delta({\bf r})}{ 2(\epsilon_{\bf k}-\mu + V^{\rm ex}({\bf r}))}+\frac{1}{\mathcal{V}}\sum_{\bf k} \frac{\Delta({\bf r})}{ 2\epsilon_{\bf k}}\nonumber\\
&=& \frac{m\Delta({\bf r})}{4\pi^2\hbar^2}\left(2 k_{\Lambda}({\bf r})-k_{\mu}({\bf r})\log\left[\frac{k_{\Lambda}({\bf r})+k_{\mu}({\bf r})}{k_{\Lambda}({\bf r})-k_{\mu}({\bf r})}\right]\right)
\end{eqnarray}
where we considered only high-energy states with energies above $E_{\Lambda}$, since $\theta(x)$ is the Heaviside stepfunction, and where we introduced the wavevectors
$k_{\mu}({\bf r})= \sqrt{2m(\mu-V^{\rm ex}({\bf r}))/\hbar^2}$ and
$k_{\Lambda}({\bf r})= \sqrt{2m(E_{\Lambda}-V^{\rm ex}({\bf r}))/\hbar^2}$.
In the second step of Eq.\ (\ref{eq:bdgtail}), we have used that for high-energy states, the gap profile $\Delta({\bf r})$ and the potential $V^{\rm ext}({\bf r})$ are slowly varying functions, so that we may use the semi-classical (i.e., WKB or local-density) approximation for the coherence factors $u_{\bf n}({\bf r})\approx u_{\bf k}({\bf r})/\sqrt{\mathcal{V}}$ and $v_{\bf n}({\bf r}) \approx v_{\bf k}({\bf r})/\sqrt{\mathcal{V}}$, where $u_{\bf k}({\bf r})$ and $v_{\bf k}({\bf r})$ are the analytically known homogeneous coherence factors evaluated with the local chemical potential $\mu - V^{\rm ex}({\bf r})$, the local order parameter $\Delta({\bf r})$, and the local wavevector ${\bf k}= \sqrt{2m(E_{\bf n}-V^{\rm ex}({\bf r}))/\hbar^2}$. These homogeneous coherence factors were given below Eq.\ (\ref{eq:nsf}) and for high energies they reduce to
\begin{equation}
2 u_{\bf k}({\bf r}) v_{\bf k}({\bf r})=\frac{\Delta({\bf r})}{\sqrt{(\epsilon_{\bf k}- \mu +V^{\rm ext}({\bf r}))^2+\Delta({\bf r})^2}}\approx\frac{\Delta({\bf r})}{\epsilon_{\bf k}- \mu +V^{\rm ext}({\bf r})},
\end{equation}
which was used in Eq.\ (\ref{eq:bdgtail}). By combining Eqs.\ (\ref{eq:bdggapfor}) and (\ref{eq:bdgtail}) we finally obtain the appropriate gap equation in the unitarity limit
\begin{eqnarray}\label{eq:bdggap}
\Delta({\bf r}) ={\sum_{\bf n}}' \frac{u_{\bf n}({\bf r}) v^*_{\bf n}({\bf r})}{T^{-1}_{\Lambda}({\bf r})}\theta(E_{\Lambda}-E_{\bf n}),
\end{eqnarray}
which, together with Eq.\ (\ref{eq:bdgeq}), gives the mean-field BdG approach. The selfconsistent solution for the imbalanced Fermi gas is presented in the next section.

\subsubsection{Results}

\begin{figure}
\begin{center}
\includegraphics[width=0.7\columnwidth]{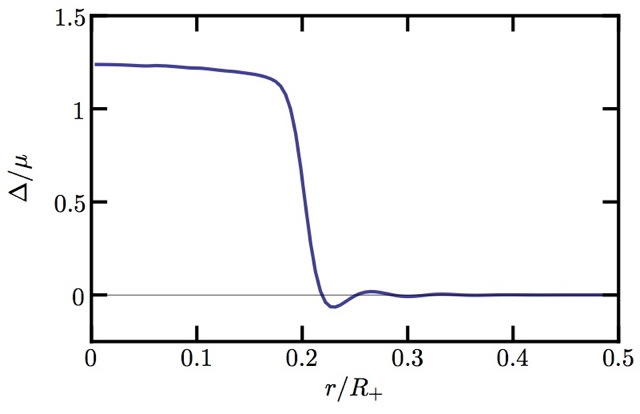}
\caption{\label{fig:oscgap}
    The order-parameter profile for a spin-polarized superfluid in the unitarity limit
    calculated with the Bogoliubov-de Gennes method at zero temperature. The figure was produced using $\mu=32\hbar\bar{\omega}$ and $h=24\hbar\bar{\omega}$, which results in a total number of particles given by $N=3\times 10^4$ and a total polarization of $P=0.95$. The horizontal axis was scaled by $R_+ =\sqrt{2\mu_+/m\bar{\omega}^2}$. The
    interface between the normal and superfluid phase is seen to have a nonzero width, while near the interface oscillations in the order parameter are observed. }
\end{center}
\end{figure}

In Fig.\ \ref{fig:oscgap}, we show the results for the order parameter profile by applying the Bogoliubov-de Gennes equations to the imbalanced Fermi gas. We used $\mu= 32 \hbar \bar{\omega}$, $h= 24 \hbar \bar{\omega}$, $E_\Lambda=50\mu$, $\varDelta x=0.05 l$, and $M=250$. The local-density approximation predicts for these chemcial potentials a sharp interface at a radius of about $r= 2 l$. Around this radius, the BdG approach is indeed seen to give a smooth interface, where the most remarkable feature is the oscillatory behavior of the order-parameter close to the interface. This oscillatory behavior of the gap has been studied for example near superconducting-ferromagnetic interfaces and is known in this case as the proximity effect. It follows from a selfconsistent study of Andreev reflection from the interface \cite{McMill68a,Demler97a,Schaey07a}, where reflected holes have a different wavevector than incoming particles due to a difference in chemical potentials.
Although the proximity effect shares similarities to LO type pairing, since both involve pairing correlations at nonzero center-of-mass wavevectors, we recall that the LO phase requires the spontaneous formation of a oscillatory superfluid, as happens below a Lifshitz point. This is in contrast to the present case, where the oscillations occur due to the presence of a sharp interface, as happens below a tricritical point.

\begin{figure}
\begin{center}
\includegraphics[width=0.7\columnwidth]{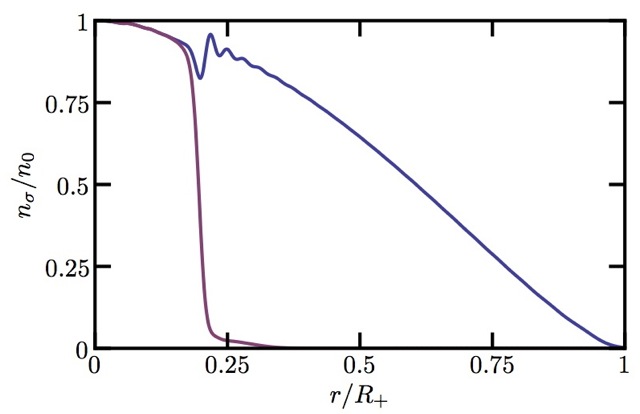}
\caption{\label{fig:oscdens}
    The majority and minority density profiles of the polarized superfluid in the unitarity limit at zero temperature
    calculated using the Bogoliubov-de Gennes method. The same parameters were used as for Fig.\ \ref{fig:oscgap}. The vertical axis is scaled using $n_{0}=n_{+}({\bf 0})$. Expecially in the majority density profile (the upper curve) clear oscillations are seen near the interface. }
\end{center}
\end{figure}

In order to study the resulting density profiles, we use the following equations for the densities
\begin{eqnarray}\label{eq:bddens}
n_{\pm}({\bf r}) &=&  \langle \phi^*_{\pm}(\tau,{\bf r}) \phi_{\pm}(\tau,{\bf r}) \rangle  \nonumber\\
 &=&{\sum_{\bf n}}' v^2_{\bf n}({\bf r}) [1-f(E_{\bf n}\pm h] +u^2_{\bf n} ({\bf r})f(E_{\bf n}\mp h),
\end{eqnarray}
which are similar to Eq.\ (\ref{eq:nsf}) for the homogeneous case. The high-energy tail of Eq.\ (\ref{eq:bddens}) can be simplified to
\begin{eqnarray}
{\sum_{\bf n}}' v^2_{\bf n}({\bf r})\theta(E_{\bf n}-E_{\Lambda})\approx\frac{1}{\mathcal{V}}\sum^{\infty}_{|{\bf k}|=k_{\Lambda}({\bf r})} \frac{\Delta({\bf r})^2}{ 4(\epsilon_{\bf k}-\mu + V^{\rm ext}({\bf r}))^2},
\end{eqnarray}
with the same cutoff $\Lambda$ as for the gap equation. The resulting integral on the right-hand side can be performed analytically.
The chemical potentials lead to a total number of particles of $N=3\times 10^4$ and a total polarization in the trap of $P= 0.95$. The resulting density profiles are shown in Fig.\ \ref{fig:oscdens}. We see that the densities also show the oscillatory behavior close to the interface.  So far, we have ignored the inclusion of self-energy effects, so that we do not have the correct equations of state for the normal state and superfluid state. As a result, the interface is not in the correct position compared to experiments. In principle, we could include the information about the interaction effects from Monte-Carlo calculations in a similar way as for the Landau-Ginzburg approach, but then the densities and the gap would have to be determined by the appropriate derivatives to the thermodynamic potential, so that Eqs. (\ref{eq:bdggap}) and (\ref{eq:bddens}) would not be valid anymore. Such a generalized BdG approach was put forward in Ref.\ \cite{Bulgac12a}, where also a DVR-based method was used to solve the BdG equations. Since the prediction for the proximity effect is so far only based on the mean-field BdG equations, the existence of the effect in the unitarity limit remains uncertain until confirmed or invalidated by experiments.

\section{Discussion and outlook}
\label{sec:fin}

In this review, we have treated the imbalanced atomic Fermi gas with strong attractive interactions, for which pairing plays an important role. We have focussed mainly on the population-imbalanced case, for which in recent years many exciting experiments and a large amount of theoretical calculations have been preformed. The large amount of interest in the topic started in the beginning of 2006, when two experimental groups obtained a full control over the spin imbalance or polarization $P$ in an ultracold atomic Fermi gas \cite{Zwierl06a,Partri06a}. This made it possible to study experimentally in detail the fundamental question: what happens to a superfluid Fermi mixture upon frustrating the pairing by increasing the spin polarization? We have seen that in the atomic Fermi gas there are only two-body attractions between particles of opposite spin, so that by increasing the polarization, also the number of particles that cannot pair increases. As a result, we argued that there must be a transition from the superfluid to the normal state somewhere between zero and full spin polarization. But when? And how? One of the experimental and theoretical findings was that these questions depend on the temperature of the Fermi gas. Namely, when the temperature is above the tricritical point, the transition is smooth (or second-order) and occurs at lower polarizations, while below the tricritical point, the transition is sharp (or first-order) and occurs at higher polarizations. We have seen that this behavior can be qualitatively explained by mean-field theory, where we mapped out the phase diagram both for the homogeneous and the trapped case. However, on a  quantitative level the mean-field critical polarization, or Chandrasekhar-Clogston limit, turned out to be very different from the experimental results.

This problem was solved by Monte-Carlo studies of the strongly-interacting normal state and the superfluid state at unitarity, which resulted also in very good quantitative agreement with experiments \cite{Lobo06a}. Although in this review we have not reviewed Monte-Carlo techniques, we have discussed several other ways to go beyond mean-field theory, such as including Gaussian order-parameter fluctuations (the random-phase approximation), diagrammatic many-body scattering calculations, and renormalization group calculations.  Although these methods are in principle less accurate than Monte-Carlo simulations, they have also some advantages. First, they require much less numerical effort. Second, they show more clearly the relevant physics involved. And third, they can be used at nonzero temperatures and polarizations, while Monte-Carlo simulations have so far been restricted to either zero temperature or to the balanced case. Together, the theoretical approaches allow for a rather complete present understanding of the equilibrium properties of strongly-interacting imbalanced Fermi gases. This holds both for the homogeneous case, and for the trapped case with the use of the local-density approximation.

In this review, we also calculated the mean-field phase diagram for the mass-imbalanced case of the experimentally relevant ${}^6$Li-${}^{40}$K mixture, which not only contains a tricritical point in the unitarity limit, but also a Lifshitz point, so that supersolidity is expected to occur. Key open questions in this respect are the precise phase structure below the Lifshitz point and a determination of the crystalline structure of the various supersolid phases that are realized. Moreover, the Sarma regime occupies a large part of the phase diagram for the $^{6}$Li-$^{40}$K mixture, so that experimental observation of a gapless superfluid with (smoothened) Fermi surfaces might be within reach. Finally, we treated the trapped inhomogeneous Fermi gas beyond the local-density approximation in order to study in more detail the superfluid-normal interface that occurs below the tricritical point. With a Landau-Ginzburg approach we calculated the surface tension of the interface, while with the Bogoliubov-de Gennes approach also order-parameter oscillations near the interface were observed due to the proximity effect. However, in order to compare these theoretical results for the interface in detail with experiments more accurate experimental data for the interface would be required.

Although significant steps have been made in understanding imbalanced Fermi gases, there is still room for more research. From the theoretical side, further improvements could be made by having also Monte-Carlo data for the imbalanced  Fermi gas at nonzero temperatures, and in particular for the tricritical point. Another theoretical improvement of interest would be a more detailed account of screening effects of the interaction by particle-hole excitations, since screening is usually either neglected in analytical approaches, or treated in simple approximations. The latter typically ignore the precise momentum and frequency dependence of the particle-hole excitations. From the experimental side, a more careful study of the superfluid-normal interface is desirable. When the local-density approximation applies, this would then allow for a more careful experimental mapping of the homogeneous phase diagram, which up to now contains very few points with large error bars \cite{Shin06a}. Seen the experimental accuracy that is currently being reached for balanced Fermi gases \cite{Nascim10a,Ku12a}, more precise data for the phase diagram of the imbalanced Fermi gas should be within reach. This would then also result in a more precise determination of the tricritical point. If the interface is studied more accurately, then it is also natural to check if the proximity effect can be observed in a cold-atom system.

Furthermore, an experimental determination of the surface tension would be a test for our knowledge of the superfluid state beyond equilibrium, since the surface tension probes the unstable barrier of the superfluid grand potential as a function of the order parameter. It would also be interesting to study various other non-equilibrium aspects for imbalanced Fermi gases. These systems contain a first-order phase transition, so that they are ideally suited to study nucleation in a quantum system. Novel topics of interest for imbalanced Fermi systems include transport properties, artificial gauge fields, spin-orbit coupling, optical lattices, $p$-wave interactions, reduced dimensions, and so on. These topics are all expected to become experimentally available for imbalanced Fermi gases, or have already recently been realized, so that imbalanced Fermi gases are expected to keep us interested for many years to come.

\section*{Acknowledgements}

Many people have directly or indirectly contributed to the work we present in this review. We want to thank the members of the Quantum Fluids Group in Utrecht, in particular Jildou Baarsma, Jeroen Diederix, Rembert Duine, Masud Haque, Pietro Massignan, and Mathijs Romans for many fruitful collaborations and stimulating discussions. More precisely, Masud Haque, Mathijs Romans, and Jildou Baarsma were directly involved in the research presented in Section \ref{sec:mf}, while much of the discussion in Section \ref{sec:inh} follows from the PhD thesis of Jeroen Diederix. Over the years we have profited enormously from the close collaboration with Randy Hulet, whose physical insight and experimental expertise have been crucial in shaping our understanding of imbalanced Fermi gases. In this respect, we also want to mention his collaborators, and in particular Guthrie Partridge. We are grateful to Wolfgang Ketterle, Martin Zwierlein, Yong-Il Shin, and Sylvain Nascimb\`{e}ne for stimulating discussions and for sharing their experimental data. We also wish to thank Carlos Lobo, Alex Gezerlis and Evgeni Burovski for sharing their Monte-Carlo data. We acknowledge many insightful discussions with Ad van der Avoird, Gerrit Groenenboom, Achilleas Lazarides, Lih-King Lim, Achim Rosch, and Bert van Schaeybroeck. We thank Gerard Meijer for general support. Finally, we thank P{\"{a}}ivi T{\"{o}}rm{\"{a}} and Leo Radzihovsky for useful comments, just as Jildou Baarsma and Randy Hulet who carefully read large parts of the manuscript. Koos Gubbels acknowledges financial support by the Alexander von Humboldt Foundation, and by the European Community's Seventh Framework Program ERC-2009-AdG under grant agreement 247142-MolChip. This work is also supported by the Stichting voor Fundamenteel Onderzoek der Materie (FOM) and the Nederlandse Organisatie voor Wetenschappelijk Onderzoek (NWO).

\bibliographystyle{elsarticle-num}
\bibliography{library_new}

\end{document}